\documentclass[a4paper]{article}

\usepackage{xcolor}

\usepackage{graphicx}
\usepackage[utf8]{inputenc}
\usepackage{float}
\usepackage{multicol}
\usepackage{geometry}
\usepackage[T1]{fontenc}
\usepackage{bookman}

\title{Appearance of Keplerian discs orbiting on both sides of a reflection-symmetric wormhole}
\author{Jan Schee and Zden\v{e}k Stuchl\'{i}k\\
     \small{Research Centre for Theoretical Physics and Astrophysics}, \\
	   \small{Institute of Physics},\\
	   \small{Silesian University in Opava}\\
	   \small{Bezru\v{c}ovo n\'{a}m. 13, CZ-746 01 Opava, Czech Republic}
}

\newcommand{\diff}{\mathrm{d}}

\begin{document}
\maketitle
\begin{abstract}	
We construct optical appearance and profiled spectral lines of Keplerian discs with inner edge at the innermost circular geodesic located on both sides of the reflection-symmetric Simpson-Visser wormhole, in dependence on its parameter and inclination angle of a distant observer. We demonstrate significant differences in appearance of the discs on the our side and the other side of the wormhole. Large part of the other-side disc is always in the dark region of the image of the disc orbiting on the our side, enabling thus a simple distinguishing in observations. The profiled spectral lines generated by the disc on the other side (our side) demonstrate strong (weak) dependence on the spacetime parameter, and weak (strong) dependence on the inclination angle; they have also different shape, giving thus other clues to clearly distinguish in observations reflection-symmetric wormholes as alternatives to black holes. 
\end{abstract}

\section{Introduction}

Recent sophisticated radio-interferometry observational systems, Event Horizon Telescope (ETH) \cite{Doe-etal:Science:2012,EHT-etal:APJ:2019,Wiel-etal:ApJ:2020} and GRAVITY \cite{GRAVITY:2017}, enable detailed insight into the innermost region of accretion discs orbiting supermassive black holes as SgrA* in the Galaxy centre, or those in the centre of M87 galaxy. Fascinating detailed pictures of accreting matter in regions close to the assumed black hole event horizon were obtained by GRAVITY in the case of SgrA* \cite{GravityCol:2018:AA:}, with possible consequences discussed in \cite{tur-etal:2020:ApJ,stu-etal:2020:UNIV}, and by EHT in the case of the central region of M87 in \cite{EHT-etal:APJ:2019}, with possible implications discussed in \cite{Wiel-etal:ApJ:2020}. These pictures reflecting the shadow of the assumed central rotating Kerr black hole enable precision tests of General Relativity in the strongest field limit, and estimations of 
possible modifications of the black hole shadow extension and shape implied by alternative theories of gravity. Of course, such detailed pictures in principle allow for discovery of the black hole mimickers, or their alternatives as superspinars \cite{Gim-Hor:2009:PhysLetB:,Stu-Hle-Tru:2011:CLAQG:,Stu-Sche:2012:CLAQG:} when the shadow qualitatively (topologically) differs from those corresponding to black holes \cite{Stu-Sche:2010:CLAQG:} -- furthermore, the superspinars demonstrate astrophysical phenomena extraordinary even in comparison with those related to black holes \cite{Stu:1980:BAC:,Stu-Sche:2013:CLAQG:,Bla-Stu:2016:PHYSR4:,Stu-Bla-Sche:2017:PHYSR4:}. Of course, a special interest represent the wormholes giving optical phenomena similar to those generated by black holes, but with an extraordinary exception related to possible observed structures located "on the other side" of the wormhole. If images of such structures are located inside the region corresponding to the shadow of the wormhole (related to the assumed black hole shadow), we obtain a clear signature to a wormhole, i.e., a horizon-less object similar to black holes except the innermost black hole dynamic region. 

The idea of wormholes was introduced by Ellis \cite{Elli:1973:JMathPhys:} for the Einstein-scalar theory with phantom scalar field. Later it was developed in the standard General Relativity by Morris and Thorne \cite{Mor-Tho:1988:AmJPhys:,Mor-Tho-Yur:1988:PhysRevLet:}, as traversable spacetime tunnels connecting distant parts of the Universe, or connecting different universes, and enabling transfer of massive objects between them. At present time, there is a large variety of wormhole models based on the General Relativity, or some alternative gravity theories. 

A special simple case of the wormhole solutions are the reflection-symmetric wormholes constructed by Visser \cite{Vis:1989:NuclPhysB:,Poi-Vis:1995:PHYSR4:} by the cut-and-paste method, connecting two Schwarzschild spacetimes by a spherical shell of extraordinary matter violating the weak energy condition for positive energy density. The stress energy tensor with negative energy density violating the weak energy condition is located at the junction of these two spacetimes, and this is the agent that guarantees the correctness of the Einstein gravitational equations for stable traversable wormholes \cite{Vis:1989:NuclPhysB:}. On the other hand, 
in the high-dimensional general relativity \cite{Svi-Tah:2020:EPJC:}, or in some alternative gravity theories \cite{Gra-Wil:2007:PHYSR4:,Har-Lob-Mak-Sus:2013:PHYSR4:}, such extraordinary form of the stress energy tensor is not necessary to construct wormhole solutions. Moreover, recent results demonstrate possibility of traversable wormholes constructed without extraordinary forms of matter or alternative gravity if fermions generate negative Casimir energy \cite{Mal-Mil-Pop:2018:arXiv:} -- such solutions were generated in the Einstein-Dirac theory, or the Einstein-Maxwell-Dirac theory \cite{Bla-Sal-Kno:2020:EPJC:,Bla-Sal-Kno-Rad:2021:PhysRevLet:}. 

Another approach was initiated recently by Simpson and Visser \cite{Sim-Vie:JCAP:2019:} with metric reflecting transitions between various kinds of spacetimes due to variation of the spacetime parameter, namely from the Schwarzschild black hole, through a regular black hole with a bounce state instead of singularity, to a traversable wormhole state; the bounce that occurs behind the black hole horizon is called black bounce, being similar to the idea of the black universe \cite{Bro-Mel-Deh:2007:GenRelGrav:}. The transitions from the black hole to the wormhole state were investigated in \cite{Chur-Stu:2020:CLAQG,Bro-Kon:2020:PHYSR4:}. 

The basic observational information on the wormhole existence is, of course except the transport effect, related to the optical phenomena. Therefore, the effects of the weak \cite{Abd-Jur-Ahm-Stu:2016:AstroSpaSci:}, or strong lensing of radiating objects by the wormhole are widely discussed, but the most efficient is the measurement of the wormhole shadow as it reflects the region of the strongest gravity being related to the photon spheres of the spacetime \cite{Wie-Hor-Vin-Abr:2020:PHYSR4:}. 

Recently, a clear wormhole signature by a double shadow was demonstrated \cite{Wie-Hor-Vin-Abr:2020:PHYSR4:} for reflection-asymmetric wormholes connecting Reissner-Nordstrom spacetimes having distinct mass and charge parameters, as there could be different photon spheres formed on our (observer) side, and the other side of the wormhole. The construction of thin disk images in Kerr-like wormhole spacetime was discussed in \cite{Paul-etal:2020:JCAP:}. Interesting analysis of observing wormhole via interaction of a body in vicinity of the wormhole in our part of the universe with the body being present on the other side of the wormhole is presented in \cite{Dai-Stoj:PhRvD:2019:}. A general overview of possible astrophysical wormhole signatures is given in \cite{Bam-Stoj:UNI:2021:}.

In the present paper we are able to demonstrate clear signatures of the standard reflection-symmetric wormholes, if we assume radiating Keplerian discs located on both sides of the wormhole. We study the case of recently proposed Simpson-Visser wormhole with Keplerian discs placed on both sides of the wormhole, and having the inner edge on the innermost stable circular geodesic of the spacetime; in the Appendix we shortly discuss properties of the stress-energy tensor related to this wormhole solution. We construct the optical appearance of both Keplerian discs, and the profiled spectral lines general by these discs, as observed by distant observers on our side of the wormhole. Our results indicate significant differences for both images and spectral lines.

\section{Wormhole geometry and its embedding diagram}

Static and spherically symmetric metric describing the reflection-symmetric Simpson-Visser traversible wormhole reads  (in geometric units with $c=G=1$) \cite{Sim-Vie:JCAP:2019:}
\begin{equation}
	\diff s^2=-f(r)\diff t^2 + \frac{1}{f(r)}\diff r^2 + h(r)\left(\diff\theta^2+\sin^2\theta\,\diff\phi^2\right)\label{wh}	
\end{equation}
where 
\begin{eqnarray}
	h(r)&=&r^2+a^2,\\
	f(r)&=&1 - \frac{2M}{\sqrt{h(r)}}
\end{eqnarray}
with the parameter $a$ satisfying condition $a>2$. We distinguish two universes - lower ($r>0$) - the our side, and upper ($r<0$) - the other side.

\begin{figure}[H]
	\begin{center}
		\includegraphics[scale=0.5]{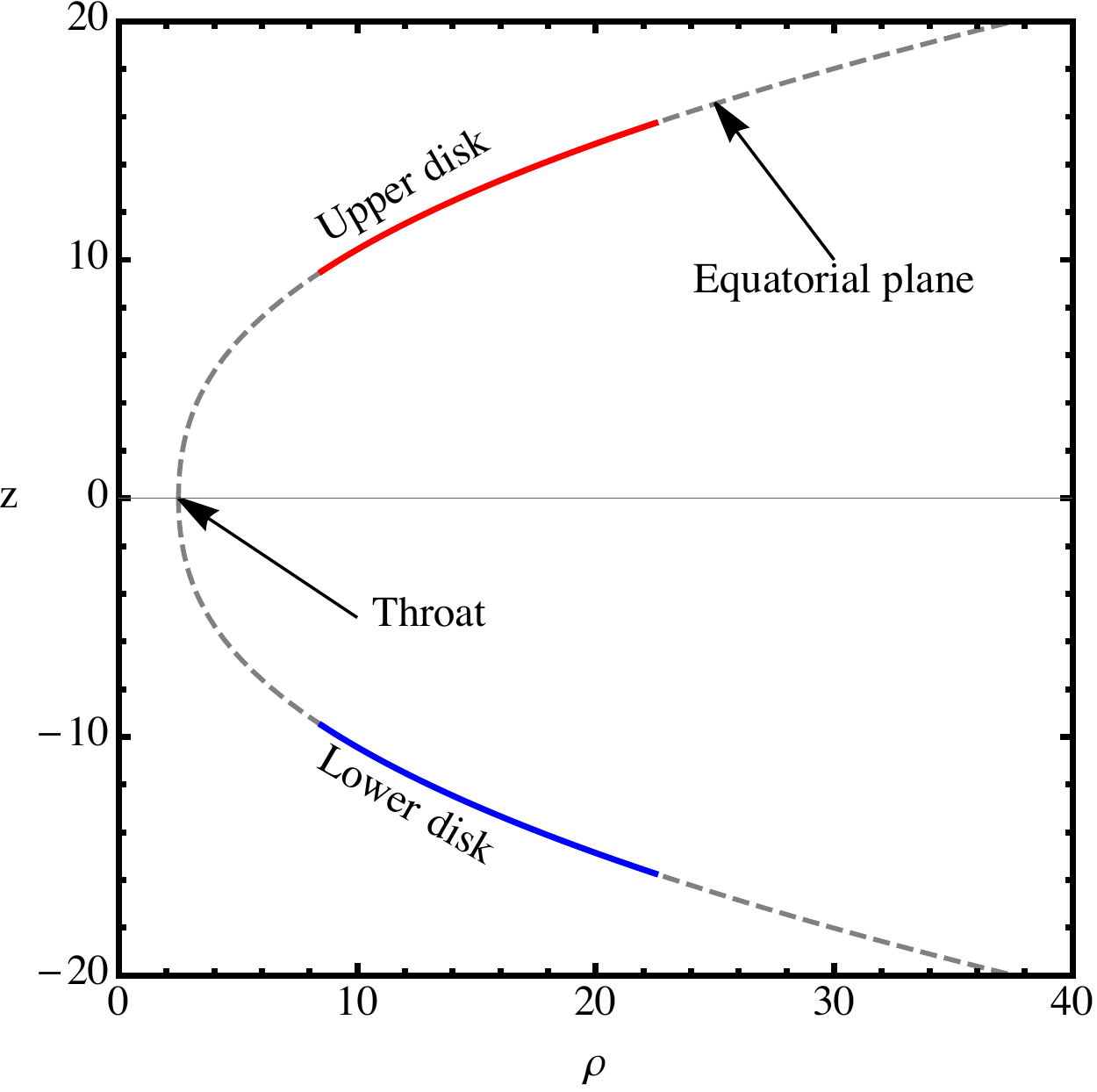}
		\caption{The plot illustrating behavior of embedding function $z=z(r)$ given by formula (\ref{embedding}) for a representative value of $a=2.5$. Dashed line represents embedded equatorial plane where two Keplerian disks are located in the upper universe (red) and lower universe (blue).\label{figembedding}}
	\end{center}
\end{figure}

The wormhole Simpson-Visser spacetime curvature can be well illustrated by the standard embedding diagram -- see e.g. its significance in \cite{Stu-Hle:1999:PHYSR4:}. We embed 2D surface, representing equatorial plane, described by interval
\begin{equation}
	\diff s^2=\frac{1}{f(r)}\diff r^2+(r^2+a^2)\diff\phi^2
\end{equation} 
into the 3D Euclid space with geometry
\begin{eqnarray}
	\diff s_E^2&=&\diff\rho^2+	\diff z^2+\rho^2\diff\alpha^2\nonumber\\
			&=&\diff\rho^2\left[1+\left(\frac{\diff z}{\diff\rho}\right)^2\right]+\rho^2\diff\alpha^2.
\end{eqnarray}
Identifying $\alpha\equiv\phi$ and $\rho^2\equiv r^2+a^2$ we get the equation governing the embedded surface
in the form 
\begin{equation}
	\left(\frac{\diff z}{\diff \rho}\right)^2=\frac{1}{f(r)}\left(\frac{\diff r}{\diff\rho}\right)^2-1
\end{equation}
that can be transformed to 
\begin{eqnarray}
	\left(\frac{\diff z}{\diff r}\right)^2&=&\frac{1}{f(r)}-\left(\frac{\diff\rho}{\diff r}\right)^2\nonumber\\
	&=&\frac{h^{3/2}(r)-r^2[h^{1/2}(r)-2]}{h(r)[h^{1/2}(r)-2]}.
\end{eqnarray} 
Integrating last formula we obtain the embedding formula $z=z(r)$ in the form
\begin{equation}
	z(r)=\int_{0}^{r}\sqrt{\frac{h^{3/2}(r)-r^2[h^{1/2}(r)-2]}{h(r)[h^{1/2}(r)-2]}}\diff r\label{embedding}
\end{equation}
which we can numerically integrate. We illustrate the behavior of the Simpson-Visser wormhole embedding formula $z(r)$ in Figure \ref{figembedding}. Naturally, in the reflection-symmetric spacetime, the Keplerian discs are distributed symmetrically. 


\section{Equations of test particle and photon motion}
The equations of motion are found using the standard Hamilton-Jacobi (H-J) method. The H-J action function $S$ satisfies the H-J equation
\begin{equation}
	-m^2=-\frac{1}{f(r)}\left(\frac{\partial S}{\partial t}\right)^2 + f(r)\left(\frac{\partial S}{\partial r}\right)^2 + \frac{1}{h(r)}\left(\frac{\partial S}{\partial \theta}\right)^2+\frac{1}{h(r)\sin^2\theta}\left(\frac{\partial S}{\partial \phi}\right)^2
\end{equation} 
where the H-J action function is connected to the test particle 4-momentum $p^\mu=\diff x^\mu/\diff\lambda$ by the relation 
\begin{equation}
	p_\mu\equiv \frac{\partial S}{\partial x^\mu}.
\end{equation}
Assuming solution in separated form, $S\equiv S_t+S_r+S_\theta+S_\phi$, together with existence of the two constants of motion determined by the spacetime symmetries, axial angular momentum $L\equiv p_\phi$ and energy $E\equiv -p_t$, the equations of motion read 
\begin{eqnarray}
	\left(p^r\right)^2&=&E^2 - f(r)\left(m^2+\frac{L^2+Q}{h(r)}\right),\\
	\left(p^\theta\right)^2&=&\frac{1}{h^2(r)}\left(Q-L^2\cot^2\theta\right),\\
	p^t&=&\frac{E}{f(r)},\\
	p^\phi&=&\frac{L}{h(r)\sin^2\theta}.
\end{eqnarray}
Q is the integration constant having the meaning of the total angular momentum of the particle. 

In case of photons the mass parameter is zero, $m=0$. Further, for the sake of numerical tractability during ray-tracing procedure we utilize new latitudinal coordinate $\mu\equiv \cos\theta$ that transforms the equations of motion to the form 
\begin{eqnarray}
	\left(p^r\right)^2&=&1 - f(r)\frac{l^2+q}{h(r)},\label{eqr}\\
	\left(p^\mu\right)^2&=&\frac{1}{h^2(r)}\left(q-\mu^2(l^2+q)\right),\label{eqm}\\
	p^t&=&\frac{1}{f(r)},\label{eqp}\\
	p^\phi&=&\frac{l}{h(r)(1-\mu^2)}\label{eqt}
\end{eqnarray}
where we also re-parametrized the equations of motion by $\lambda\rightarrow E\,\lambda$, and introduced the standard impact parameters $l\equiv L/E$ and $q\equiv Q/E^2$.

\section{Structure of circular geodesics}
Without loss of generality we consider motion of test particles in the equatorial plane ($\theta=\pi/2$, $\mu=0$). We define the effective potential for massive test particles
\begin{equation}
	V^m_{\mathrm{eff}}(r)=f(r)\left[m^2+\frac{L^2}{h(r)}\right]
\end{equation}
where $m>0$ for massive particles and $m=0$ for photons.
The circular orbits are found from the conditions 
\begin{equation}
	\frac{\diff V^m_{\mathrm{eff}}}{\diff r}=0.
\end{equation}
The stability of the orbit is determined by the sign of $V''_{\mathrm{eff}}\equiv\diff^2V_{\mathrm{eff}}/\diff r^2$
\begin{equation}
	V''(r_0)
	\left\{\begin{array}{cc}
		<0 & \textrm{stable}\\
		>0 & \textrm{unstable}\\
		=0 & \textrm{marginally stable}
		\end{array}\right.
\end{equation}
\begin{figure}[H]
	\begin{center}
		\includegraphics[scale=0.5]{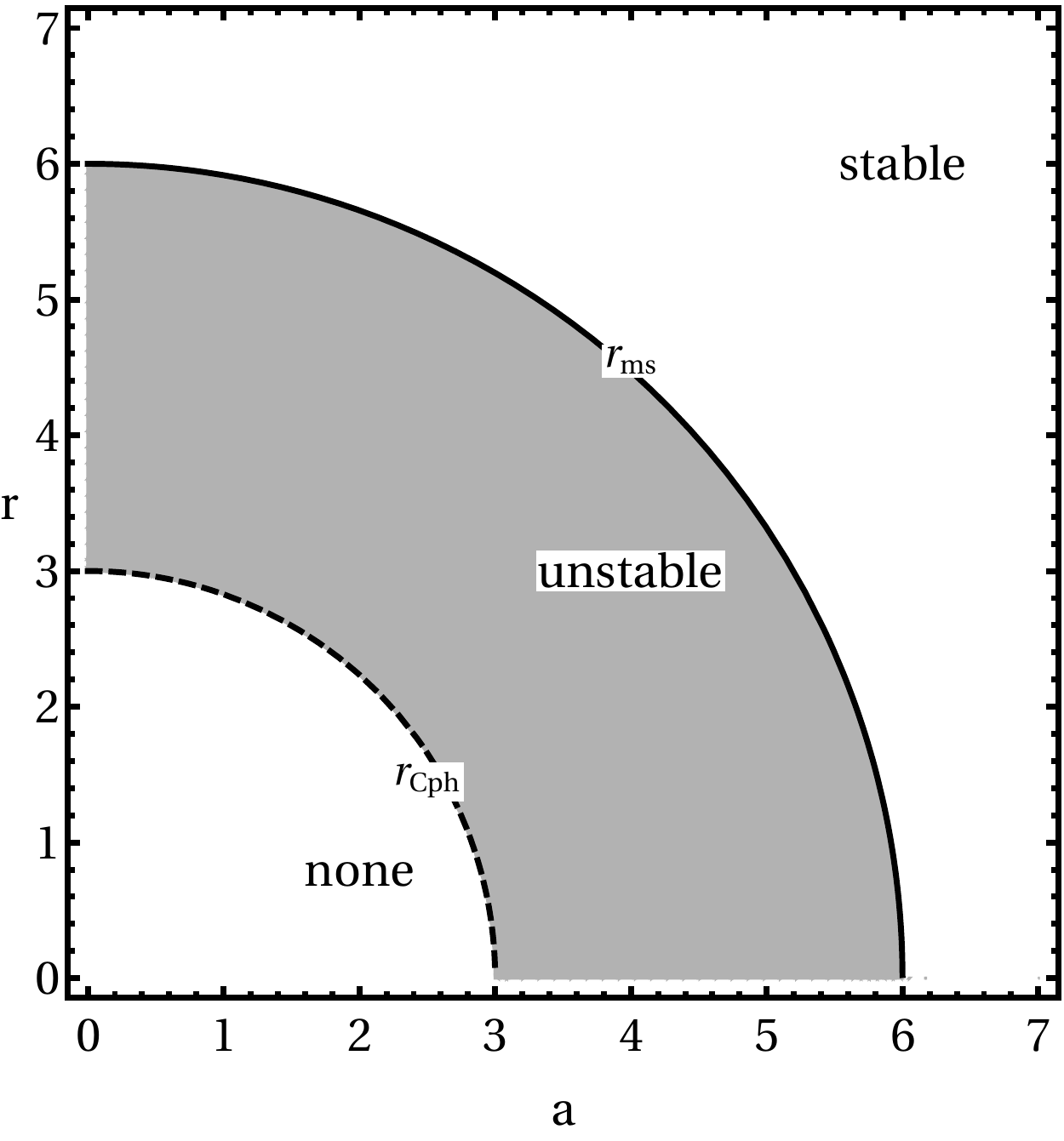}
		\caption{The structure of circular orbits in Wormhole spacetime.}\label{circorbs}
	\end{center}
\end{figure}

In the discussion we consider $r>0$, in the case of $r<0$ the discussion is equivalent. Let's introduce angular frequency by $\Omega\equiv u^\phi/u^t$. The Keplerian circular orbit angular frequency is then given by formula
	\begin{equation}
	\Omega_K^2(r)=\frac{M}{\left(r^2+a^2\right)^{3/2}}.\label{OmegK}
\end{equation}
Corresponding energy and angular momentum of the circular orbit reads
\begin{equation}
E^2=\frac{f^2(r)h_{,r}(r)}{f(r)h_{,r}-f_{,r}(r)h(r)}
\end{equation} 
and
\begin{equation}
L^2=\frac{f_{,r}(r)h^2(r)}{f(r)h_{,r}-f_{,r}(r)h(r)}.
\end{equation} 
The circular orbits exist where $E^2>0$ and $L^2>0$. There are two curves, $r_{\textrm{ms}}$ and $r_{\textrm{cph}}$, dividing the $r-a$ parameter space into three regions (see Fig.\ref{circorbs}). The first curve $r_{\textrm{ms}}$ represents marginally stable orbits determined from the conditions $V'_{\textrm{eff}}=V''_{\textrm{eff}}=0$ reading
\begin{equation}
r_{\mathrm{ms}}=\pm\sqrt{36M^2-a^2}\label{rms}
\end{equation}
with corresponding energy
\begin{equation}
E_{\mathrm{ms}}=\sqrt{8/9}
\end{equation}
and corresponding angular momentum 
\begin{equation}
L_{\mathrm{ms}}=2\sqrt{3}.	
\end{equation}
Notice that both the energy and angular momentum of the marginally stable circular geodesics in independent of the parameter $a$, being equal to the value corresponding to the Schwarzschild black hole. 

The second curve, $r_{\textrm{cph}}$ follows from the condition 
\begin{equation}
	f(r)h_{,r}-f_{,r}(r)h(r)=0
\end{equation}
which defines maximum of the effective null particle ($m=0$) effective potential and is also implies that massive particle angular momentum and covariant energy are both infinite. The quick calculation reveals explicit formula for $r_{\textrm{cph}}$ reading
\begin{equation}
	r_{\mathrm{cph}}=\pm\sqrt{9M^2-a^2}\label{cph}.
\end{equation}
Clearly, there exist photon circular orbits only for parameter $a<3M$ and the corresponding impact parameter reads
\begin{equation}
	l^2_{ph}=\frac{h(r_{\mathrm{cph}})}{f(r_{\mathrm{cph}})}=27 M^2
\end{equation}
which is independent on parameter $a$ and has the same form as for photon circular orbit around Schwarzchild black hole.

Now, the region where \emph{stable} circular orbits are located is bounded by curve $r_{\mathrm{ms}}=r_{\mathrm{ms}}(a)$ (formula (\ref{rms})). The \emph{unstable} circular orbits region lays between curves $r_{\mathrm{ms}}$ and $r_{\mathrm{cph}}$  (formula (\ref{cph})). There are no circular orbits below  curve $r_{\mathrm{cph}}$.

\section{Radiating Keplerian disc}

In order to reflect the optical phenomena related to the discs distributed on both sides of the reflection-symmetric wormhole, we model the radiating thin disc by Keplerian disc \cite{Nov-Tho:1973:BlaHol:} which consists of Keplerian rings with radii in the interval $[r_{\mathrm{ms}},r_{\mathrm{out}}]$. The radial profile of the Keplerian disc angular velocity is given by the formula (\ref{OmegK}).

In order to have representative demonstrations of the wormhole influence on the optical phenomena, we construct the bolometric flux image of the Keplerian discs on both sides of the wormhole, the map of the frequency shift of the discs images, and finally we construct profiled spectral line generated by the both discs. In order to realize these goals, we have to summarize the basic optical effects governing the constructions of the mentioned optical phenomena (see, e.g., \cite{Sche-Stu:2009:IJMPD:}). 

\subsection{Motion of the radiated photons and their frequency shift}

Photons emitted by the Keplerian discs follow null geodesics that are governed by the equations of motion (\ref{eqr})-(\ref{eqt}). For purpose of the numerical integration we turn equations (\ref{eqr}) and (\ref{eqm}) into second order differential equations for $r$ and $m$, having the form  
\begin{eqnarray}
	\frac{\diff p^r}{\diff \lambda}&=&-\frac{(l^2+q)}{2h^2(r)}\left(f'(r)h(r)-f(r)h'(r)\right),\label{eqr2}\\
	\frac{\diff p^\mu}{\diff\lambda}&=&-\frac{(l^2+q)m}{h^2(r)}-\frac{h'(r)}{h(r)}p^r p^\mu\label{eqm2} . 
\end{eqnarray}
In such an arrangement, we do not need to take care about the turning points of the photon motion. The set of differential equations (\ref{eqp}), (\ref{eqt}), (\ref{eqr2}), and (\ref{eqm2}) are solved for initial conditions
\begin{eqnarray}
	p^r_0&=&-\sqrt{1-\frac{f(r_o)}{h(r_o)}(l^2+q)},\\
	p^\mu_0&=&\pm\sqrt{q-\mu_o^2(l^2+q)}/h(r_o)
\end{eqnarray}
setting observer (the initial point of the integration) at $r_o,\mu_o$  and setting photon impact parameters to $l,\,q$. The initial values of temporal and azimuthal null geodesics coordinates are $t=\phi=0$. 

For the given solution the intersection between null geodesics and equatorial plane is found. If the corresponding radial coordinate  value of this intersection $r_e$ falls into interval $[r_{\textrm{ms}},r_{\textrm{out}}]$ the frequency shift $g$ is determined. The frequency shift function $g$ for the photons radiated by the discs and finished in the apparatus of a distant observer is defined by 
\begin{equation}
	g\equiv \frac{E_o}{E_e}=\frac{\left.k_\alpha\,u^\alpha\right|_o}{\left.k_\alpha\,u^\alpha\right|_e}.
\end{equation} 
The observer is assumed to be static relative to coordinates $r,\theta,\phi$, i.e., his 4-velocity vector reads
\begin{equation}
	\vec{u}_o=u^t_o\,\vec{e}_t
\end{equation} 
while emitter is, of course, following circular Keplerian orbit with 4-velocity vector
\begin{equation}
	\vec{u}_e=u^t_e\,\vec{e}_t+u^\phi_e\,\vec{e}_\phi.
\end{equation}
The frequency shift of photon emitted from the orbiting frame and received in the static distant frame reads
\begin{equation}
	g=\frac{\sqrt{f(r_e)-(r_e^2+a^2)\Omega_e^2}}{1-l\,\Omega_e}
\end{equation}
where we use the Keplerian frequency of the circular  orbit at radius $r_e$, i.e. $\Omega_e=\Omega_K(r_e)$ (formula (\ref{OmegK})).


\subsection{Novikov-Thorne discs}

First, we consider the Keplerian discs to be modeled in the framework developed by Novikov and Thorne \cite{Nov-Tho:1973:BlaHol:} as a thermal radiation governed locally by friction effects in the Keplerian discs. The continuum with time averaged flux (or just flux) from the disk surface can be given in the form (\cite{Nov-Tho:1973:BlaHol:,Tho-Page:1974:ApJ:})
\begin{equation}
	F(r)=\frac{\dot{M}}{4\pi}\frac{1}{\sqrt{r^2+a^2}}\Phi(r)
\end{equation}
where 
\begin{equation}
	\Phi(r)\equiv -\frac{\Omega_{,r}}{(E-\Omega\,L)}\int^r_{r_{ISCO}}(E-\Omega\,L)\,L_{r}\diff r
\end{equation}
The observer detect the flux modified by the frequency shift parameter $g$, given in the form
\begin{equation}
	F_o(r)=g^4\,F(r).
\end{equation}
We also give the map of the frequency shift alone, i.e., its distribution across the disc area. 

\subsection{Profiled spectral lines}
Second, we model the radiation as monochromatic, corresponding e.g. to the Fe spectral line, distributed across the whole disc and assume that its emissivity changes according to empirical power law
\begin{equation}
	I_e(r,g)=I_{e(0)}(g)r^{-p}
\end{equation}
where we assume the Gaussian bell profile for $I_{e(0)}(g)$, i.e.,
\begin{equation}
	I_{e(0)}(g)=I_0\exp\left(-\frac{1}{\sigma^2}(1-g)^2\right).
\end{equation}
The profiled spectral line is defined as a plot of specific flux $F_\nu$ vs observed frequency or frequency shift $g$. It is then given by the relation 
\begin{equation}
	F_\nu(g)=\int_{\Delta\Pi}g^3 I_e(r,g)\diff\Pi\approx \sum_{i}^{\mathrm{all\,\,images}} g_i^3I_e(r_i,g_i)\Delta\pi
\end{equation}
where $\Delta\Pi$ is the solid angle subtended by the disk and $\Delta\pi$ is solid angle subtended by single image (or by a pixel). To construct profiled spectral lines we apply the method presented in \cite{Sche-Stu:2009:GenRelGrav}. 

\section{Results}
Here we present results of our numerical calculations giving construction of the Keplerian disc images as seen by a distant observer in the lower (our side) universe (having such an observer in the upper universe, we would obtain the same results due to the reflection--symmetry of the wormhole and the Keplerian discs). We assume two disc settings. First, the disc is located in the upper universe; second, the disc is located in the lower universe. In both cases the disc extends from $r_{\mathrm{ms}}$ (formula (\ref{rms})) to $r_{\mathrm{out}}=\pm100$M (the plus sign refers to the lower universe, the minus sign refers to the upper universe).

\begin{figure}[H]
	\begin{center}
		\begin{tabular}{cc}
			\includegraphics[scale=0.45]{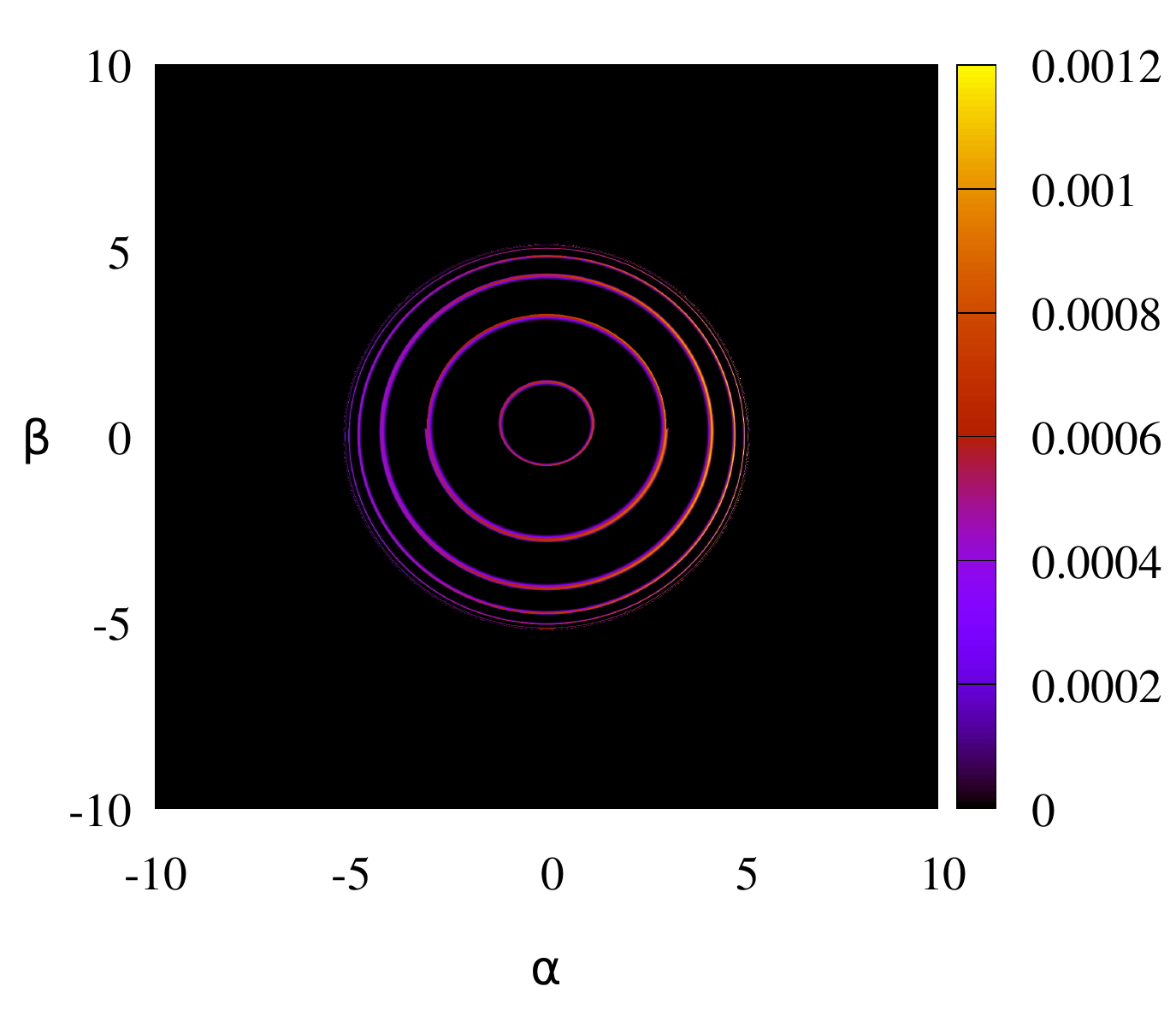} &	\includegraphics[scale=0.45]{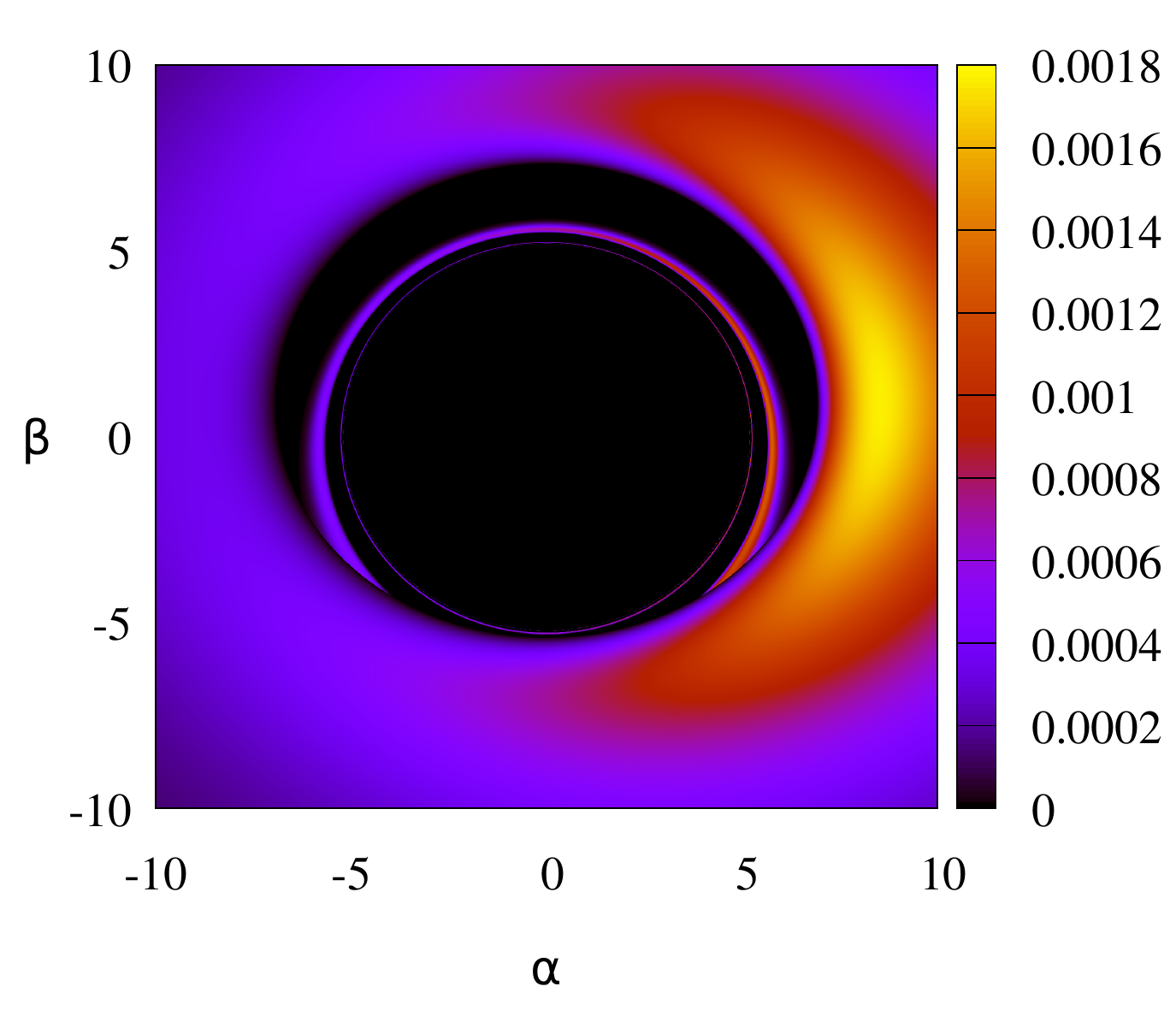}\\
			\includegraphics[scale=0.45]{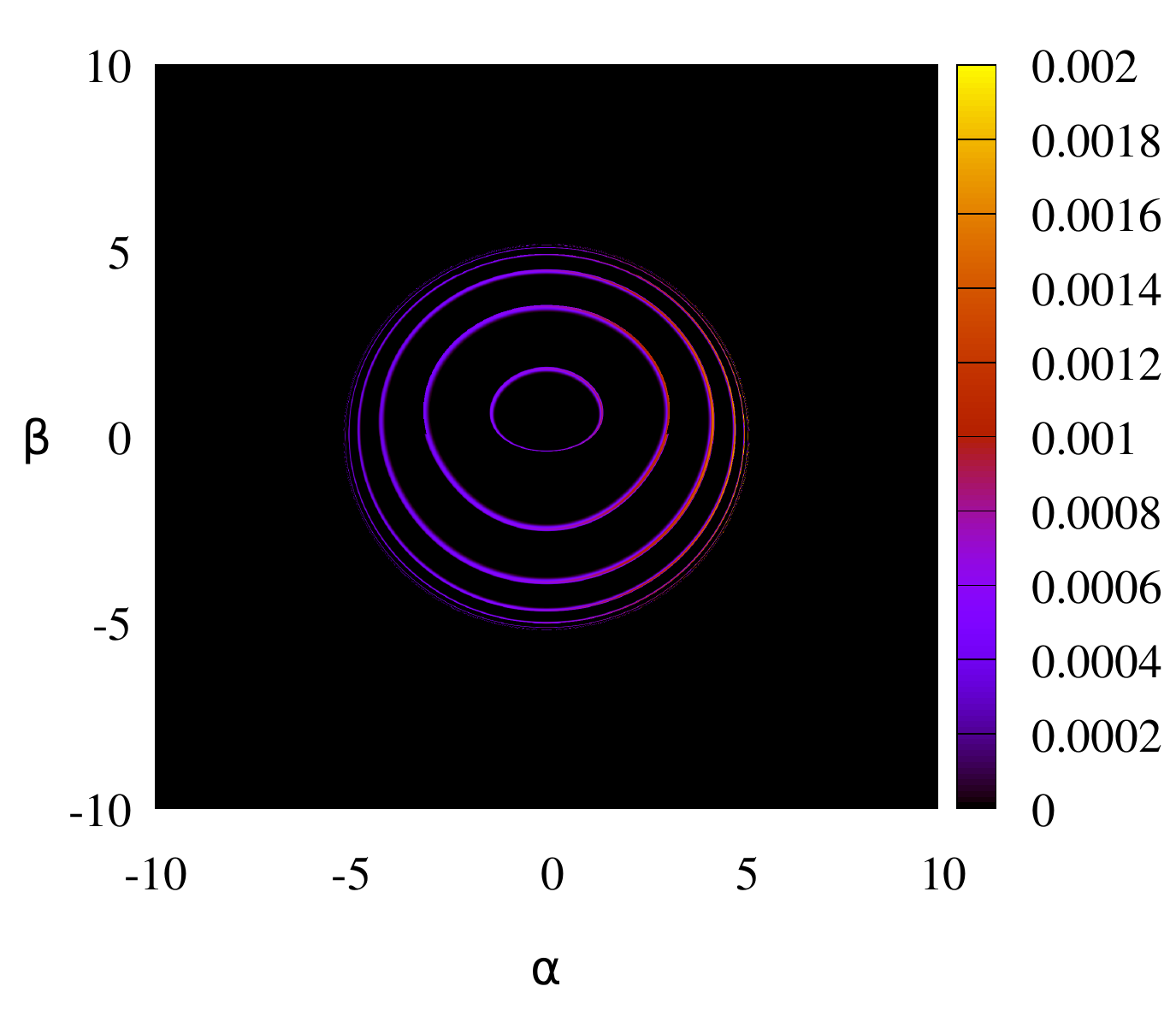} &	\includegraphics[scale=0.45]{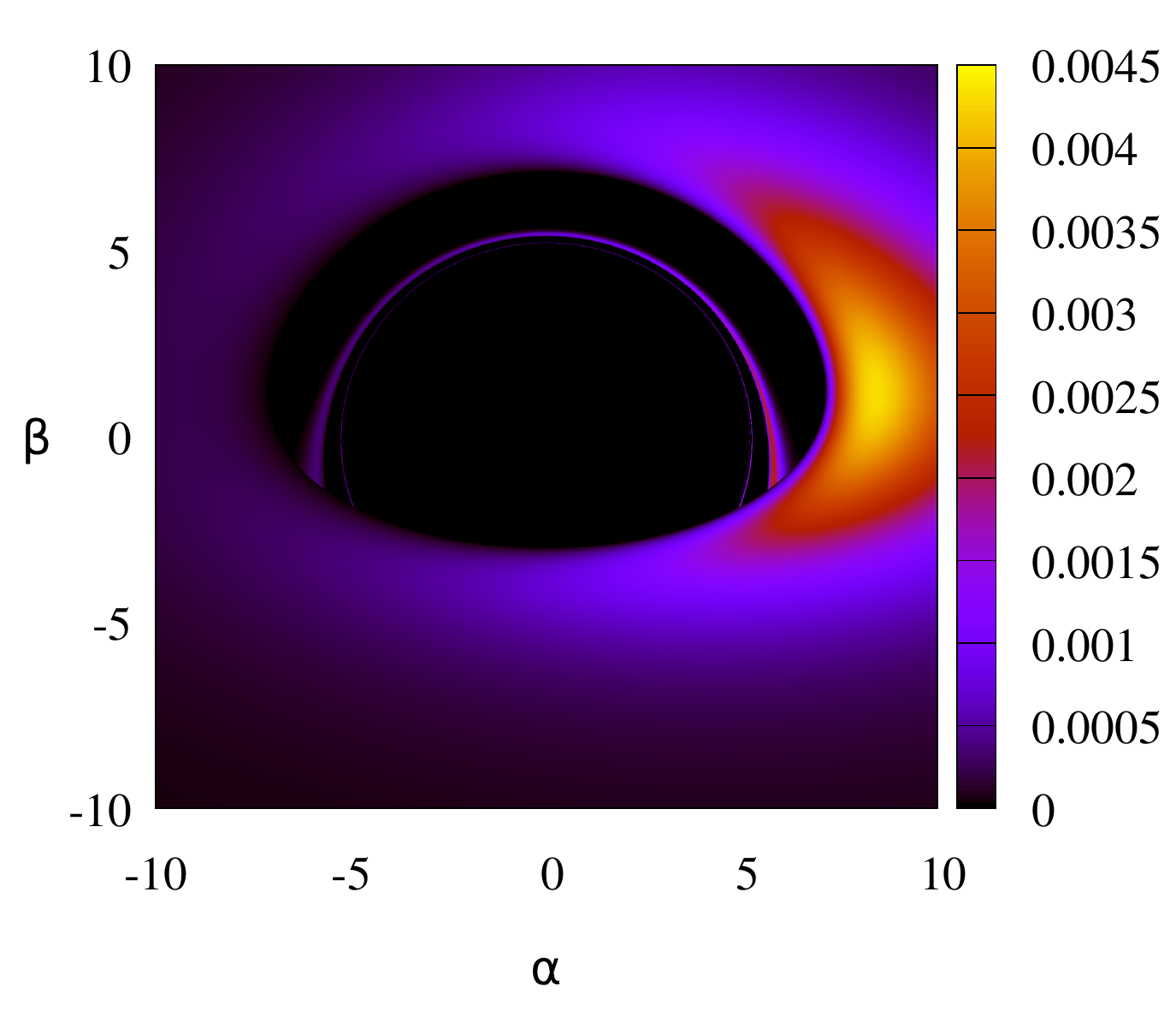}\\
			\includegraphics[scale=0.45]{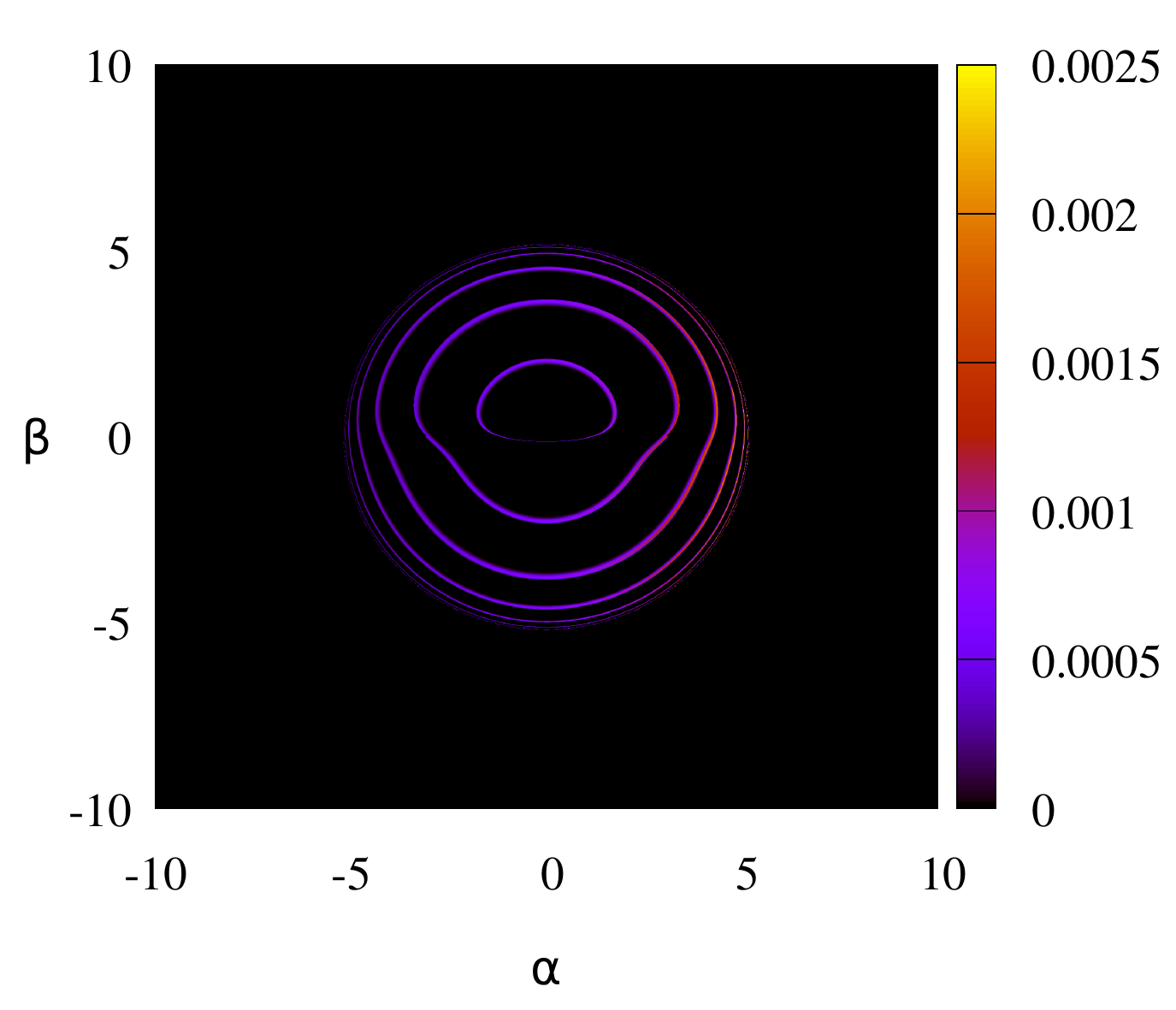} &	\includegraphics[scale=0.45]{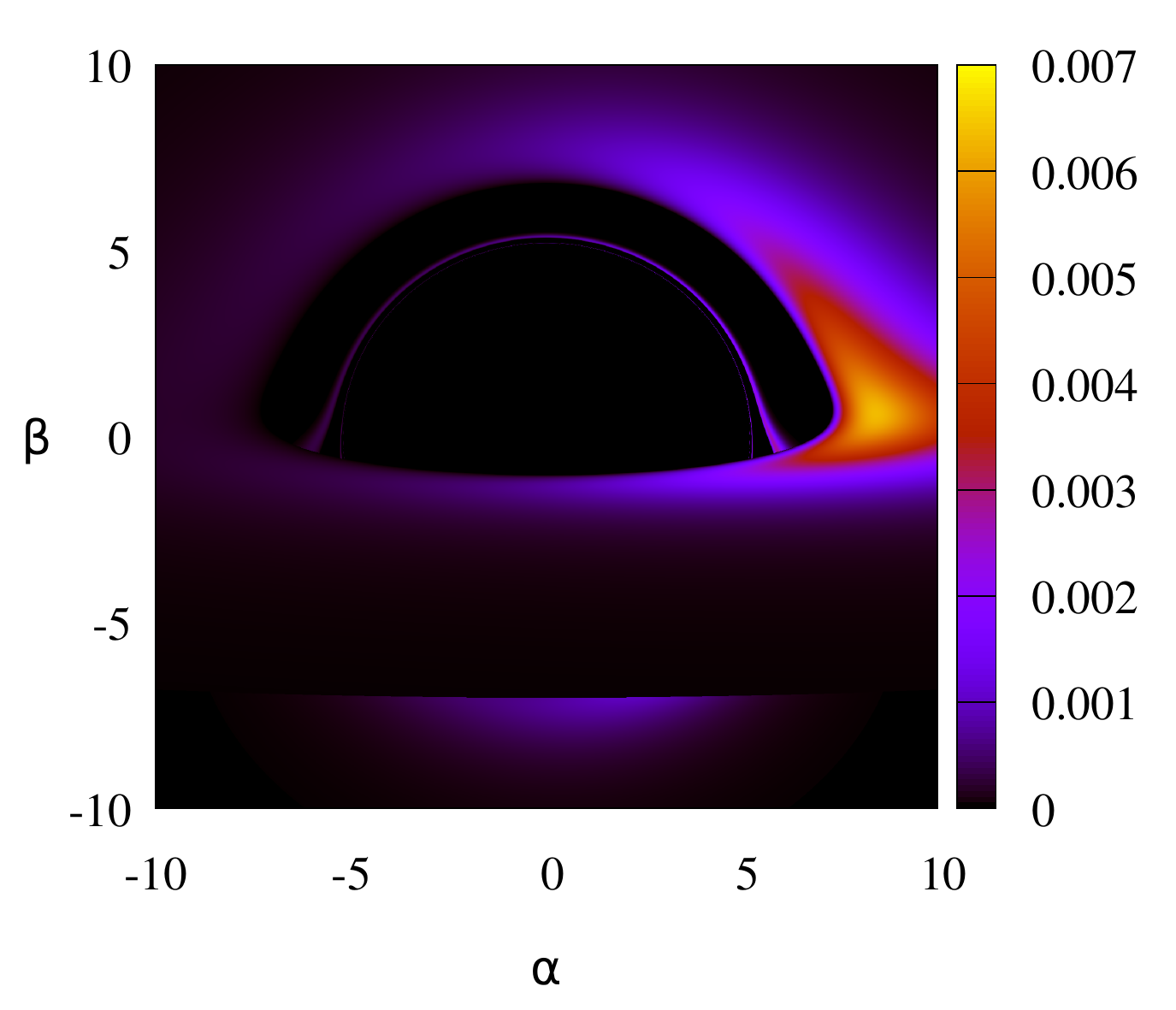}
		\end{tabular}
	\caption{The images of Keplerian disks rotating around wormhole on the other side (left), rep. on our side (right). The inclination of the observer, located at $r_o=10^3$M is $\theta_o=30^\circ$ (top), $60^\circ$ (middle) and $80^\circ$ (bottom). The wormhole parameter $a=2.1$. The colors of the plot represent bolomeric flux radiation from the disk and we assume thermal spectrum of the emitted photons. \label{bolo1}}
	\end{center}
\end{figure}

\begin{figure}[H]
	\begin{center}
		\begin{tabular}{cc}
			\includegraphics[scale=0.45]{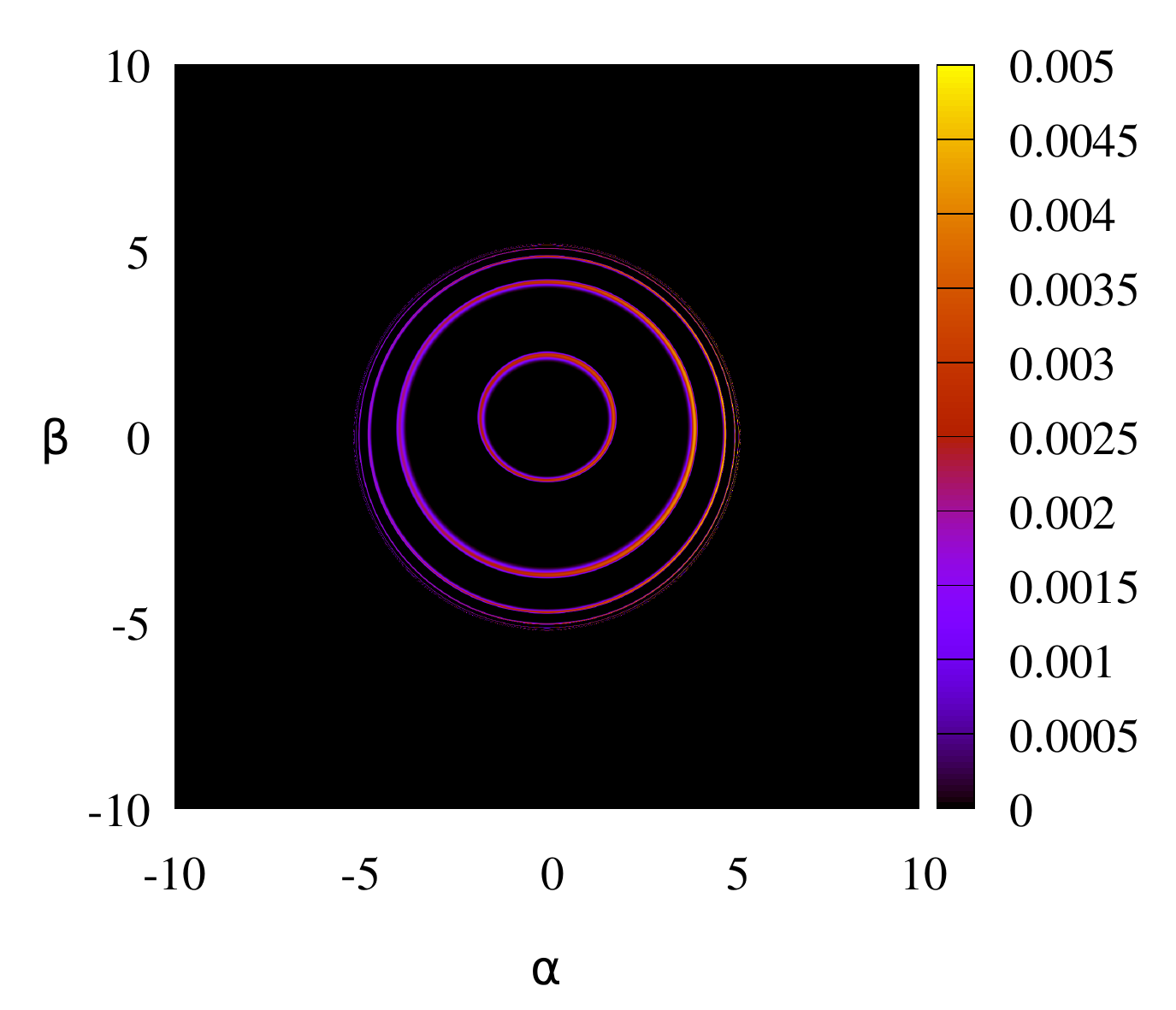} &	\includegraphics[scale=0.45]{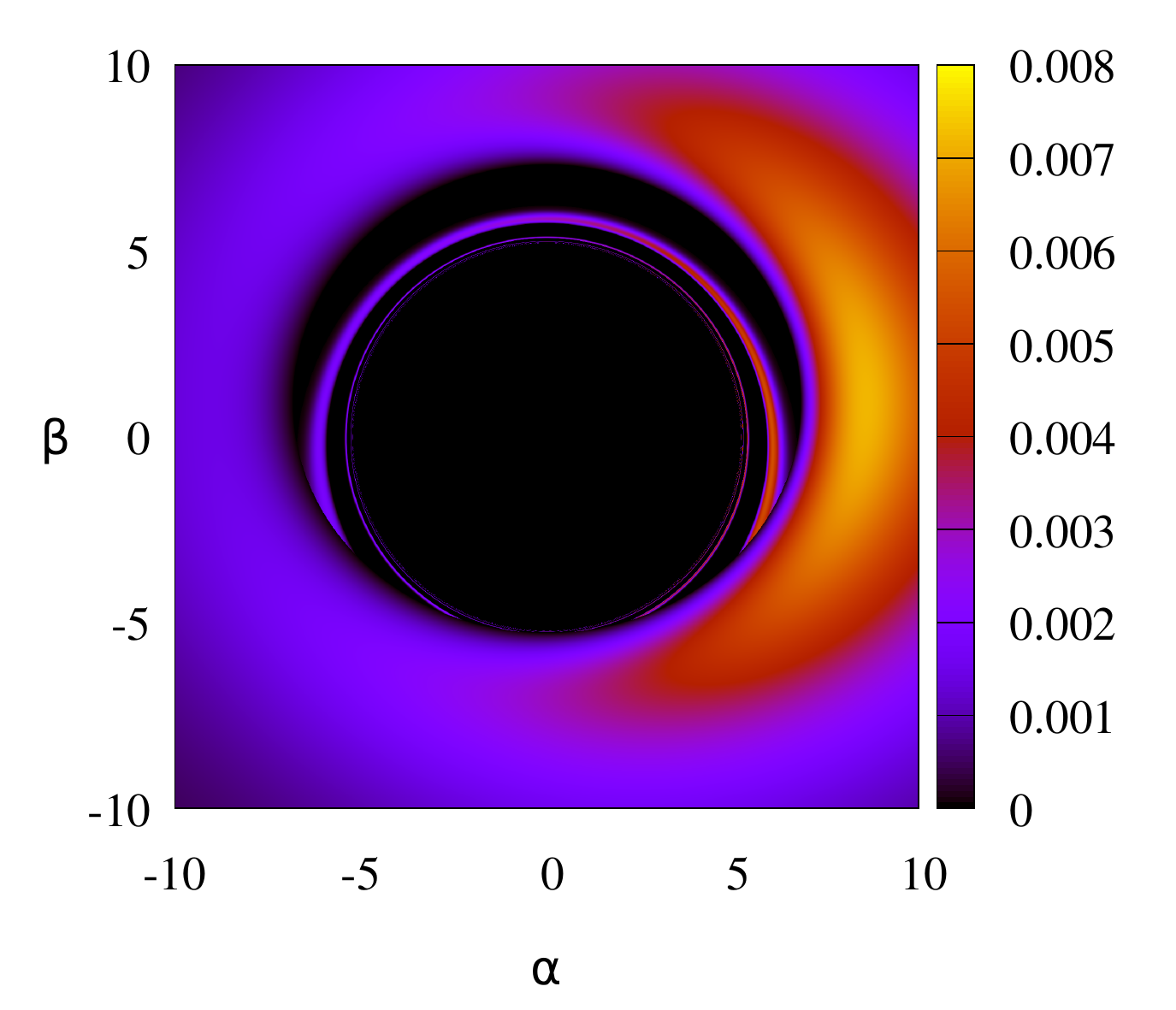}\\
			\includegraphics[scale=0.45]{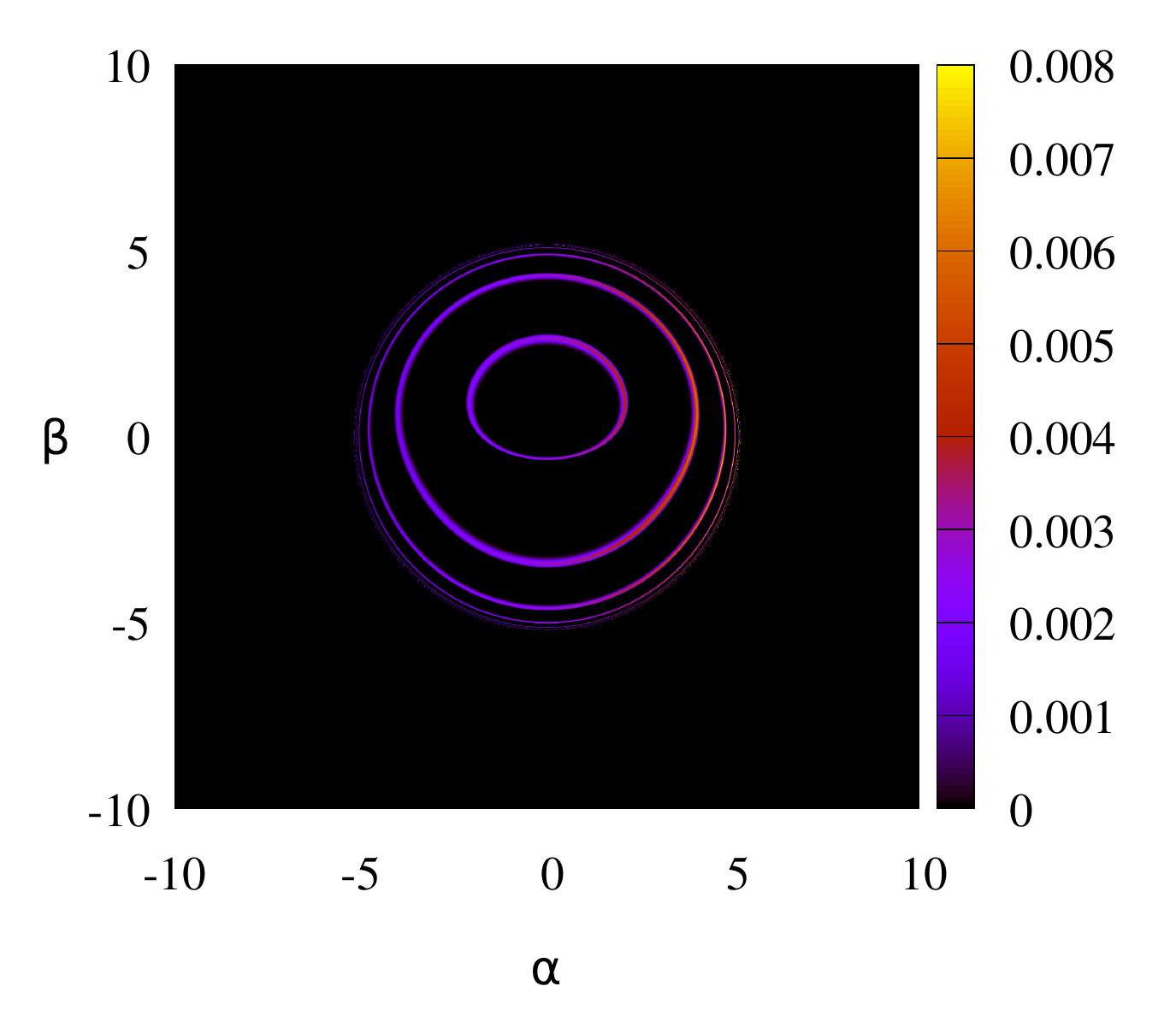} &	\includegraphics[scale=0.45]{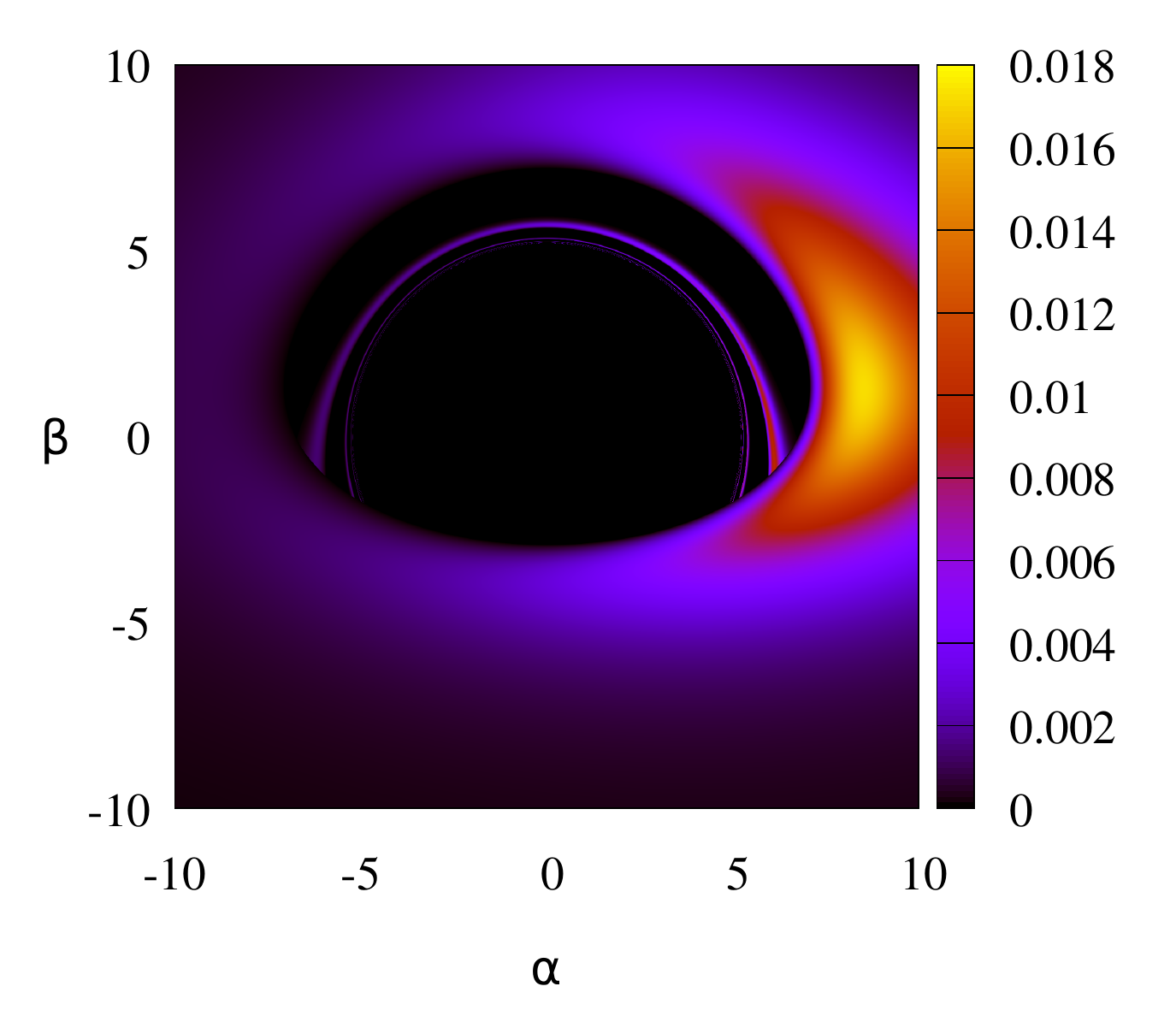}\\
			\includegraphics[scale=0.45]{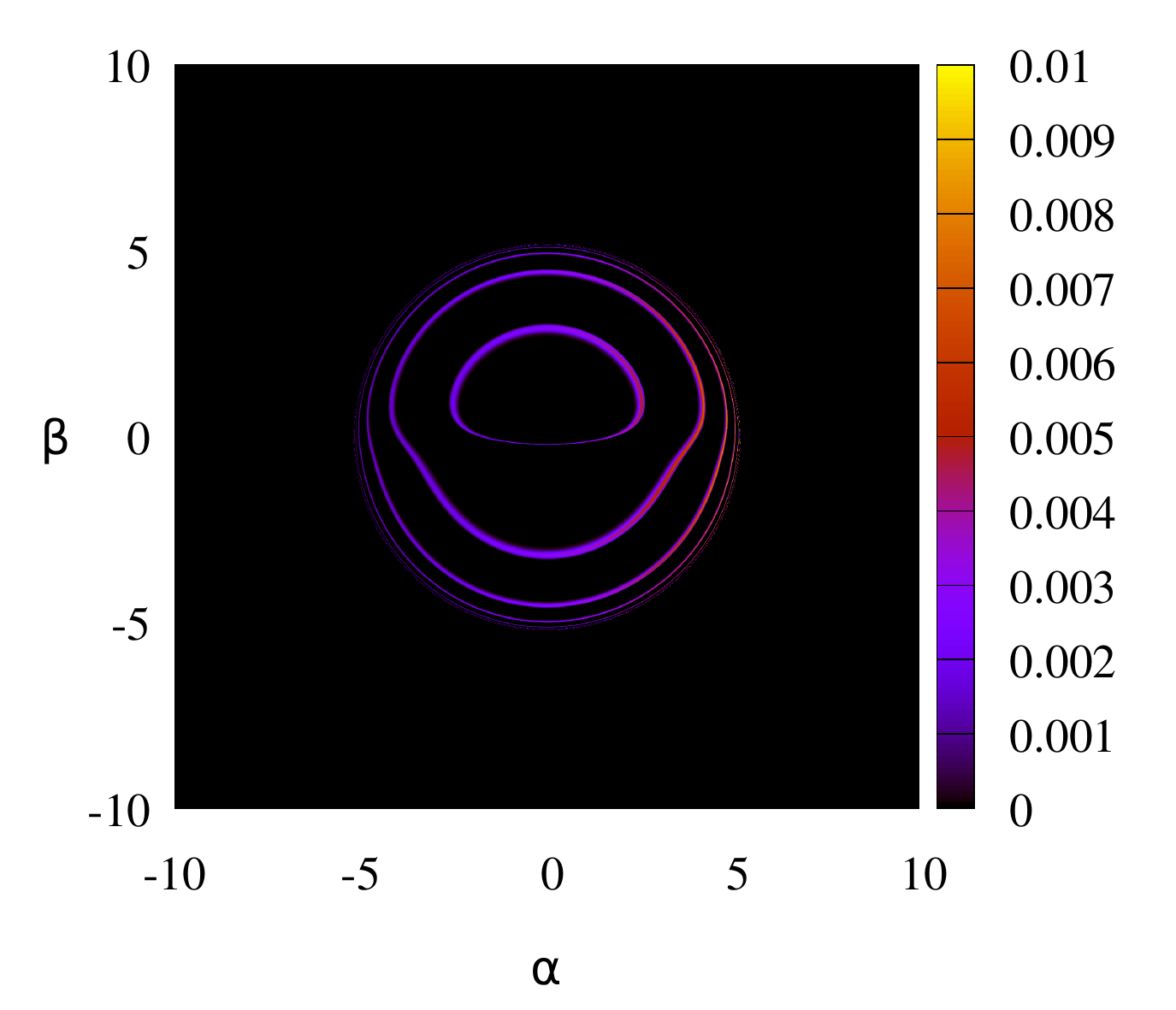} &	\includegraphics[scale=0.45]{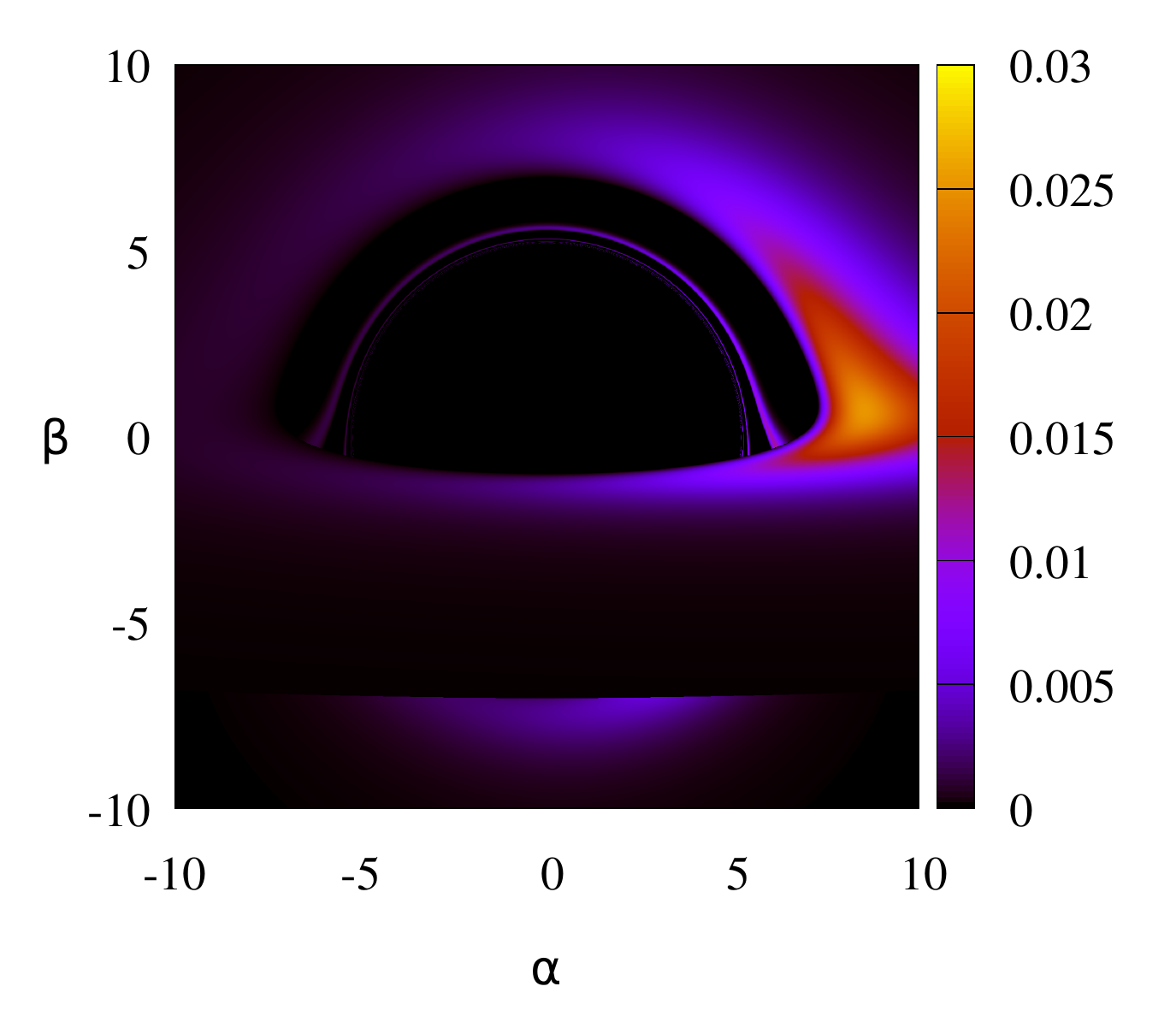}
		\end{tabular}
		\caption{The images of Keplerian disks rotating around wormhole on the other side (left), rep. on our side (right). The inclination of the observer, located at $r_o=10^3$M is $\theta_o=30^\circ$ (top), $60^\circ$ (middle) and $80^\circ$ (bottom). The wormhole parameter $a=3.1$. The colors of the plot represent bolomeric flux radiation from the disk and we assume thermal spectrum of the emitted photons. \label{bolo2}}
	\end{center}
\end{figure}

\begin{figure}[H]
	\begin{center}
		\begin{tabular}{cc}
			\includegraphics[scale=0.45]{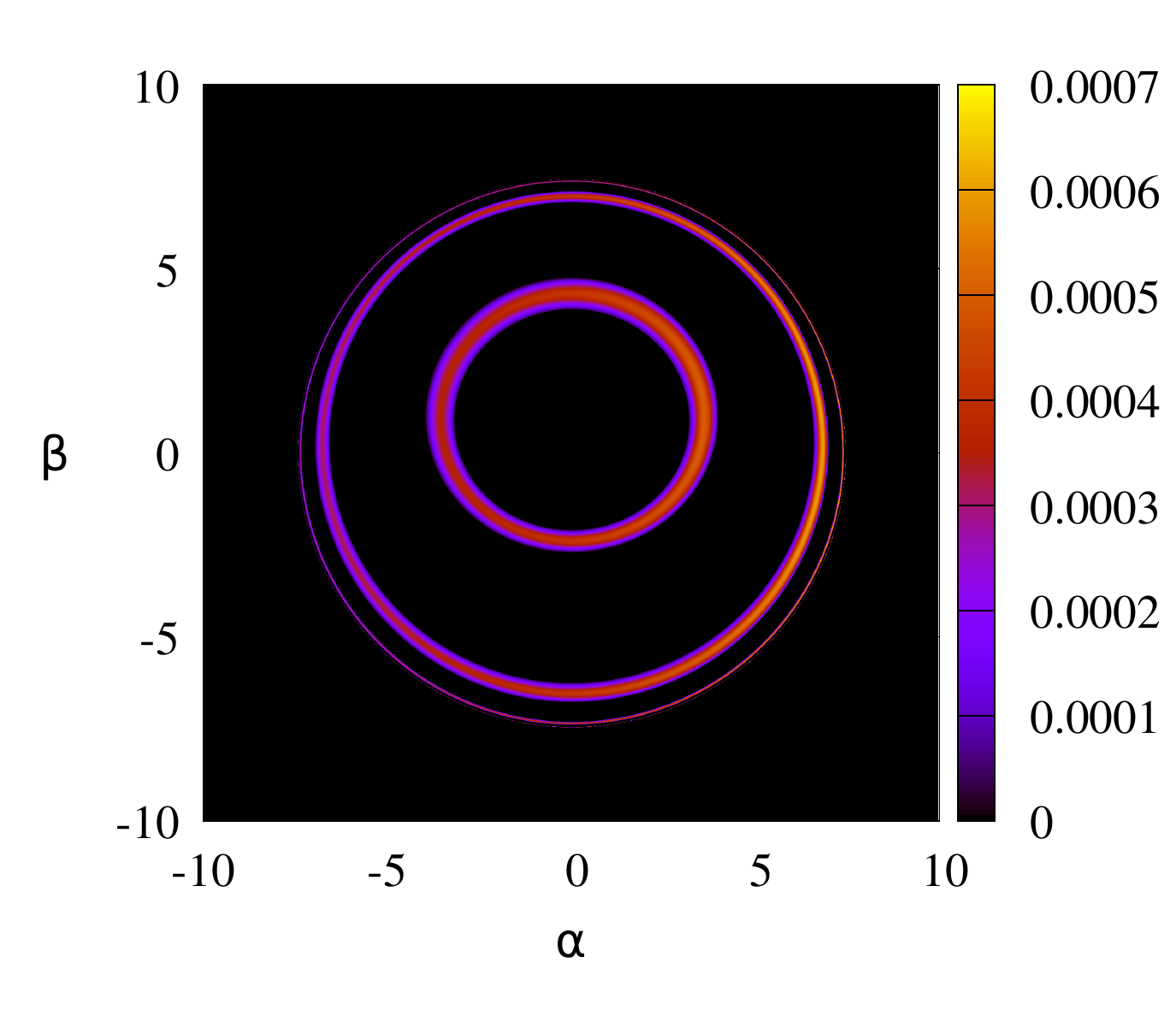} &	\includegraphics[scale=0.45]{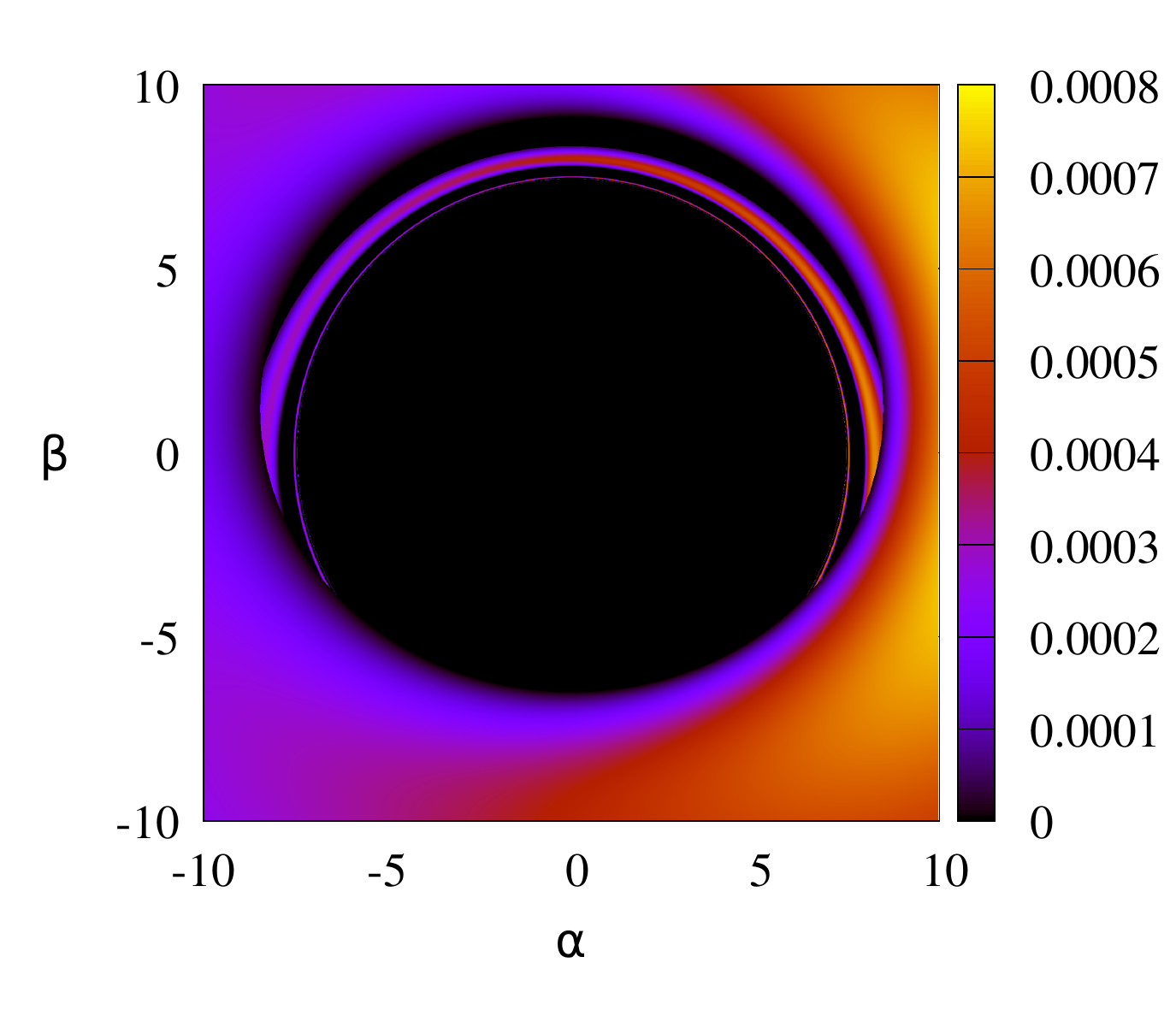}\\
			\includegraphics[scale=0.45]{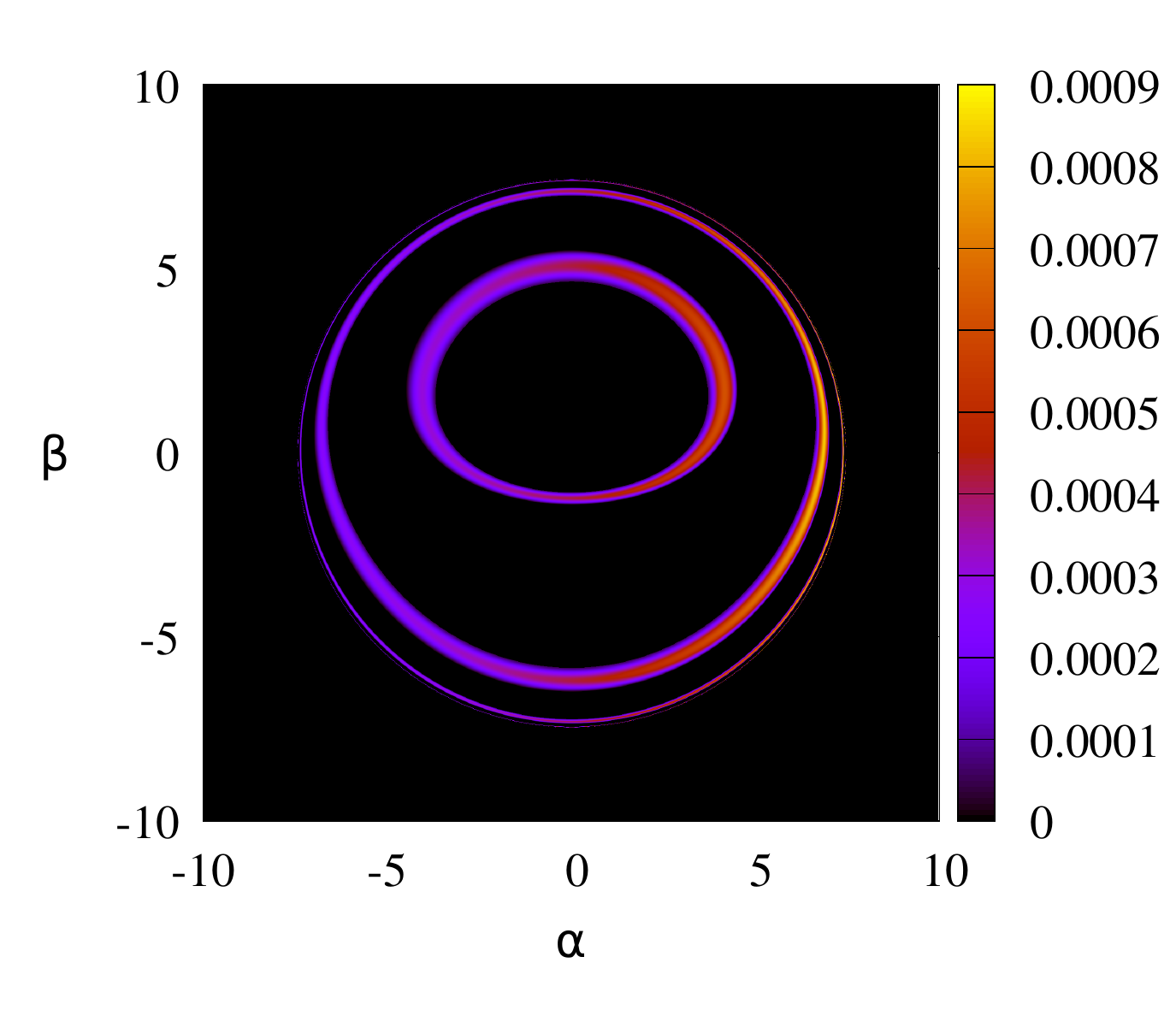} &	\includegraphics[scale=0.45]{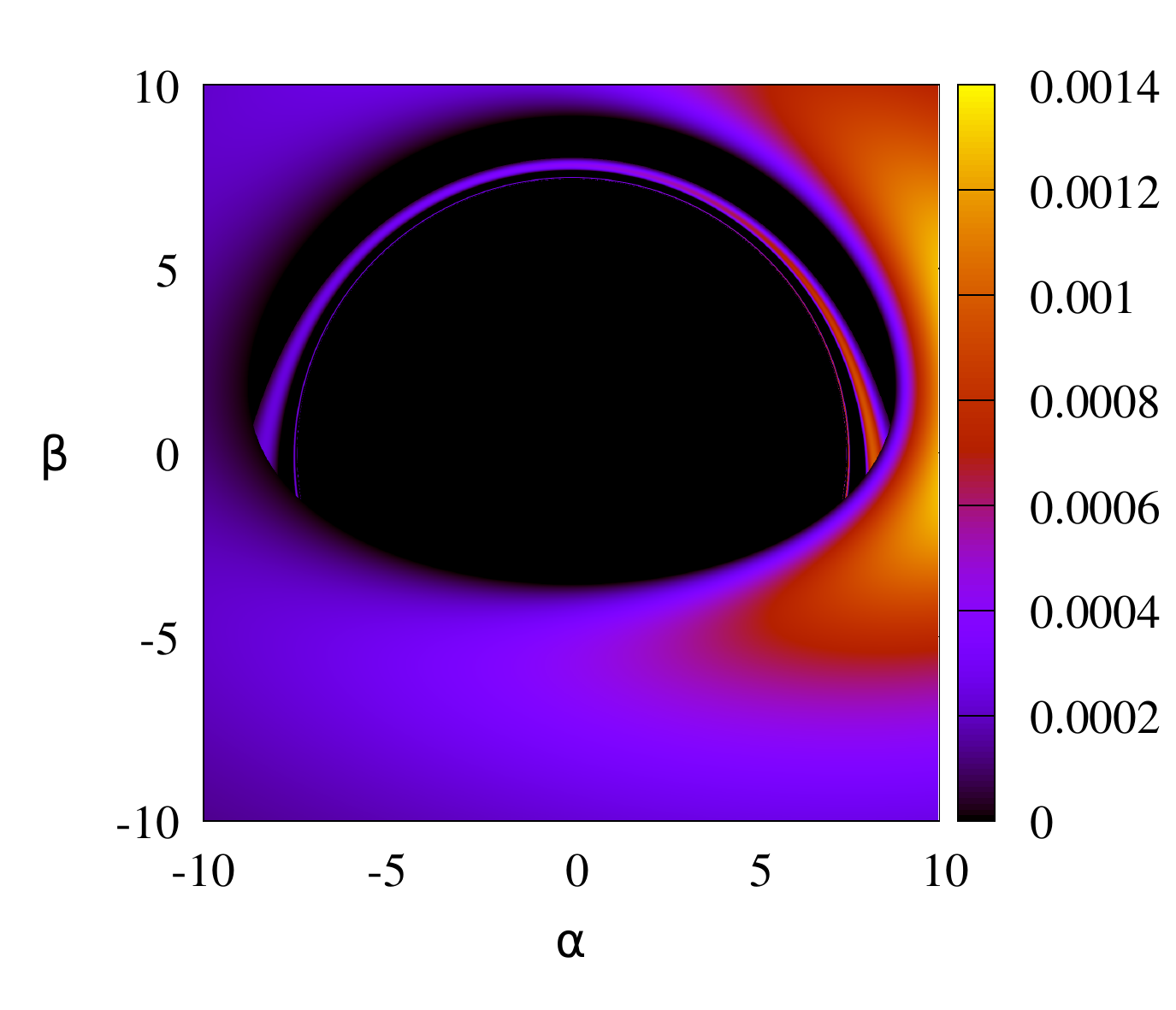}\\
			\includegraphics[scale=0.45]{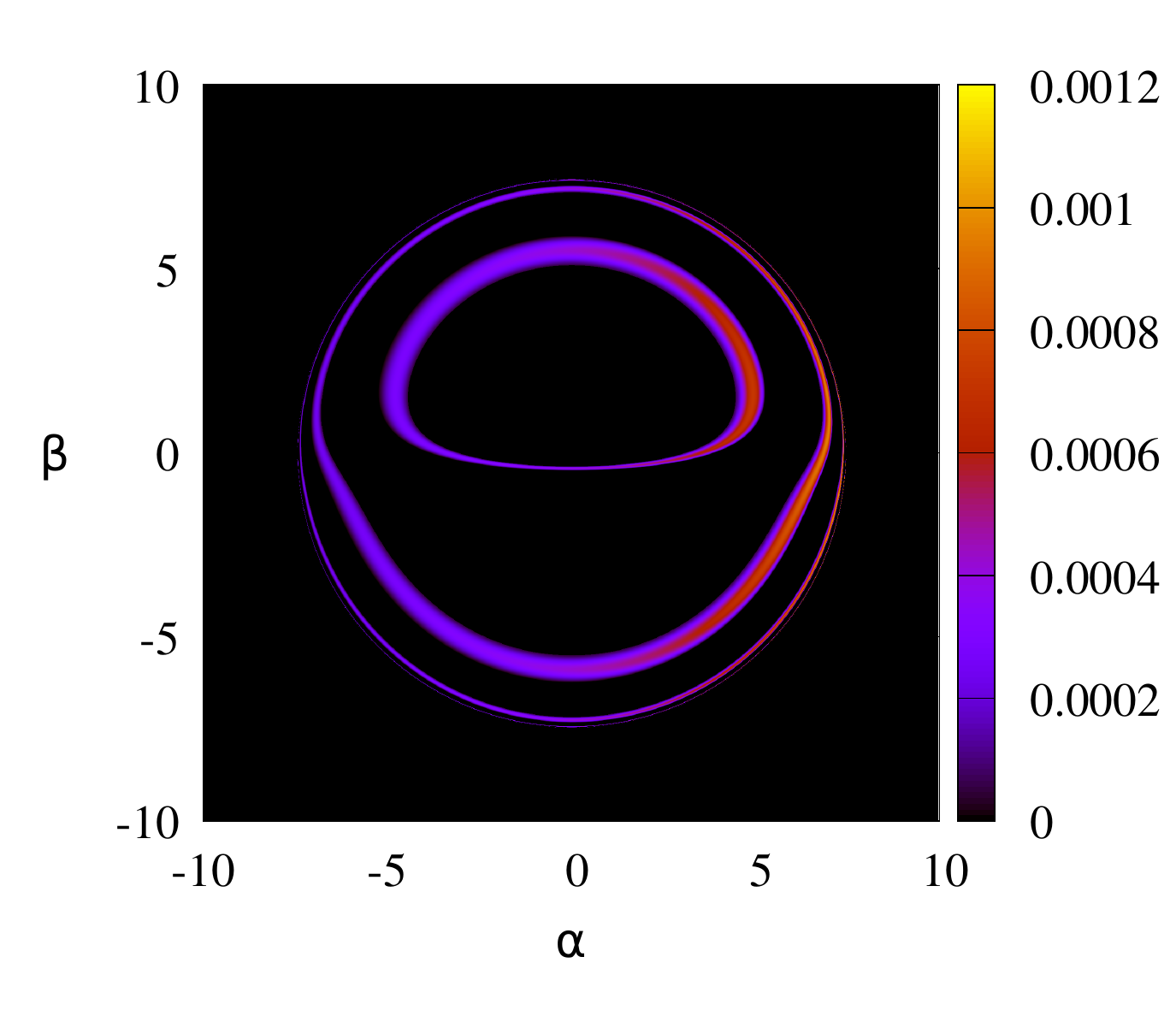} &	\includegraphics[scale=0.45]{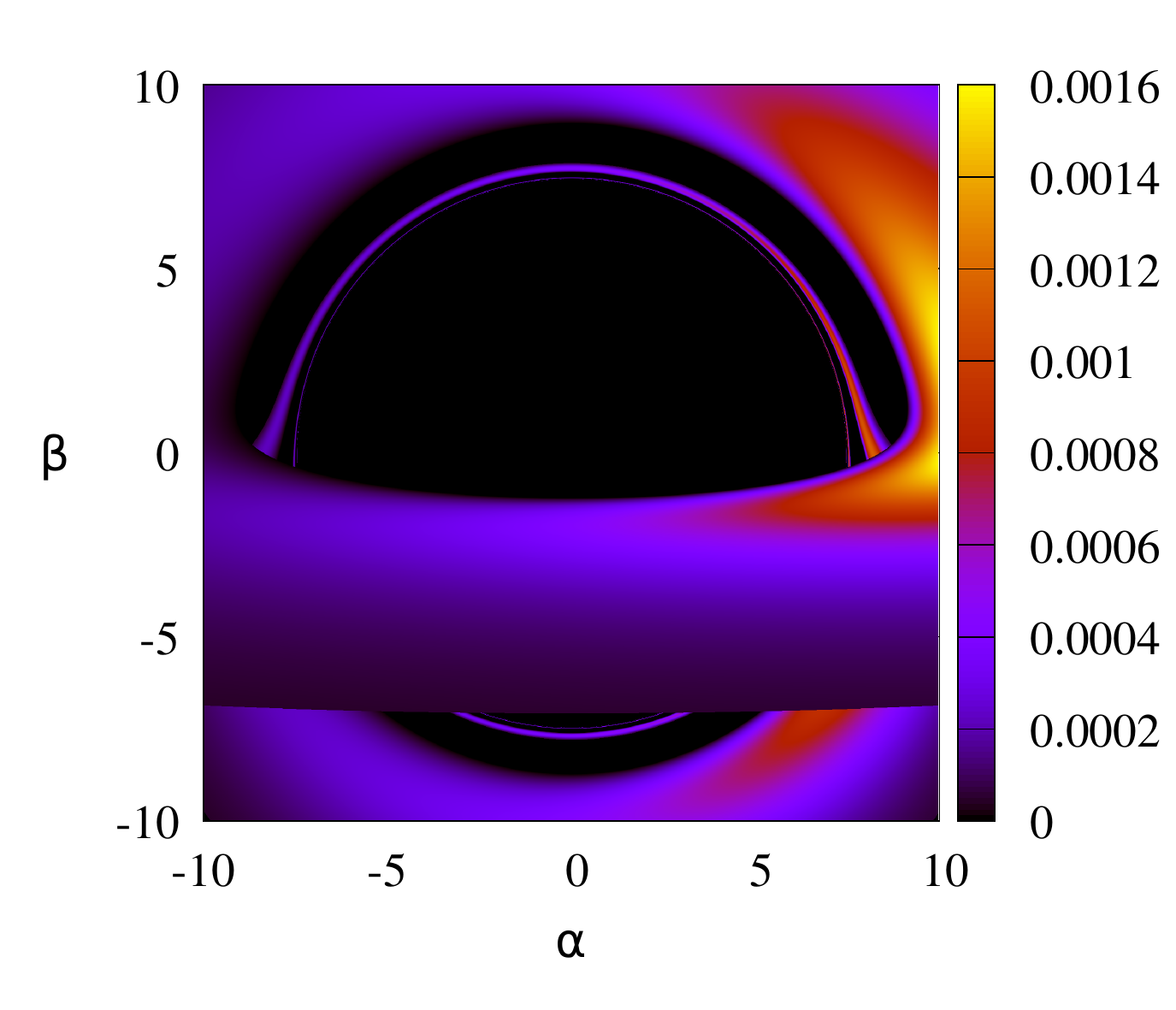}
		\end{tabular}
		\caption{The images of Keplerian disks rotating around wormhole on the other side (left), rep. on our side (right). The inclination of the observer, located at $r_o=10^3$M is $\theta_o=30^\circ$ (top), $60^\circ$ (middle) and $80^\circ$ (bottom). The wormhole parameter $a=6.1$. The colors of the plot represent bolomeric flux radiation from the disk and we assume thermal spectrum of the emitted photons. \label{bolo3}}
	\end{center}
\end{figure}

\begin{figure}[H]
	\begin{center}
		\begin{tabular}{cc}
			\includegraphics[scale=0.45]{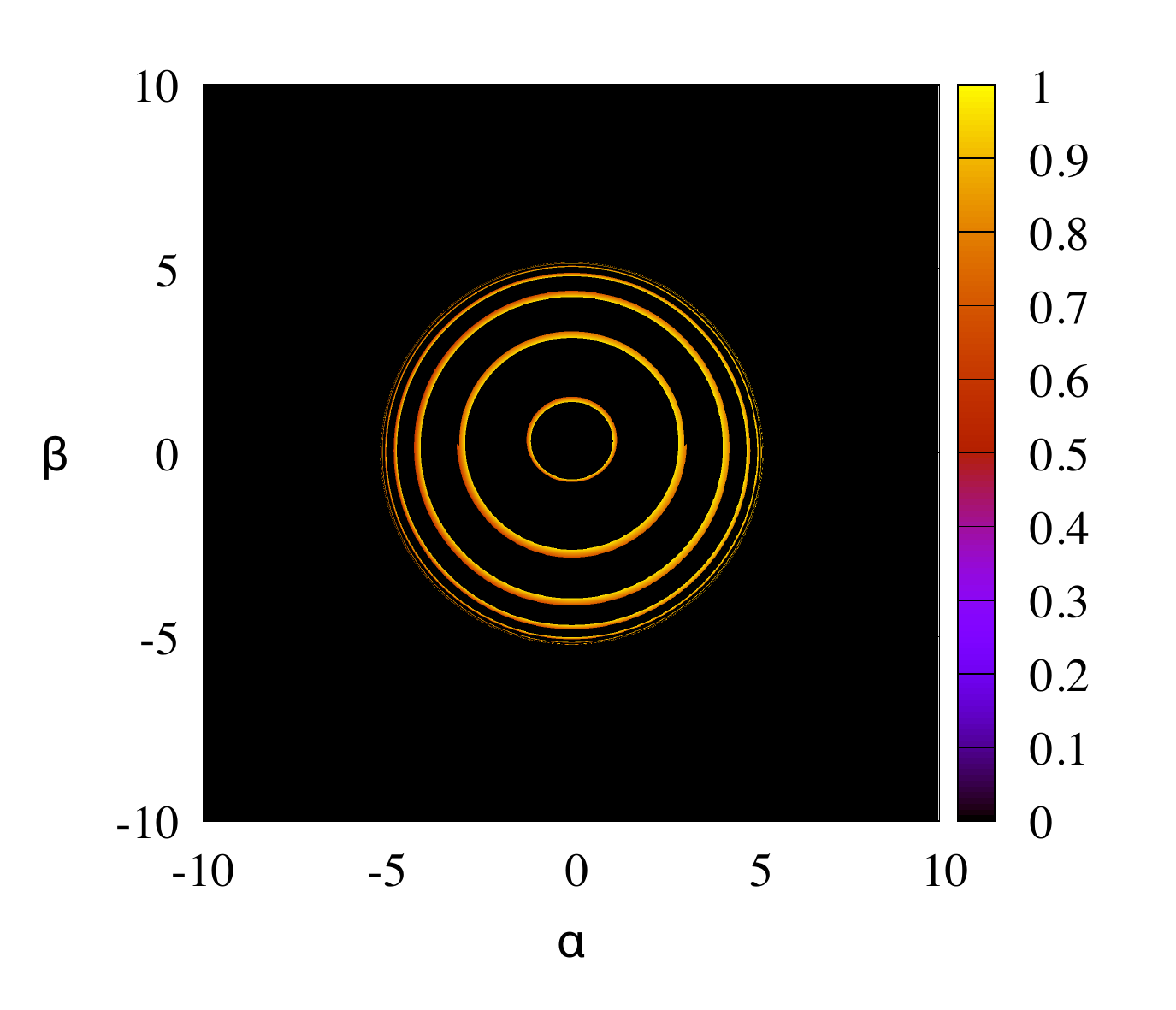} &	\includegraphics[scale=0.45]{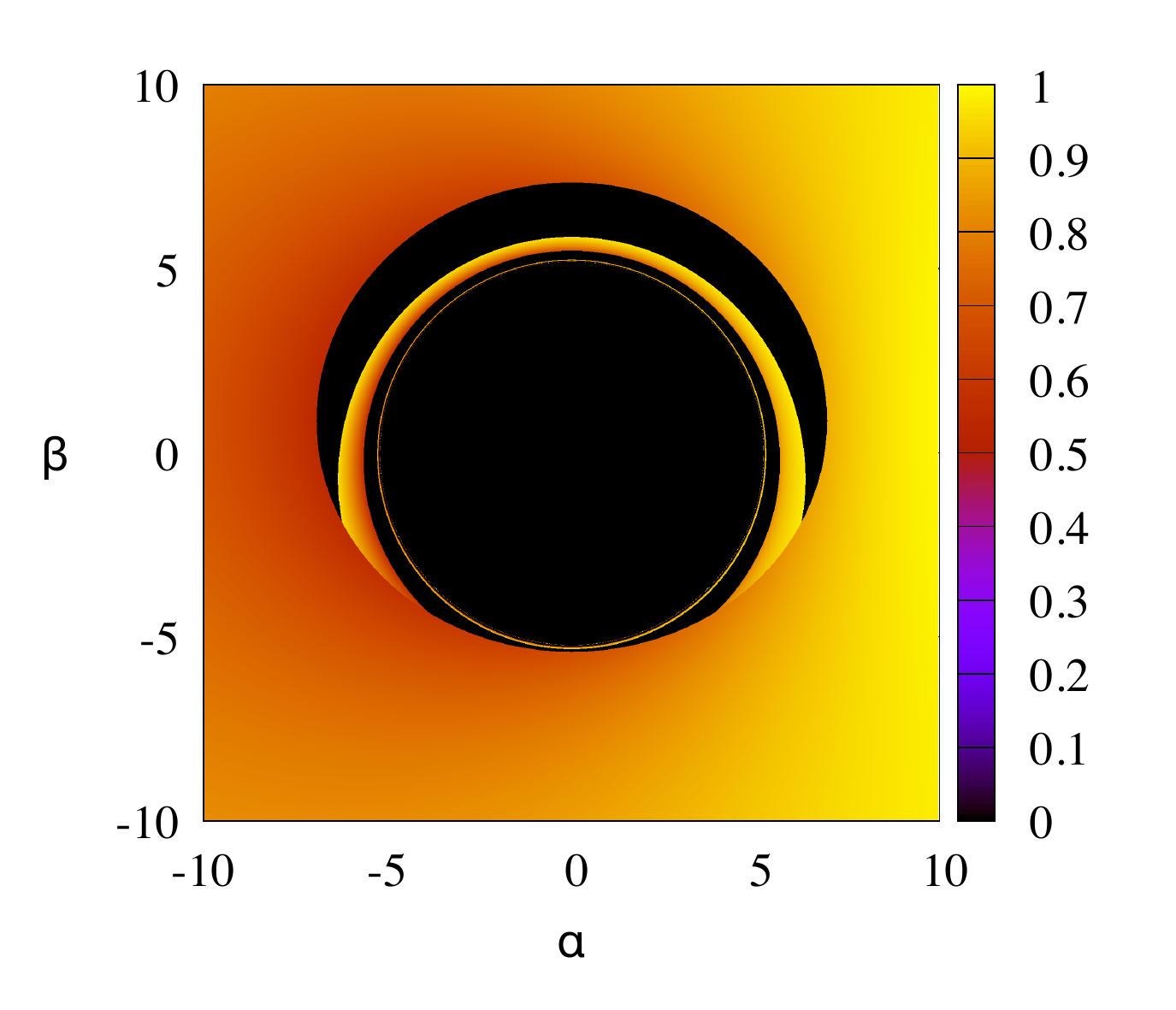}\\
			\includegraphics[scale=0.45]{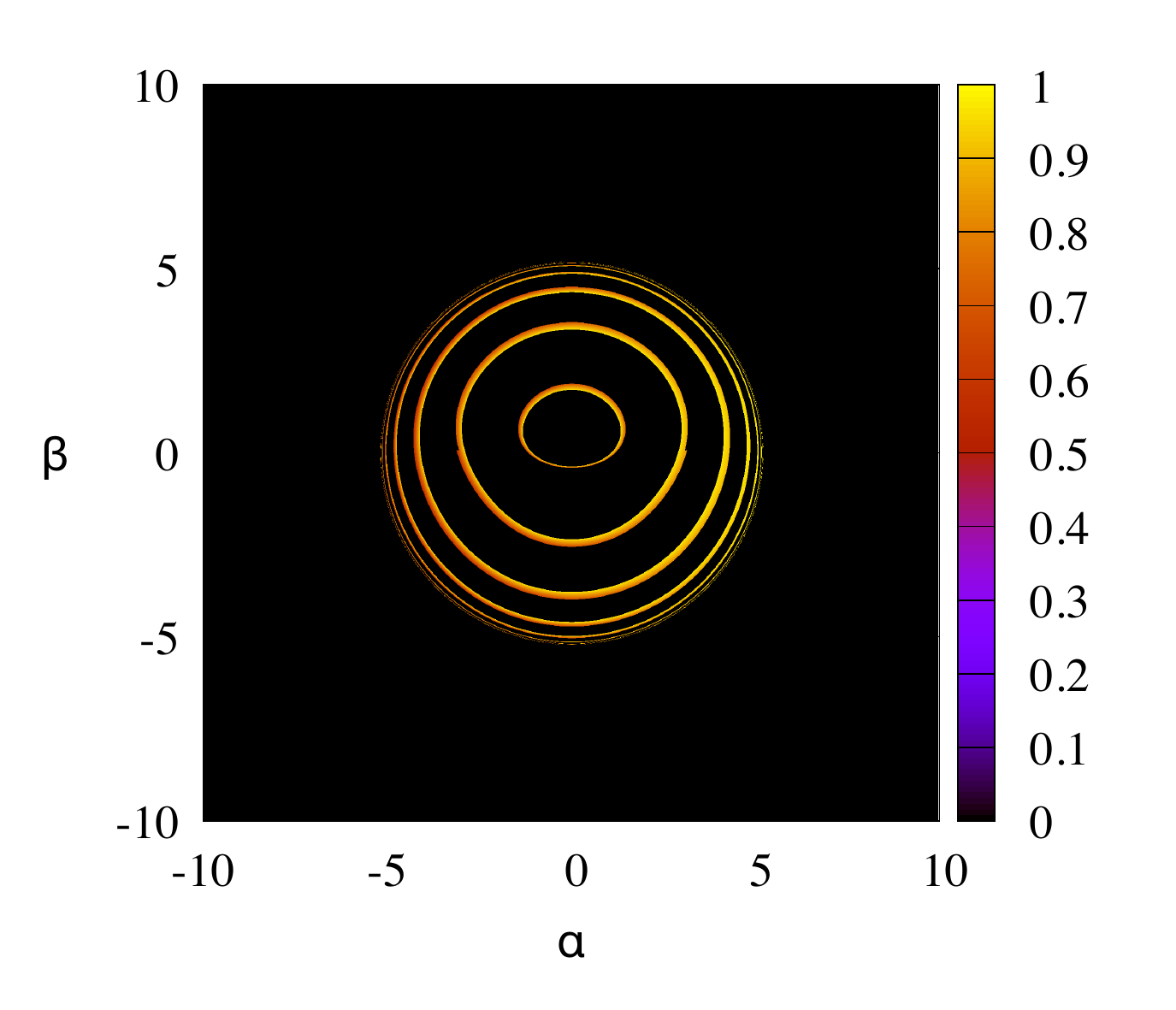} &	\includegraphics[scale=0.45]{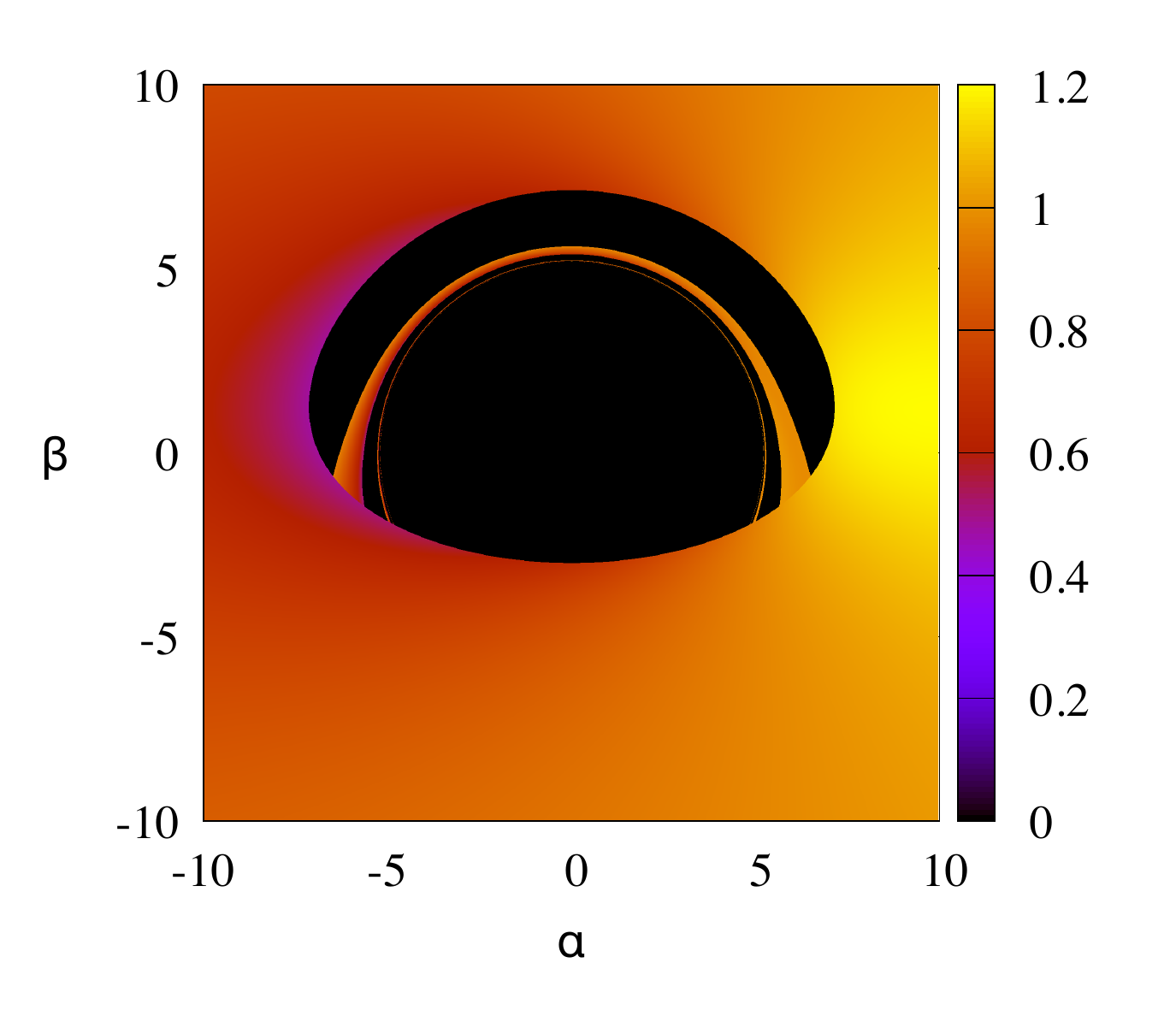}\\
			\includegraphics[scale=0.45]{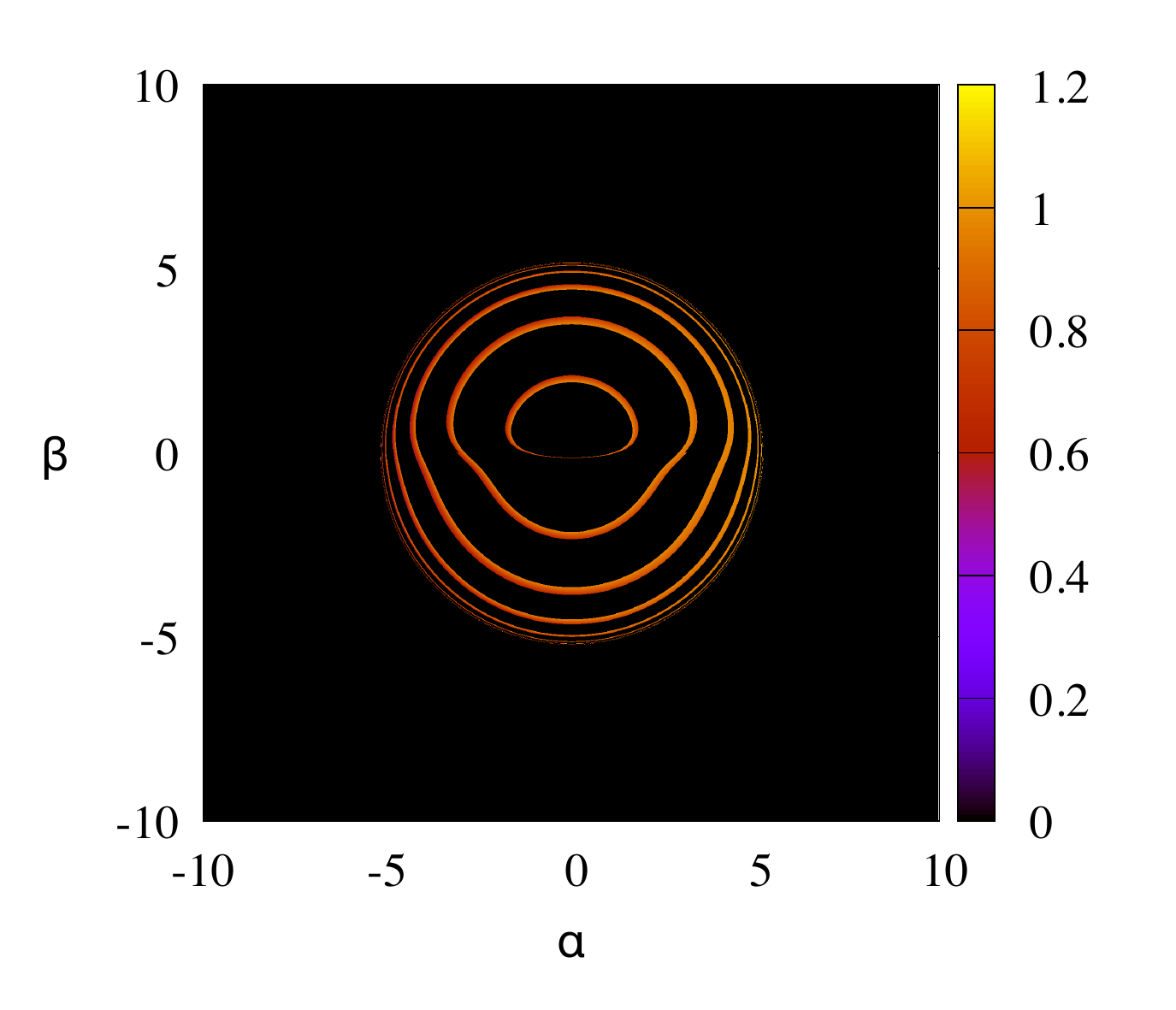} &	\includegraphics[scale=0.45]{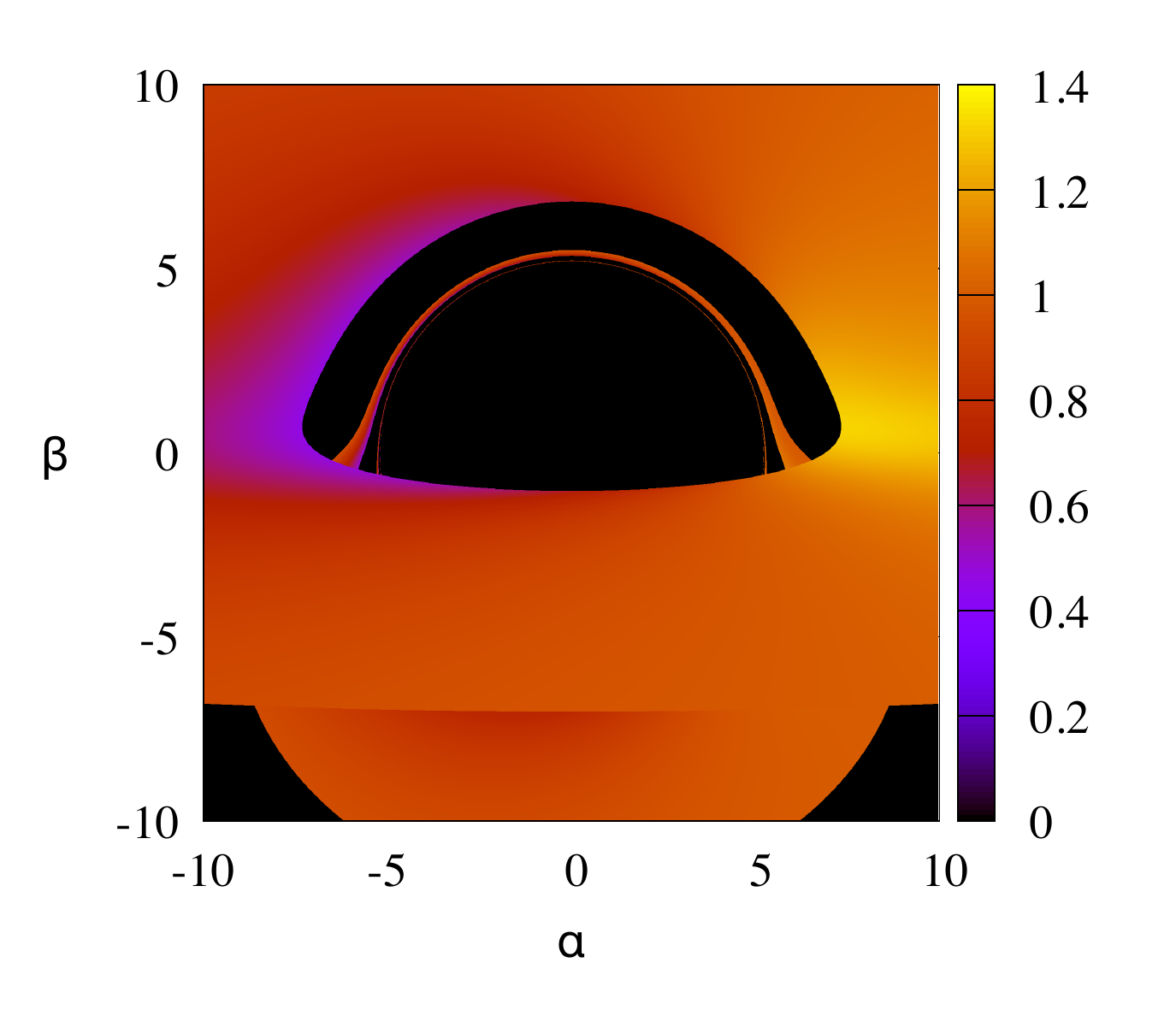}
		\end{tabular}
	\caption{The images of Keplerian disks rotating around wormhole on the other side (left), rep. on our side (right). The inclination of the observer, located at $r_o=10^3$M is $\theta_o=30^\circ$ (top), $60^\circ$ (middle) and $80^\circ$ (bottom). The wormhole parameter $a=2.1$. The colors of the plot represent frequency shift of radiation from the disk. \label{g1img}}
	\end{center}
\end{figure}

\begin{figure}[H]
	\begin{center}
		\begin{tabular}{cc}
			\includegraphics[scale=0.45]{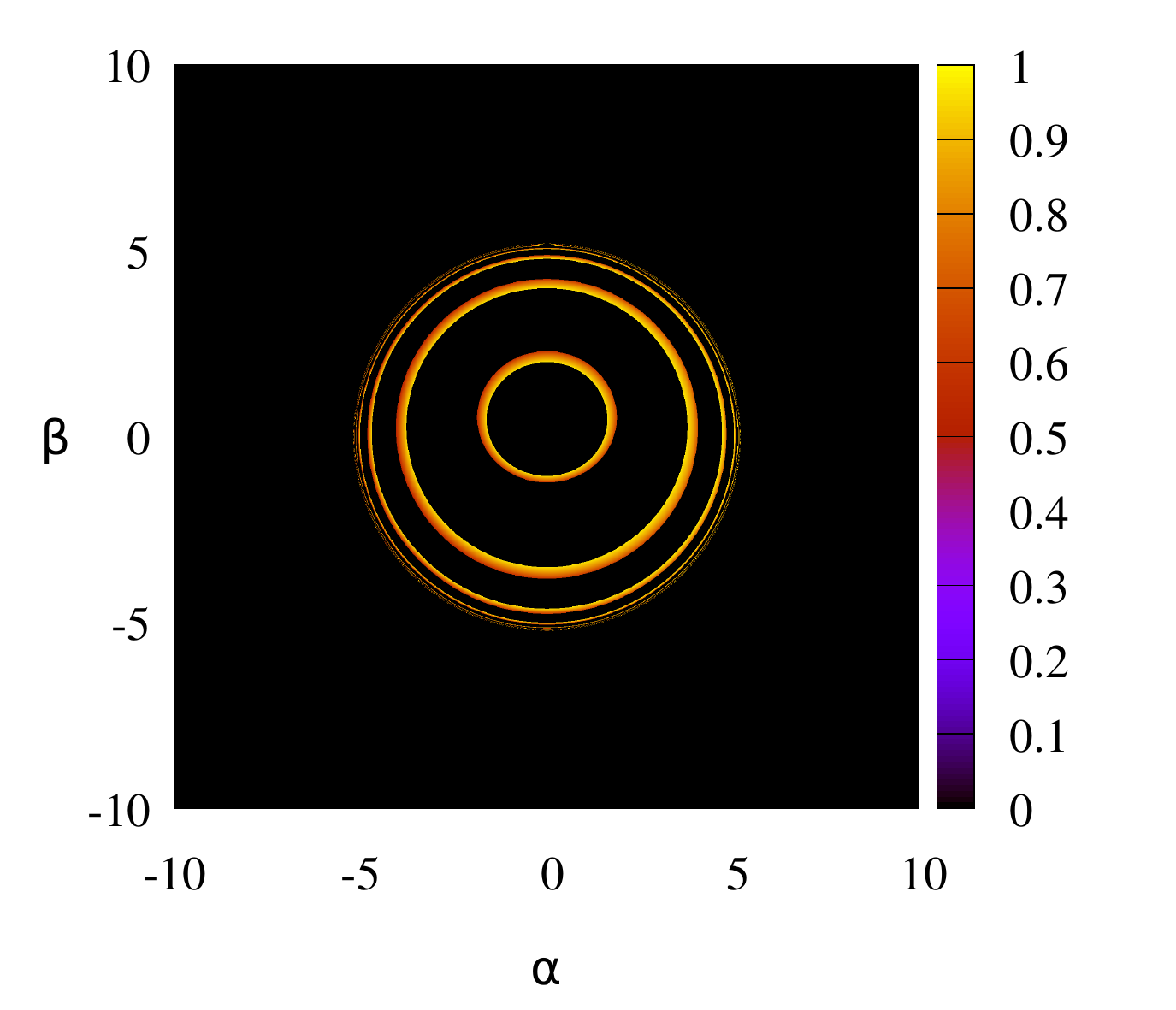} &	\includegraphics[scale=0.45]{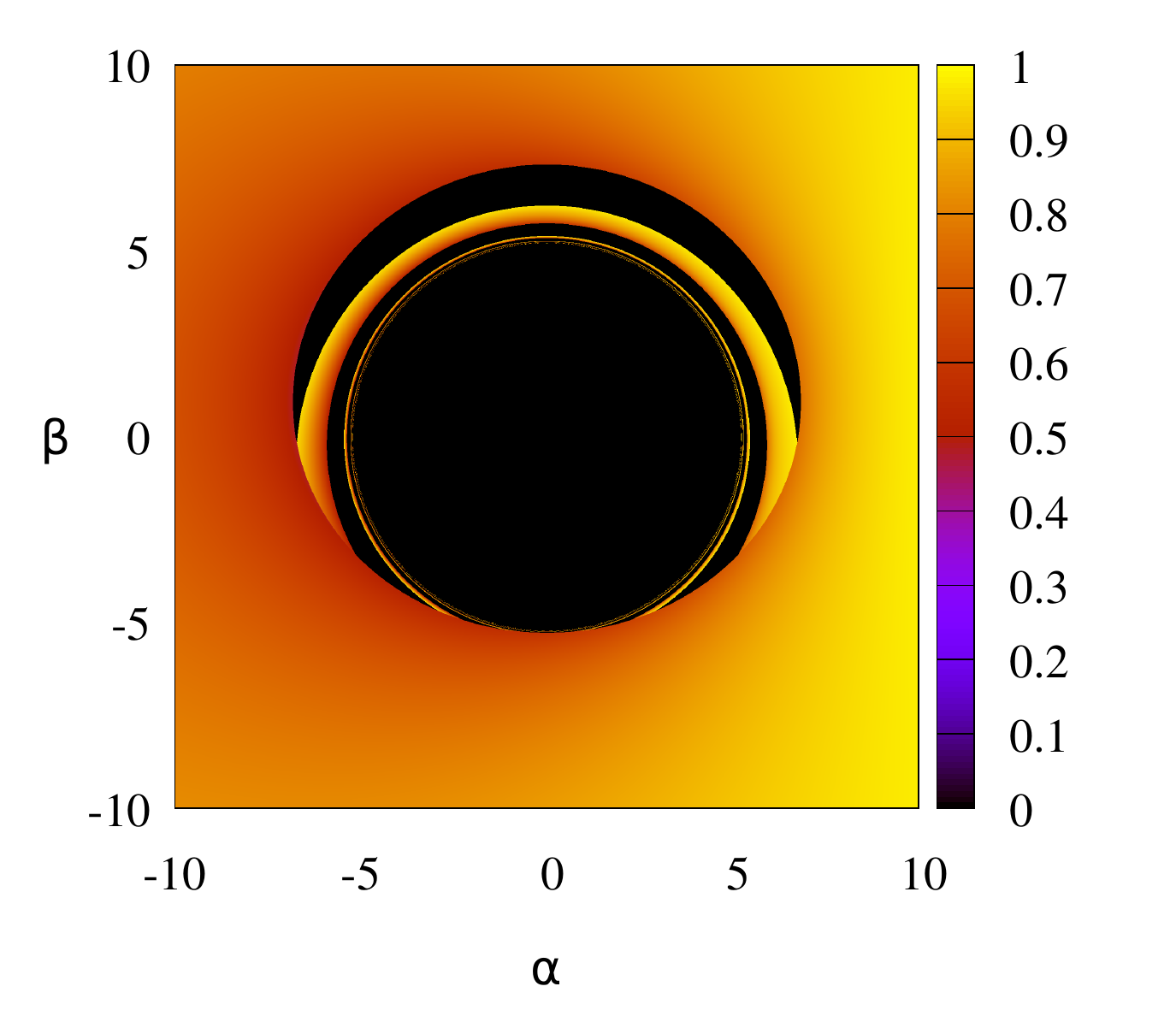}\\
			\includegraphics[scale=0.45]{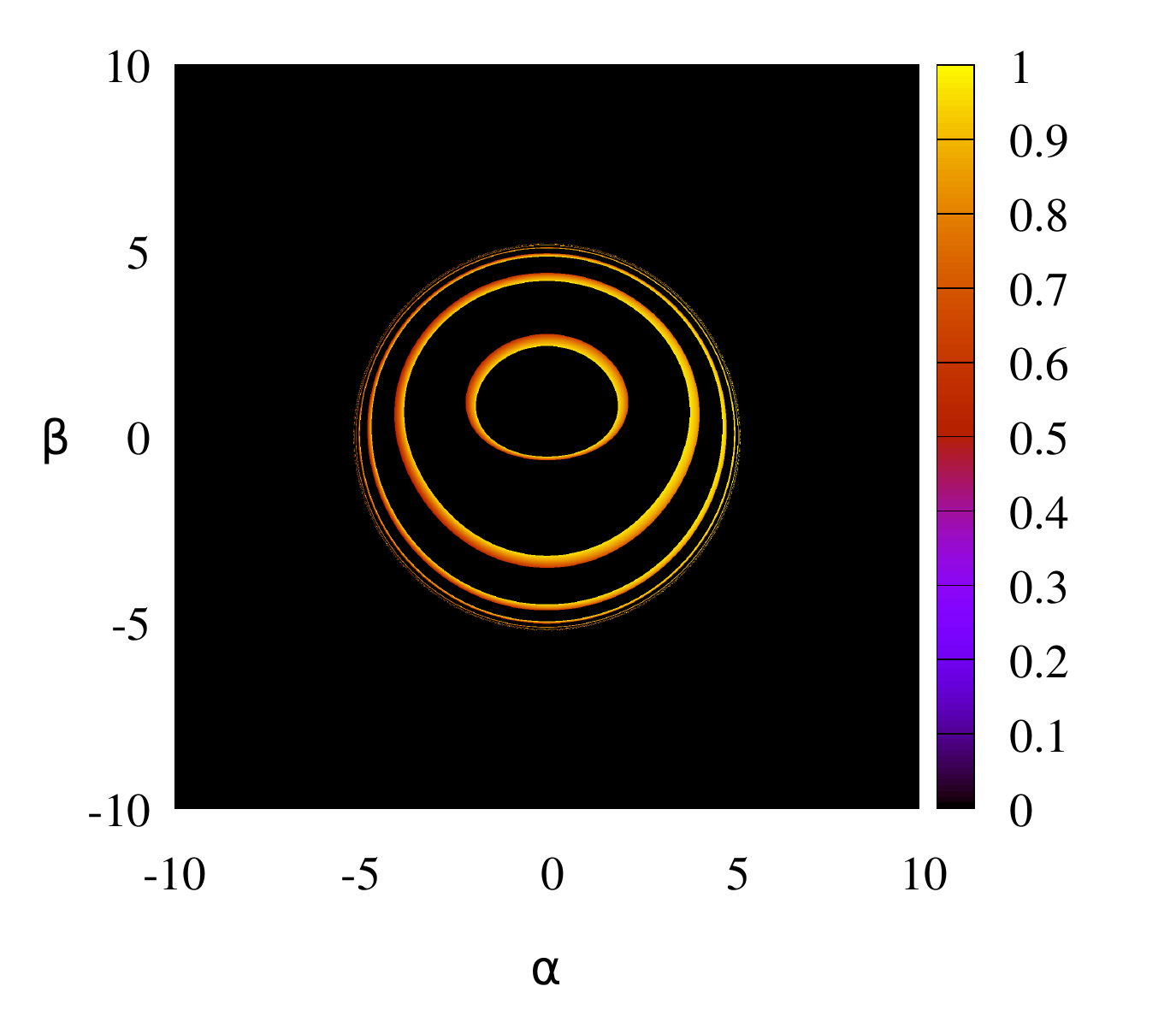} &	\includegraphics[scale=0.45]{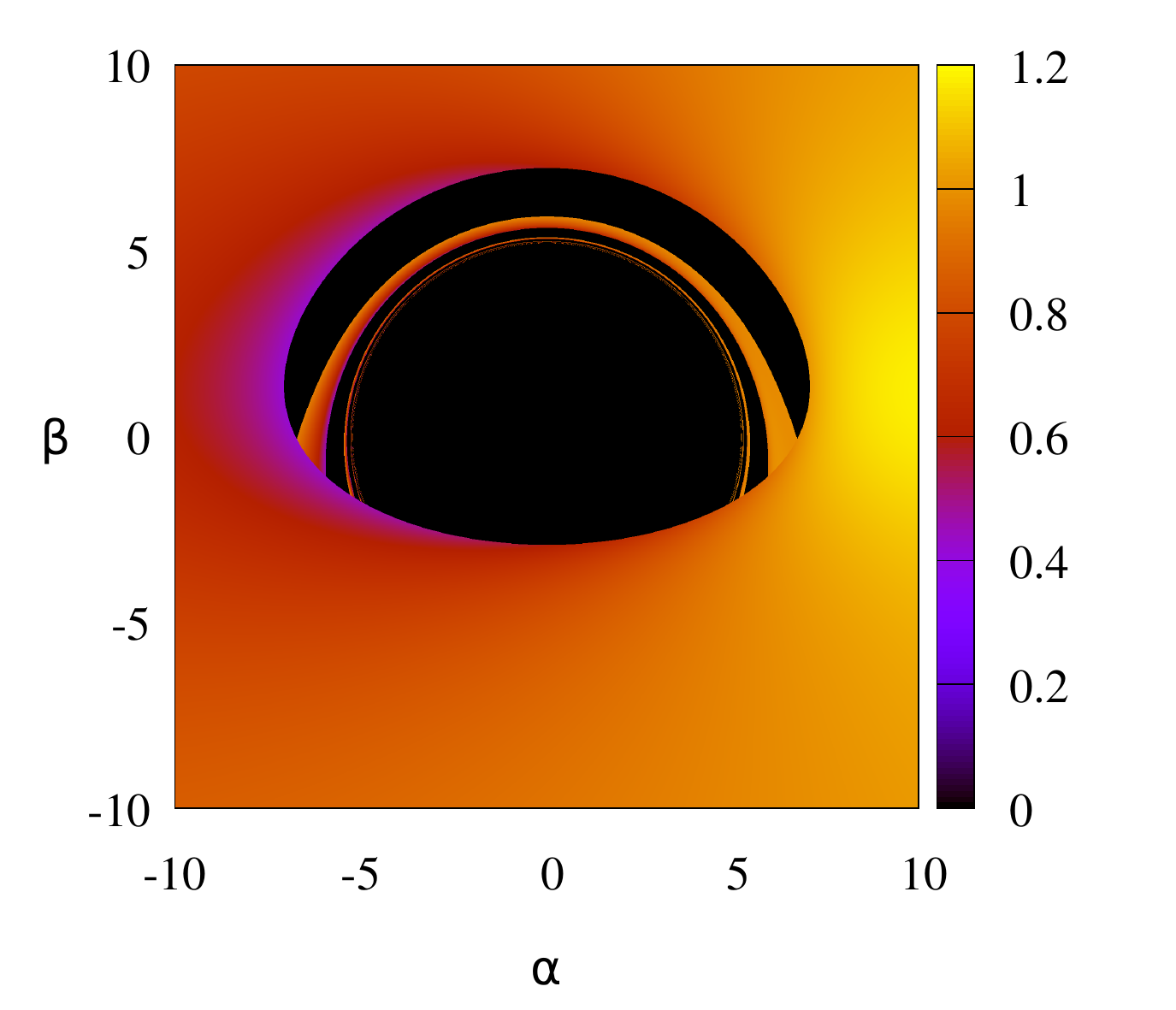}\\
			\includegraphics[scale=0.45]{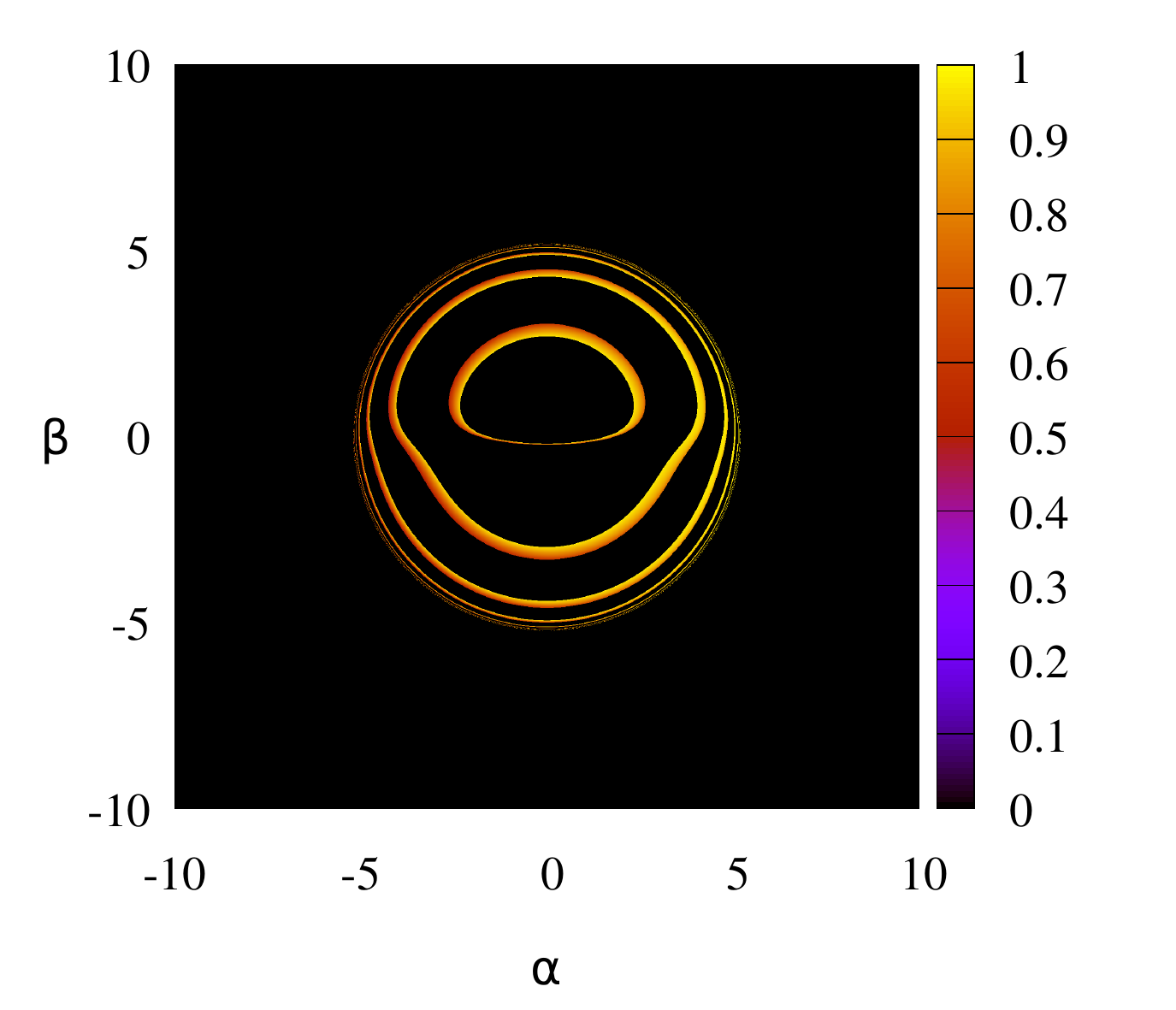} &	\includegraphics[scale=0.45]{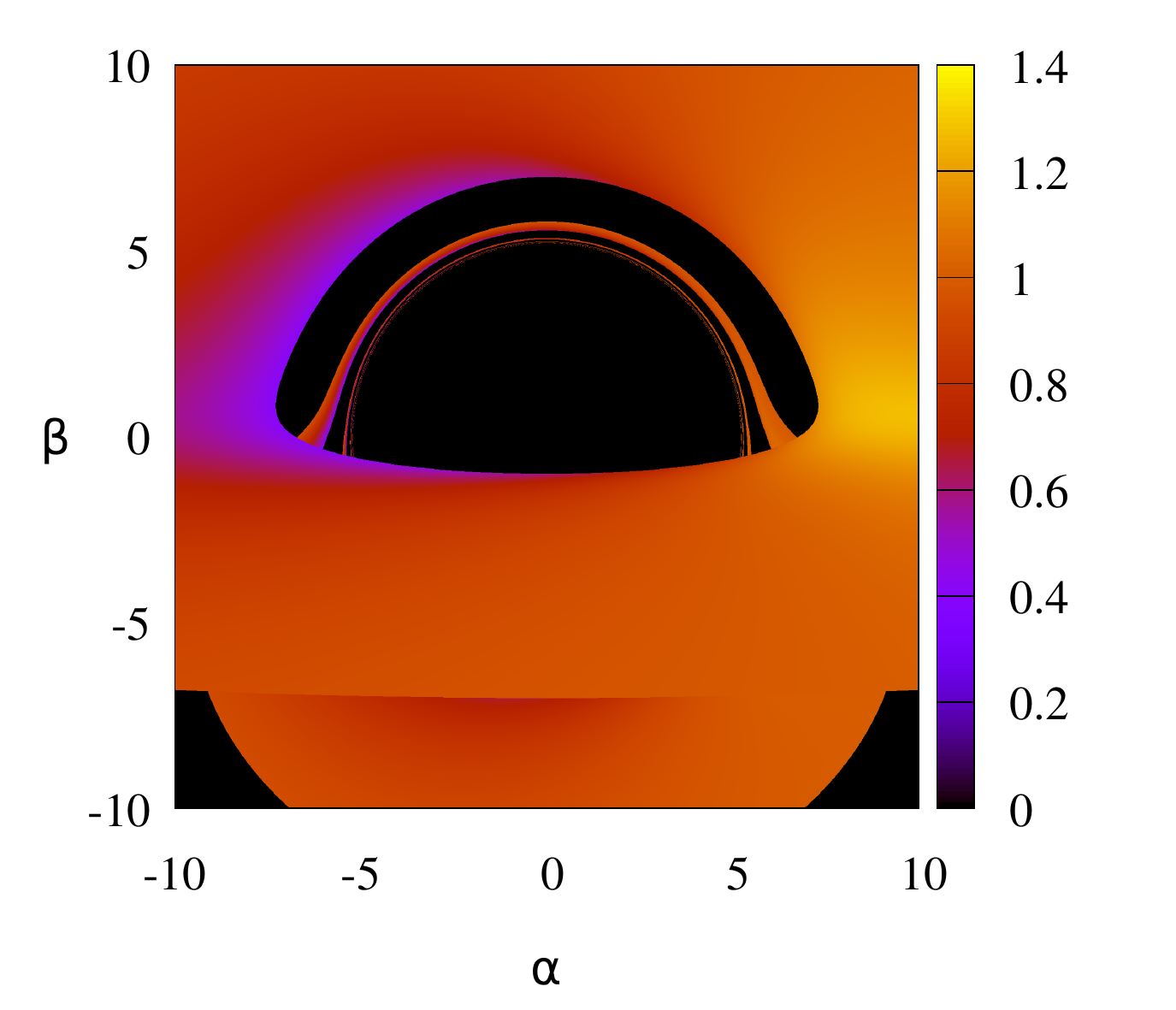}
		\end{tabular}
		\caption{The images of Keplerian disks rotating around wormhole on the other side (left), rep. on our side (right). The inclination of the observer, located at $r_o=10^3$M is $\theta_o=30^\circ$ (top), $60^\circ$ (middle) and $80^\circ$ (bottom). The wormhole parameter $a=3.1$. The colors of the plot represent frequency shift of radiation from the disk. \label{g2img}}
	\end{center}
\end{figure}

\begin{figure}[H]
	\begin{center}
		\begin{tabular}{cc}
			\includegraphics[scale=0.45]{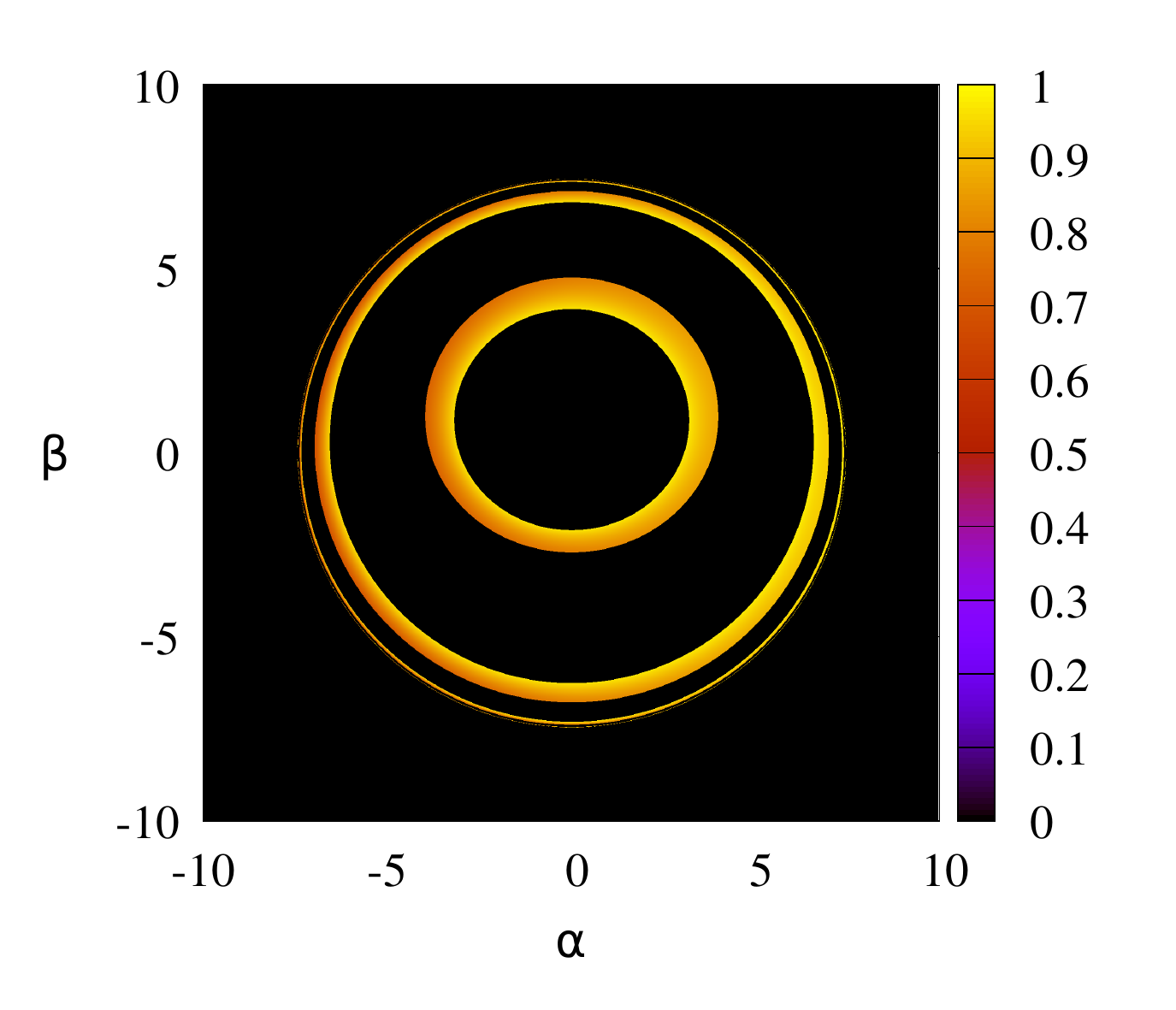} &	\includegraphics[scale=0.45]{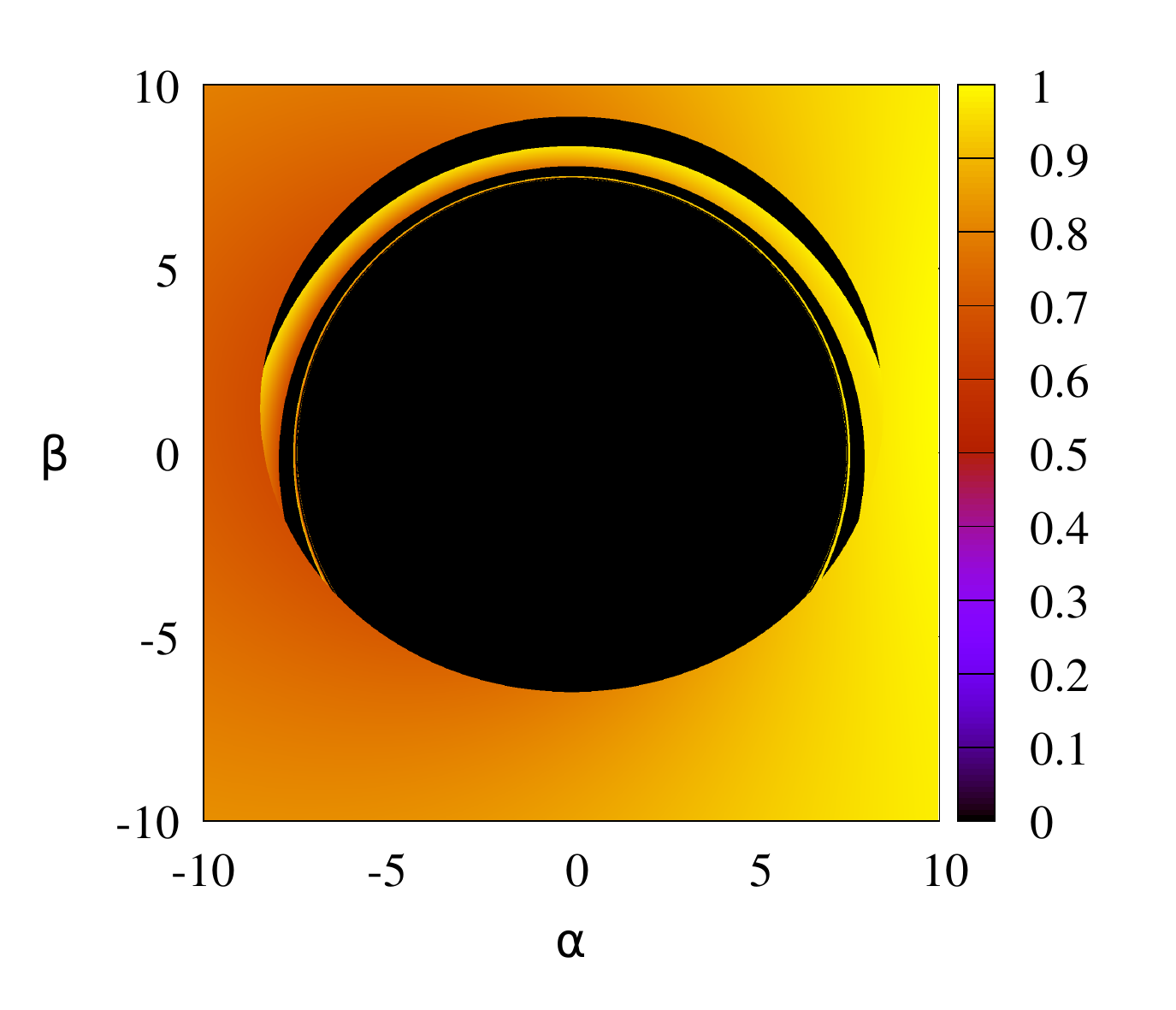}\\
			\includegraphics[scale=0.45]{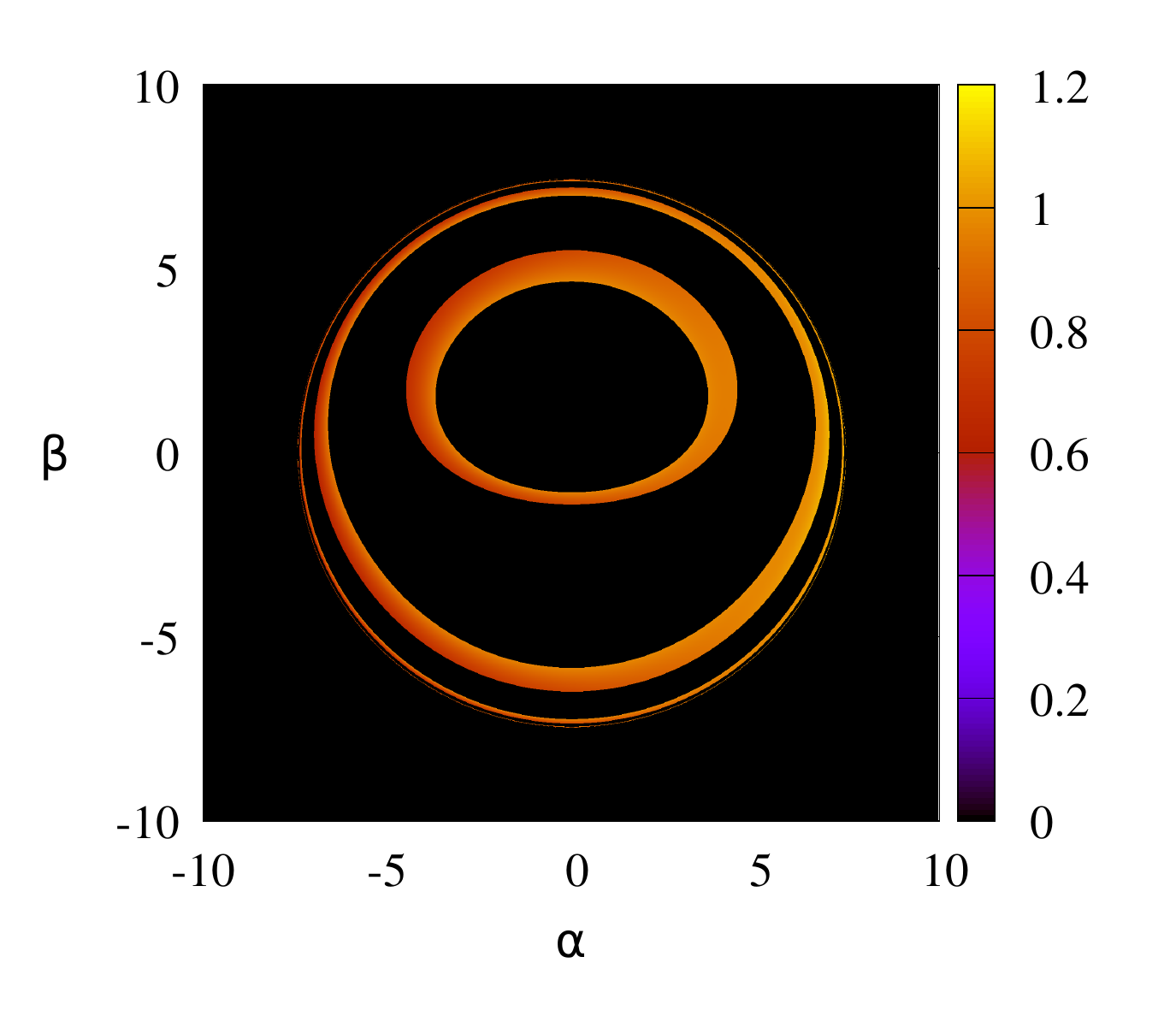} &	\includegraphics[scale=0.45]{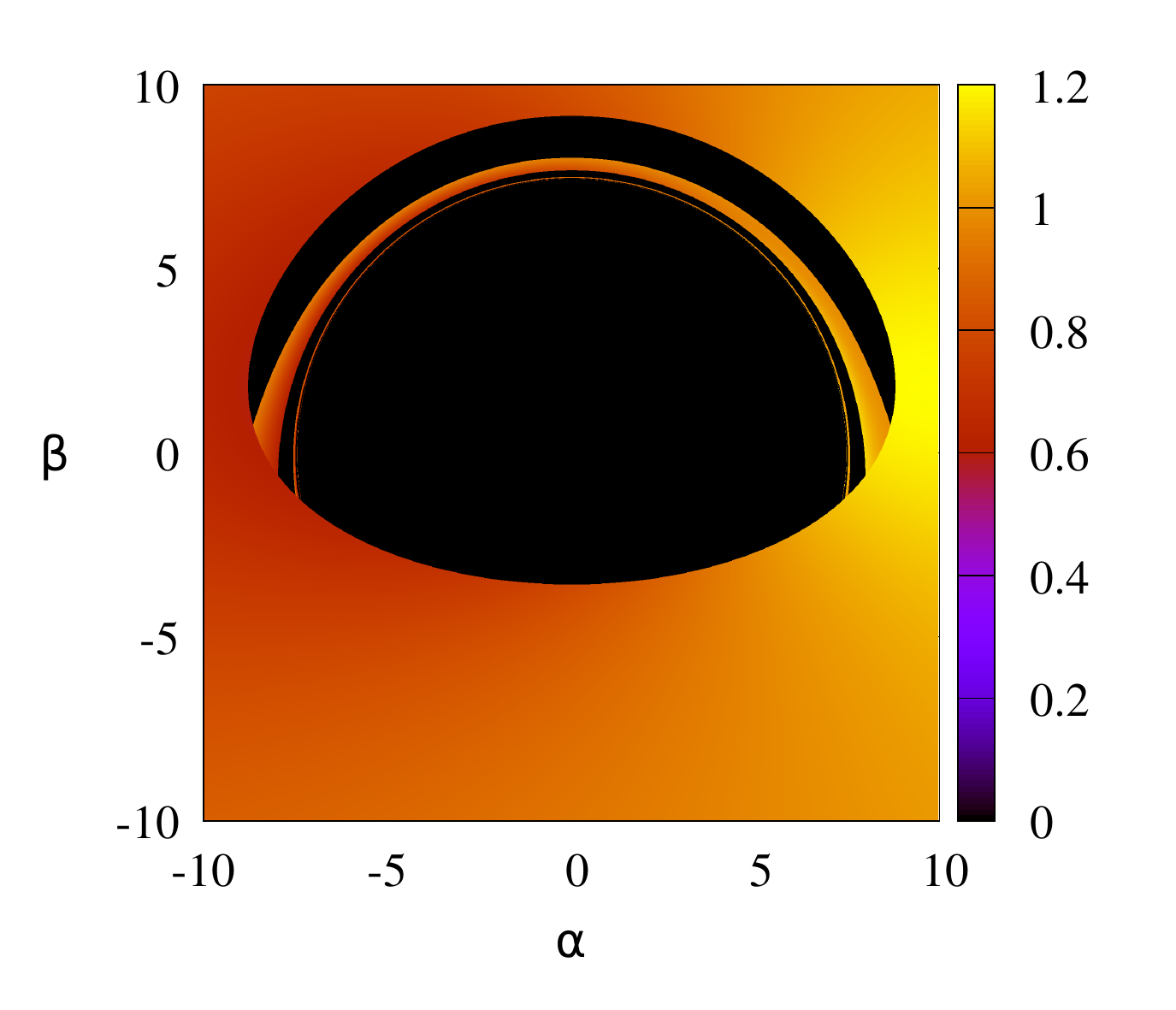}\\
			\includegraphics[scale=0.45]{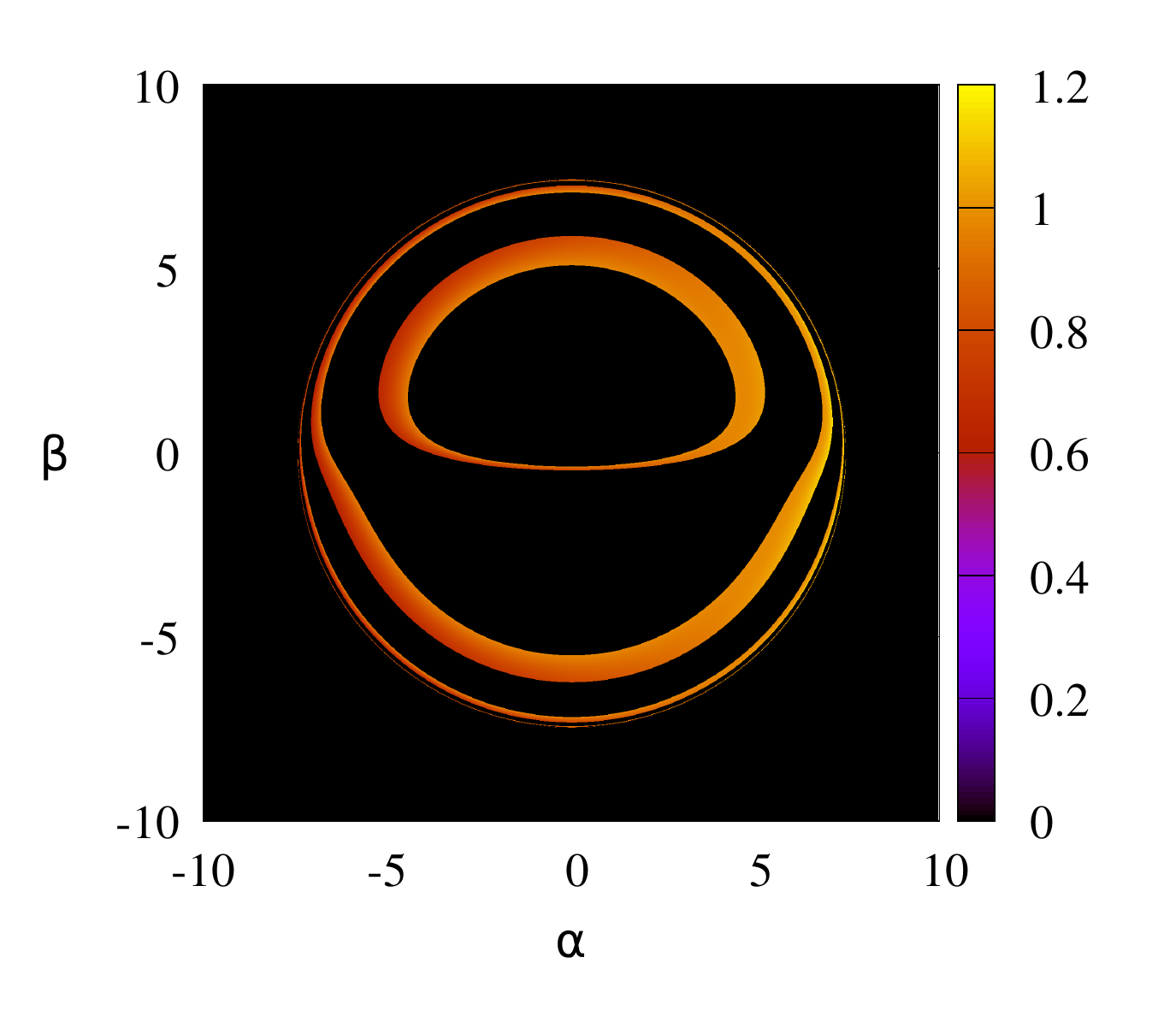} &	\includegraphics[scale=0.45]{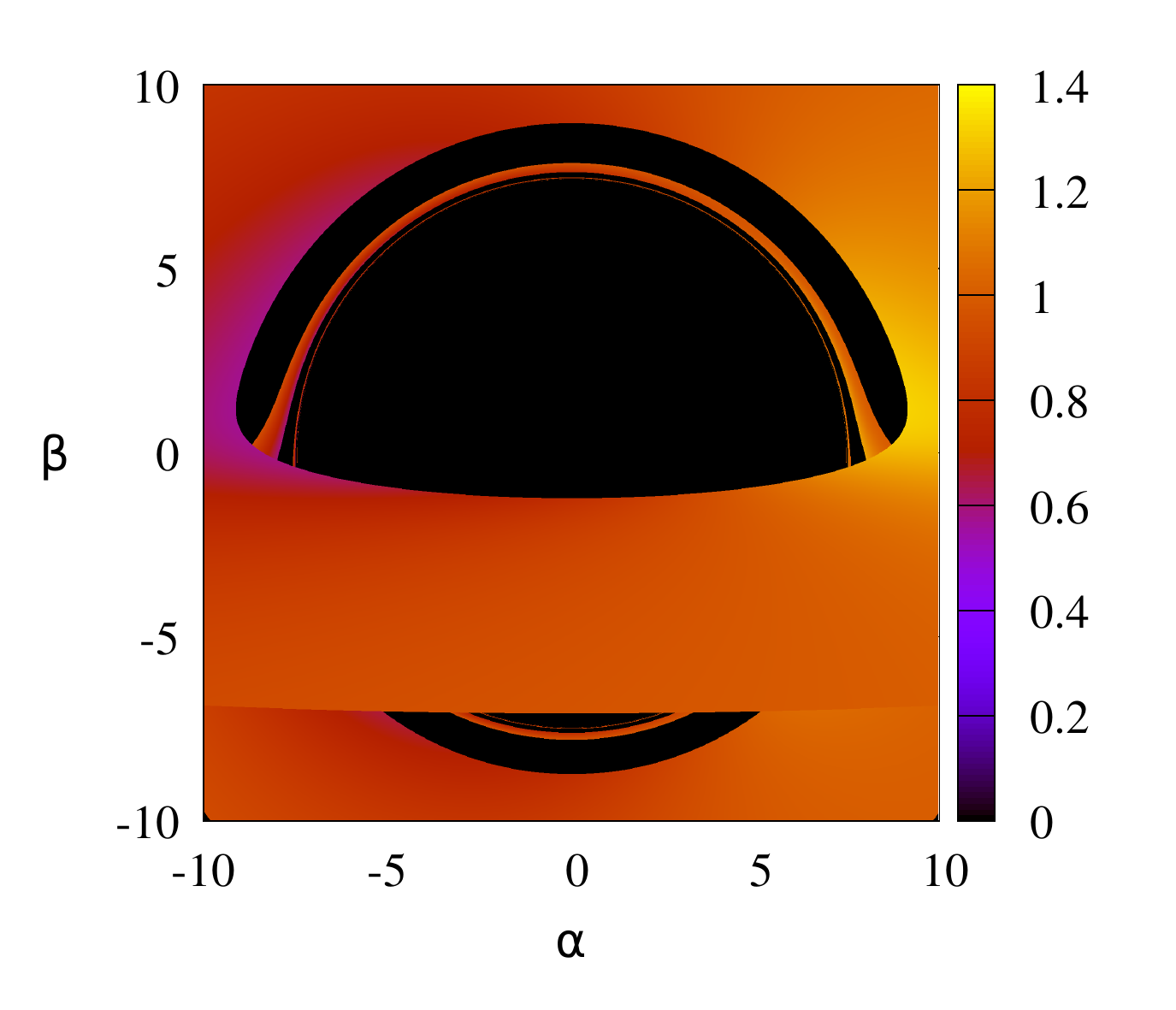}
		\end{tabular}
		\caption{The images of Keplerian disks rotating around wormhole on the other side (left), rep. on our side (right). The inclination of the observer, located at $r_o=10^3$M is $\theta_o=30^\circ$ (top), $60^\circ$ (middle) and $80^\circ$ (bottom). The wormhole parameter $a=6.1$. The colors of the plot represent frequency shift of radiation from the disk. \label{g3img}}
	\end{center}
\end{figure}

\begin{figure}[H]
	\begin{center}
		\begin{tabular}{cc}
			\includegraphics[scale=0.55]{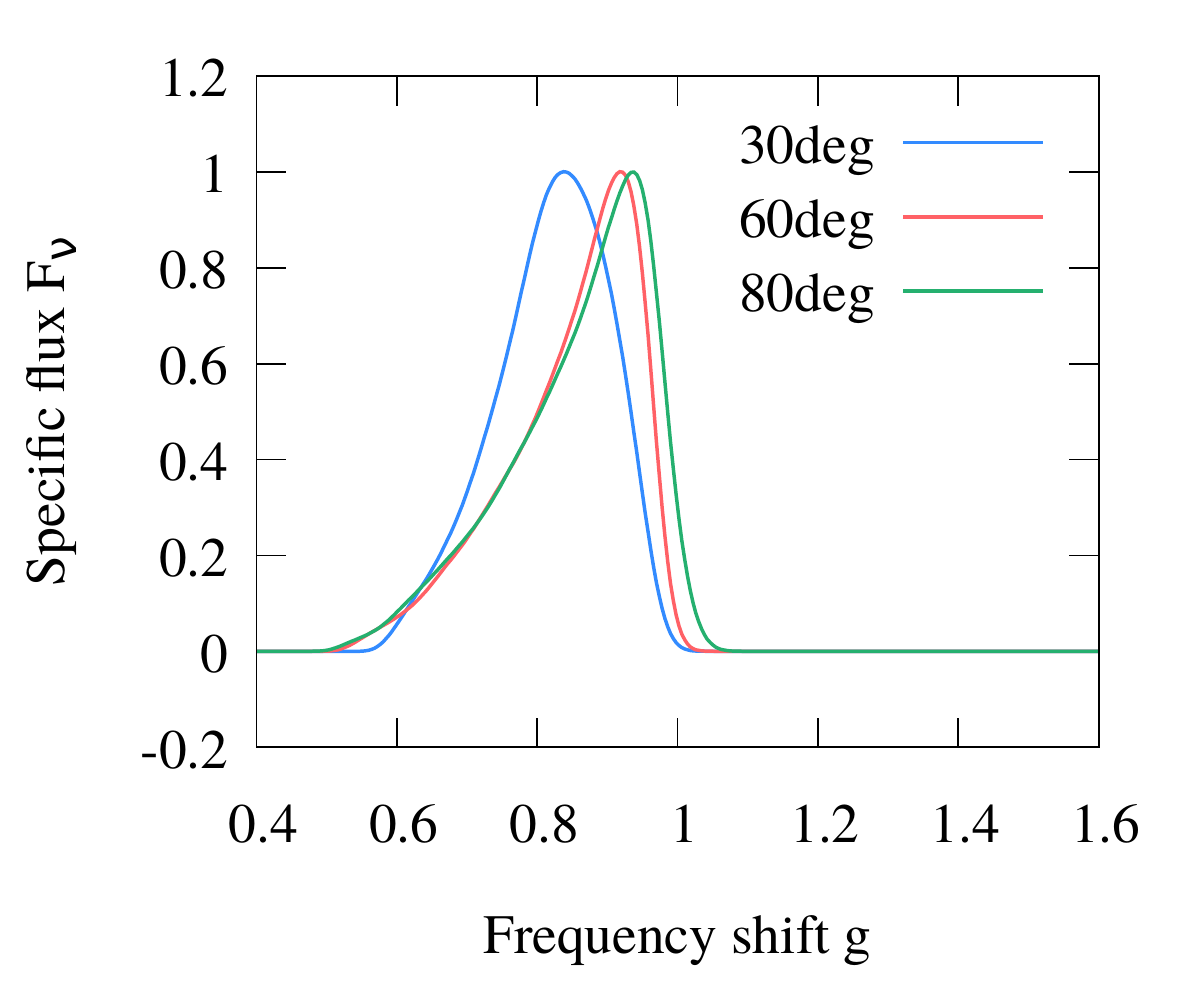}&
			\includegraphics[scale=0.55]{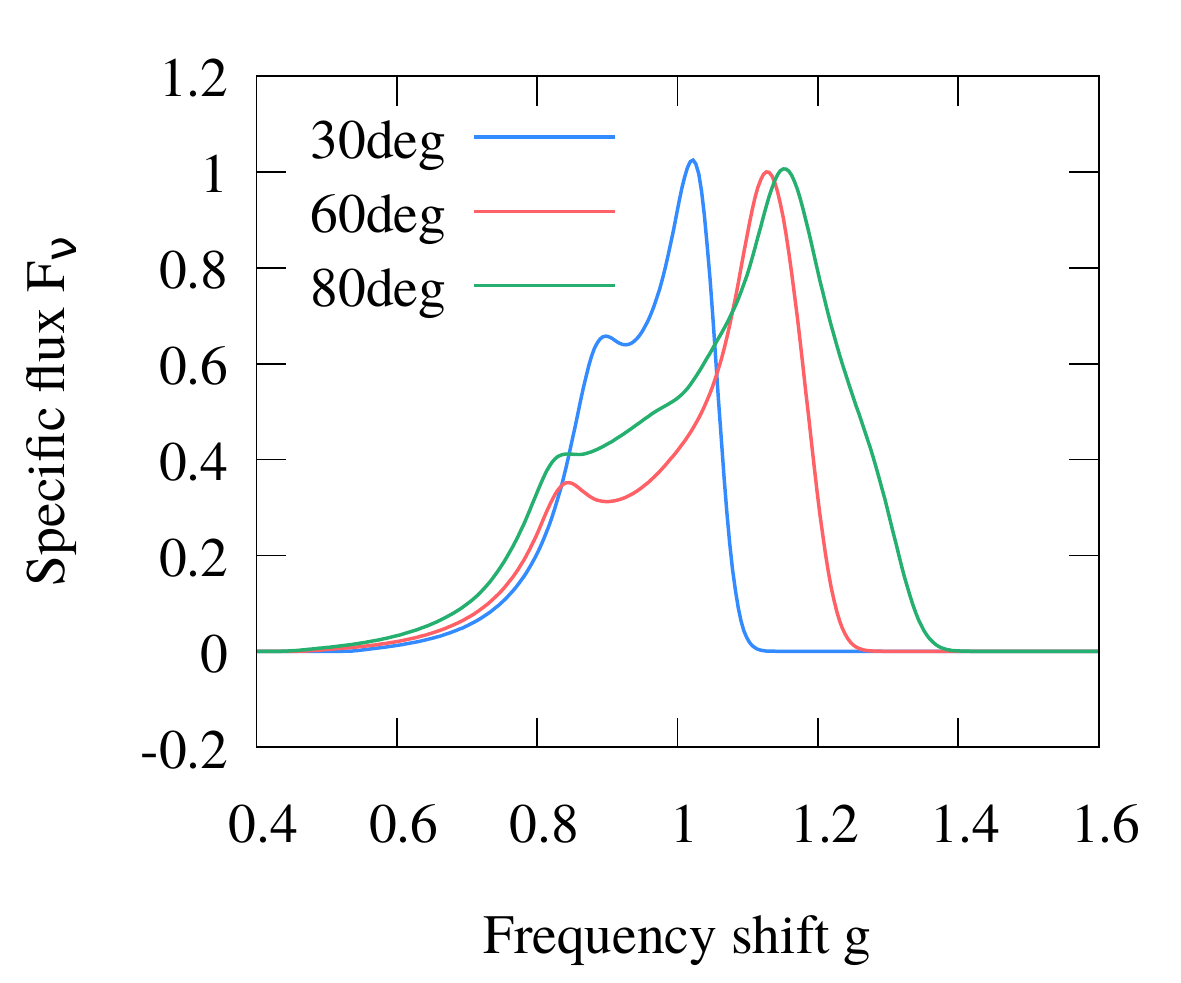}\\
			\includegraphics[scale=0.55]{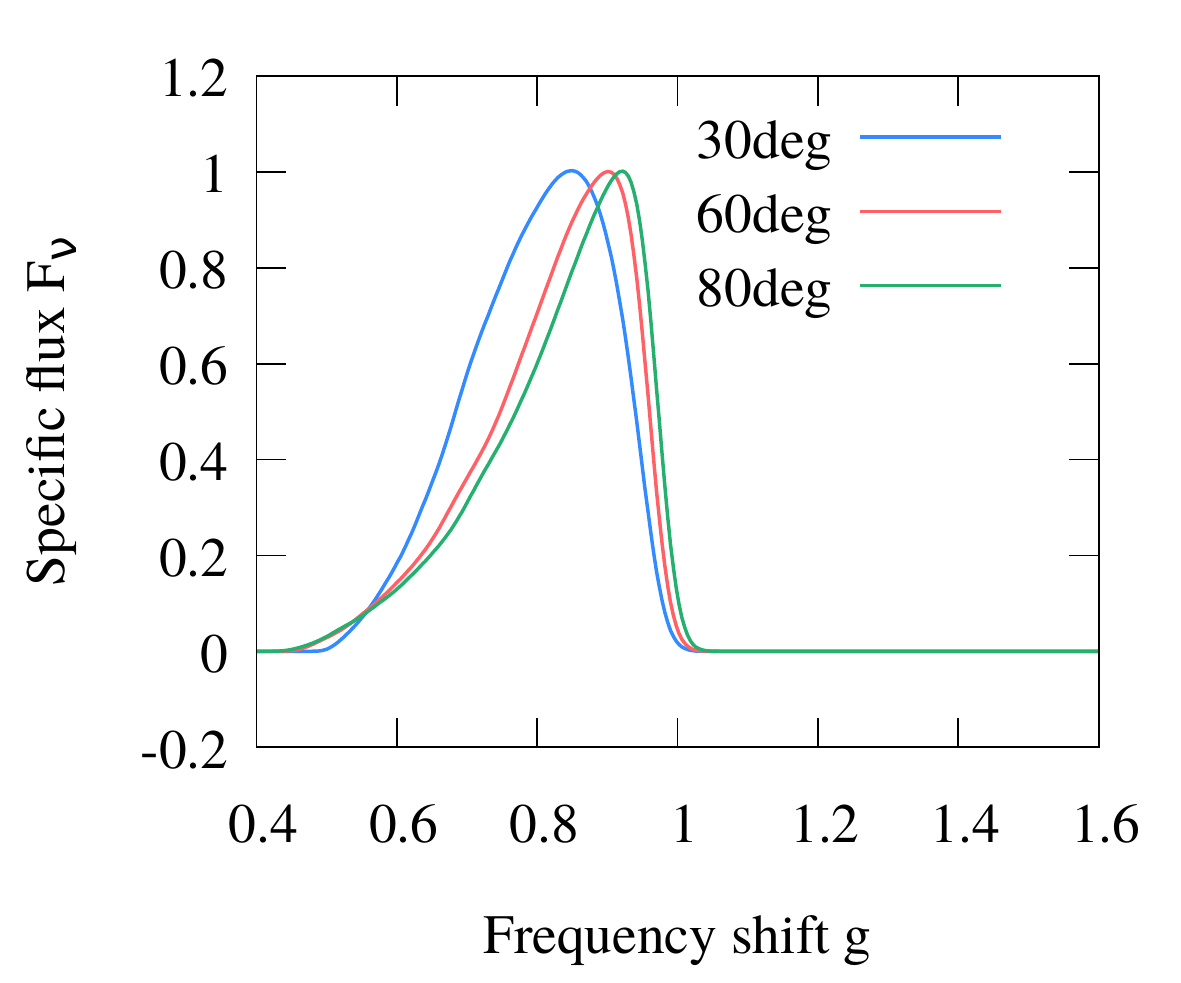}&
			\includegraphics[scale=0.55]{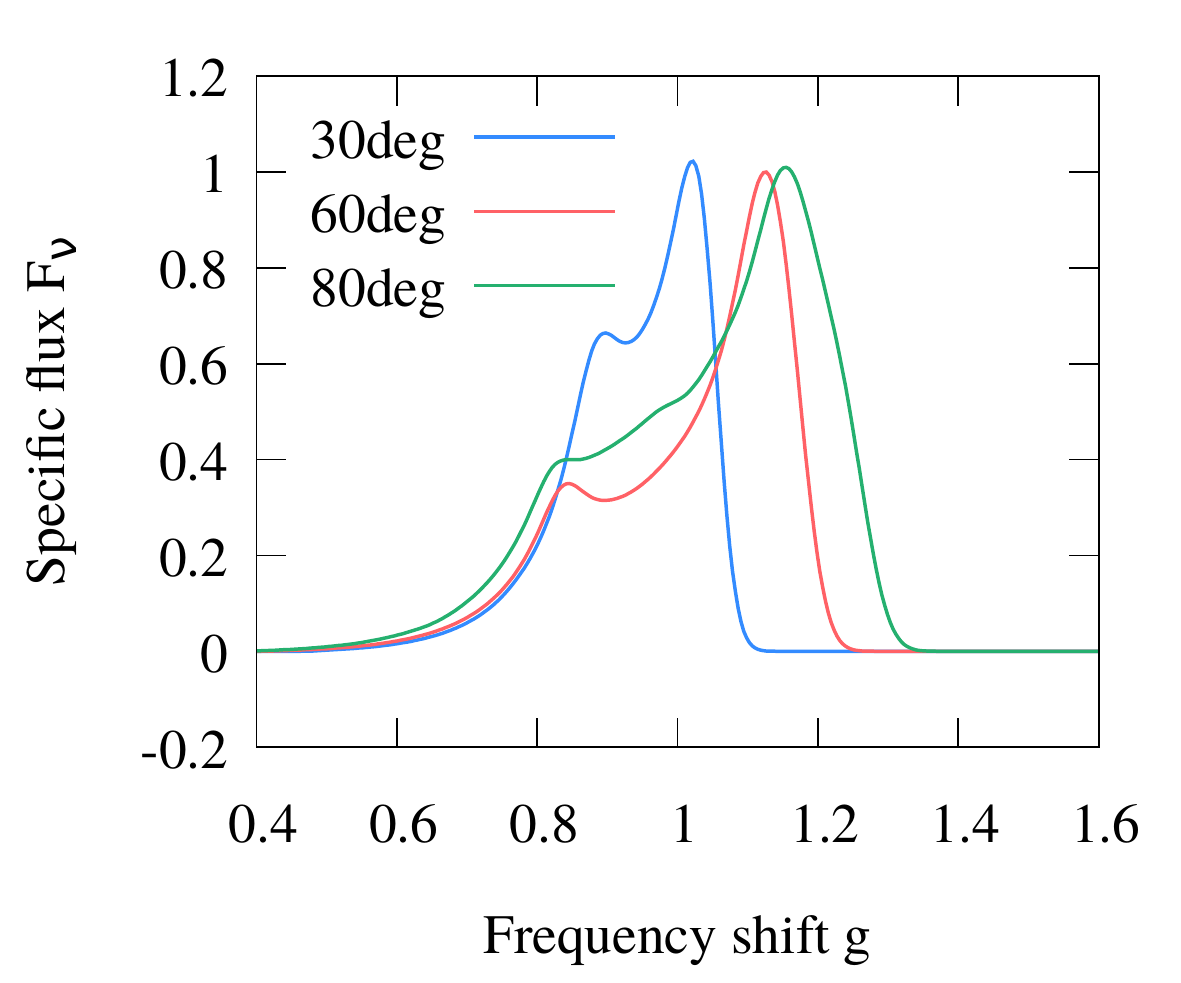}\\
			\includegraphics[scale=0.55]{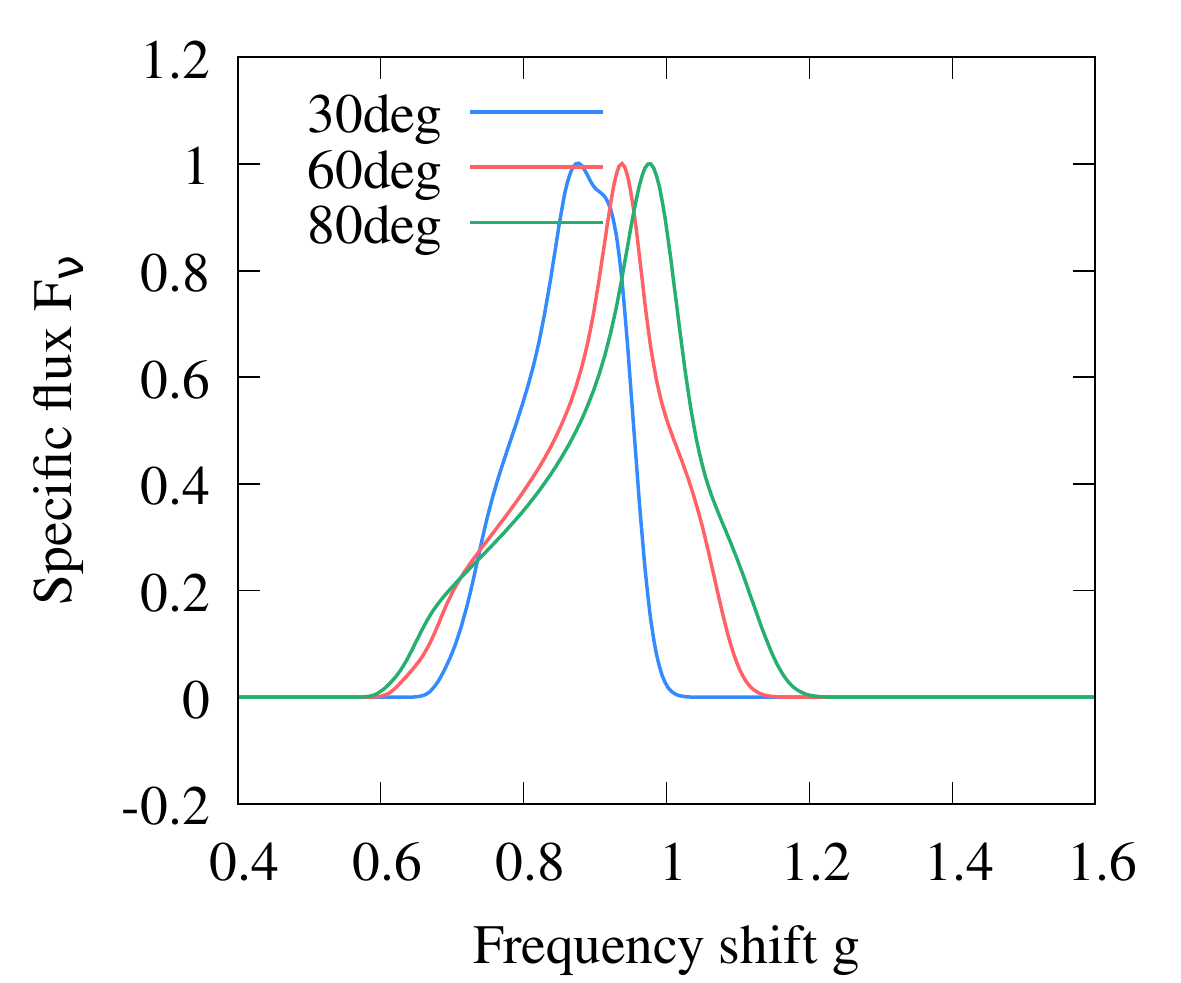}&
			\includegraphics[scale=0.55]{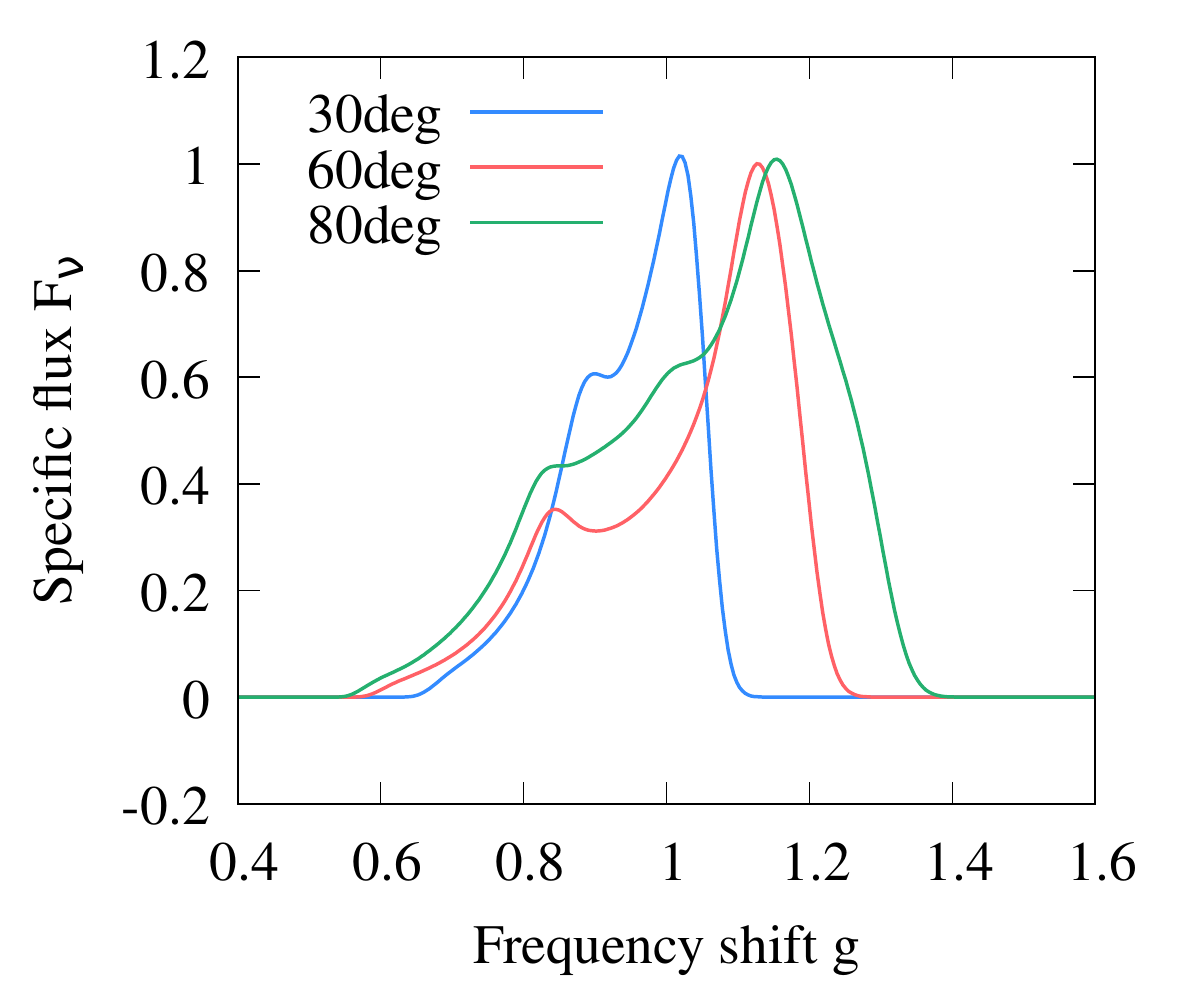}
		\end{tabular}
	\caption{Profiled line of radiation from Keplerian disk in our part of the universe (right) and in the outer part of the universe (left) generated for three representative values of wormhole parameter $a=2.1$ (top), $3.1$ (middle), and $6.1$ (bottom) and three representative values of observer inclination $\theta_o=30^\circ$, $60^\circ$, and $80^\circ$.\label{pl1}}
	\end{center}
\end{figure}

\begin{figure}[H]
	\begin{center}
		\begin{tabular}{cc}
			\includegraphics[scale=0.55]{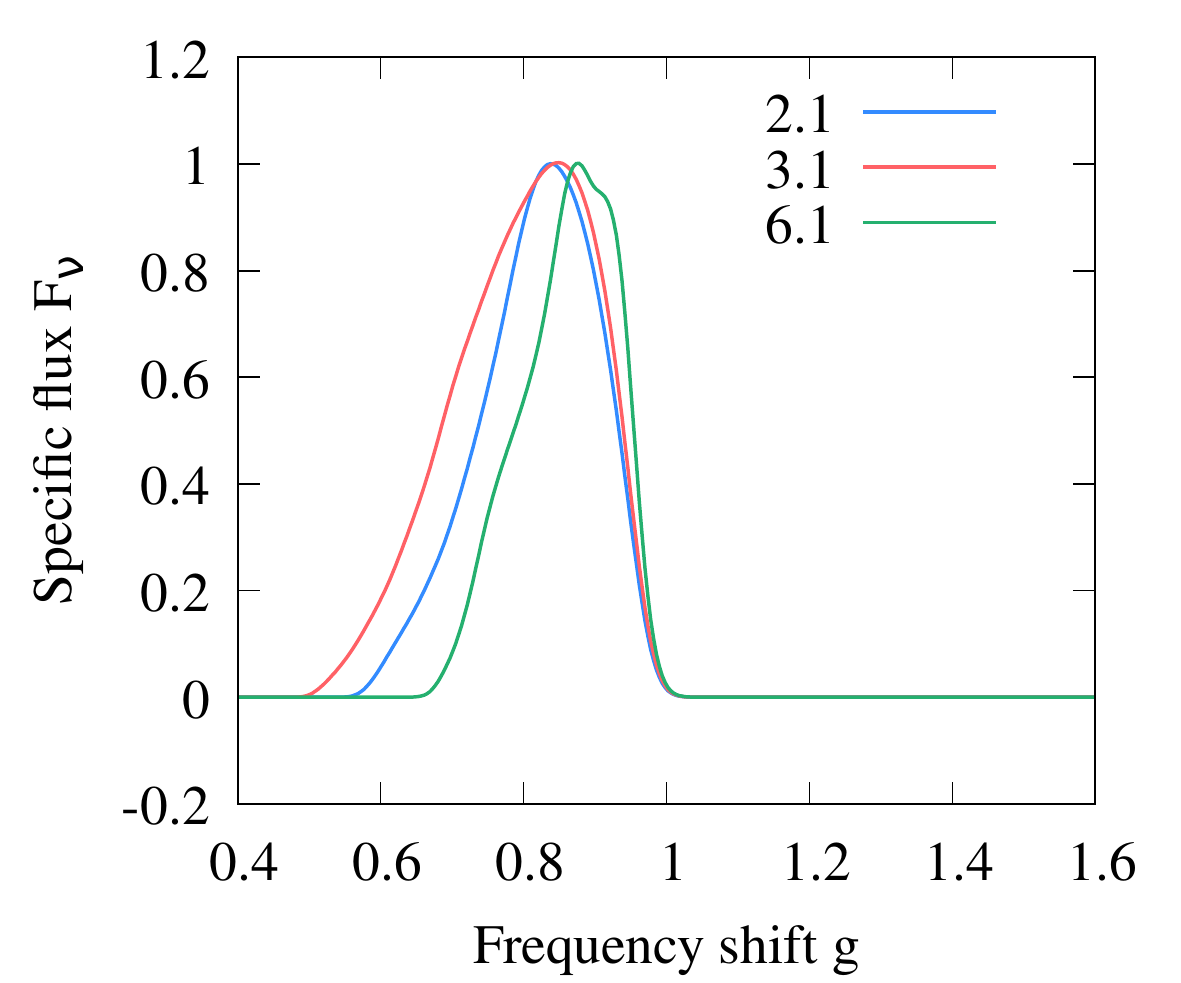}&
			\includegraphics[scale=0.55]{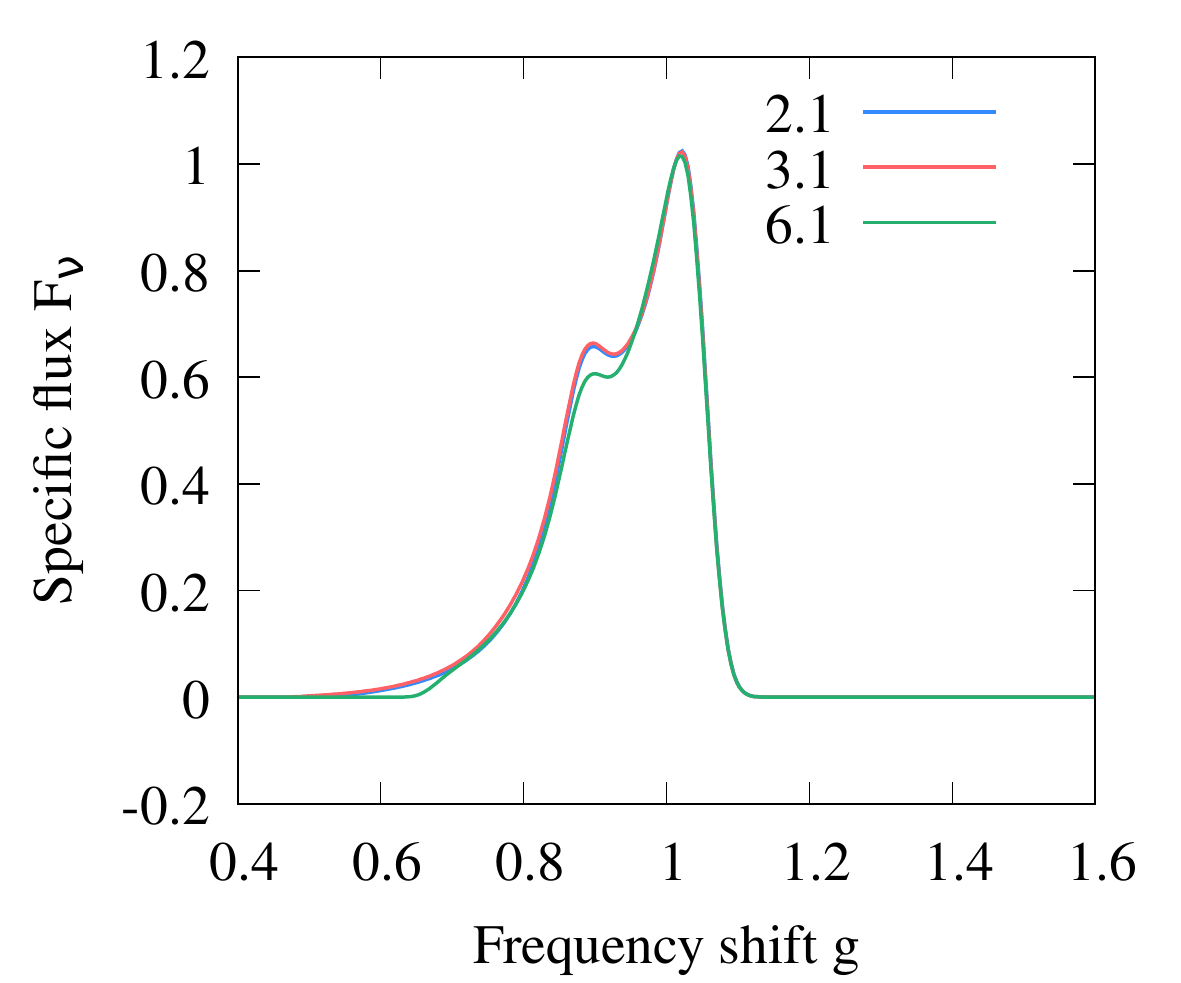}\\
			\includegraphics[scale=0.55]{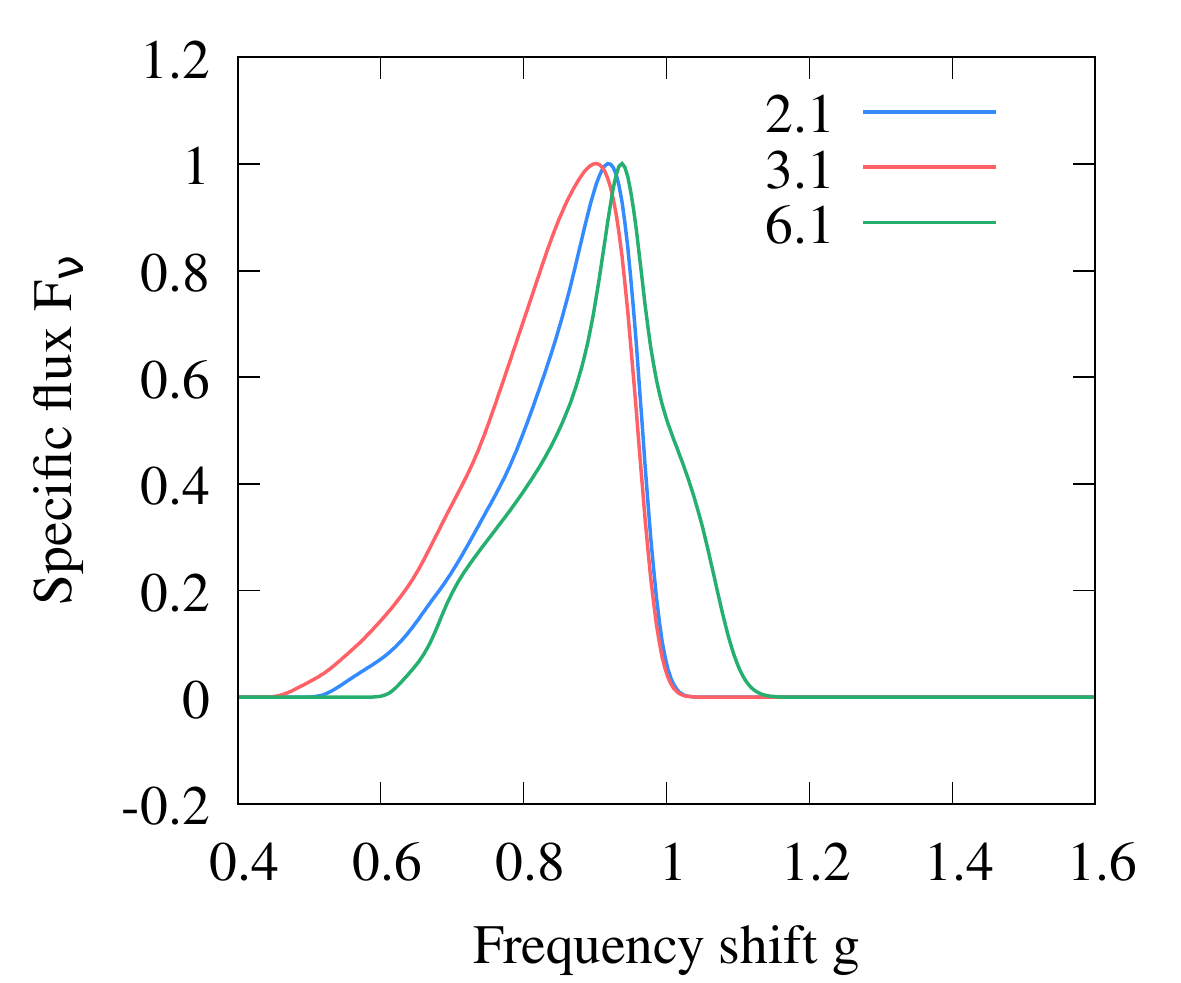}&
			\includegraphics[scale=0.55]{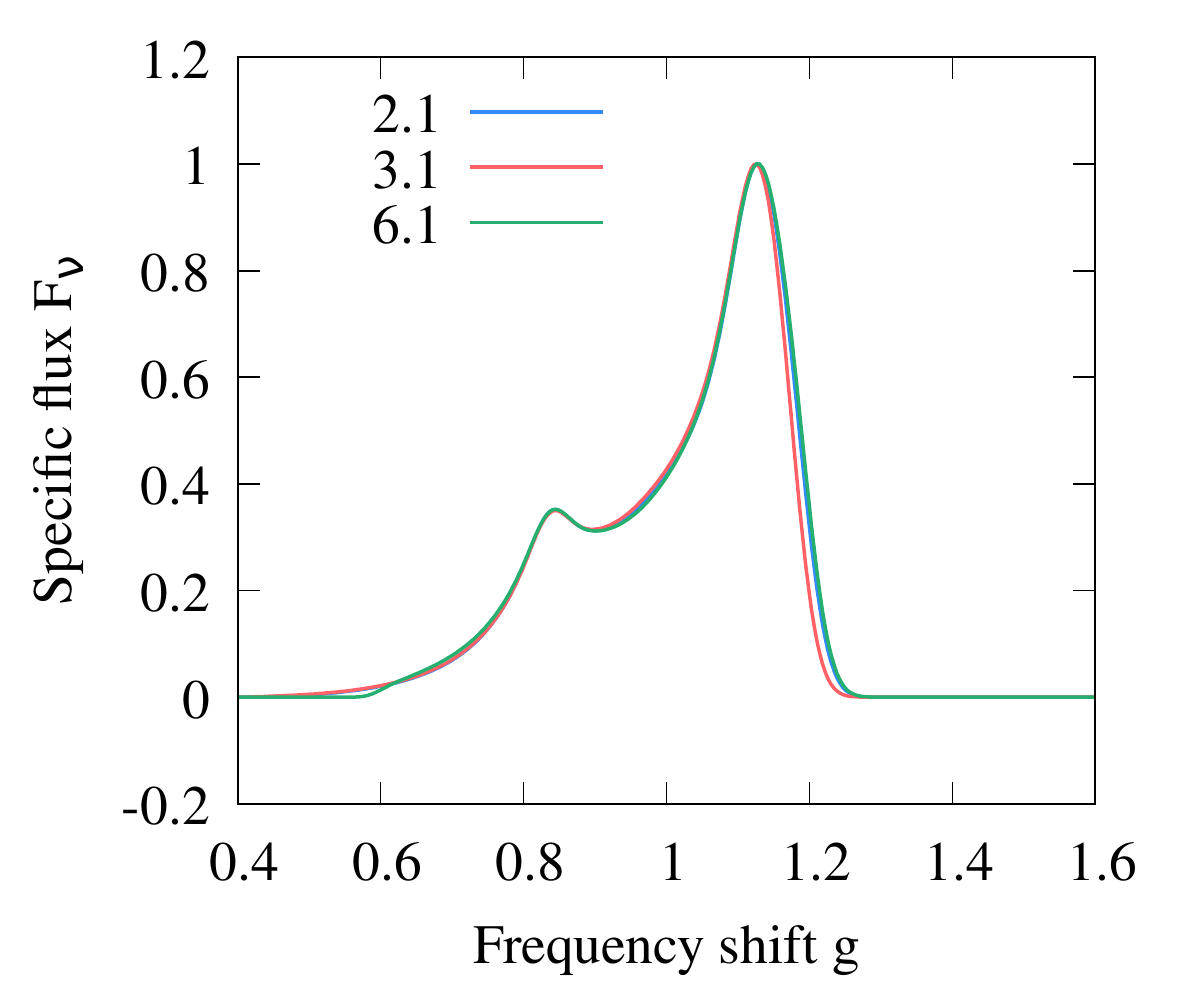}\\
			\includegraphics[scale=0.55]{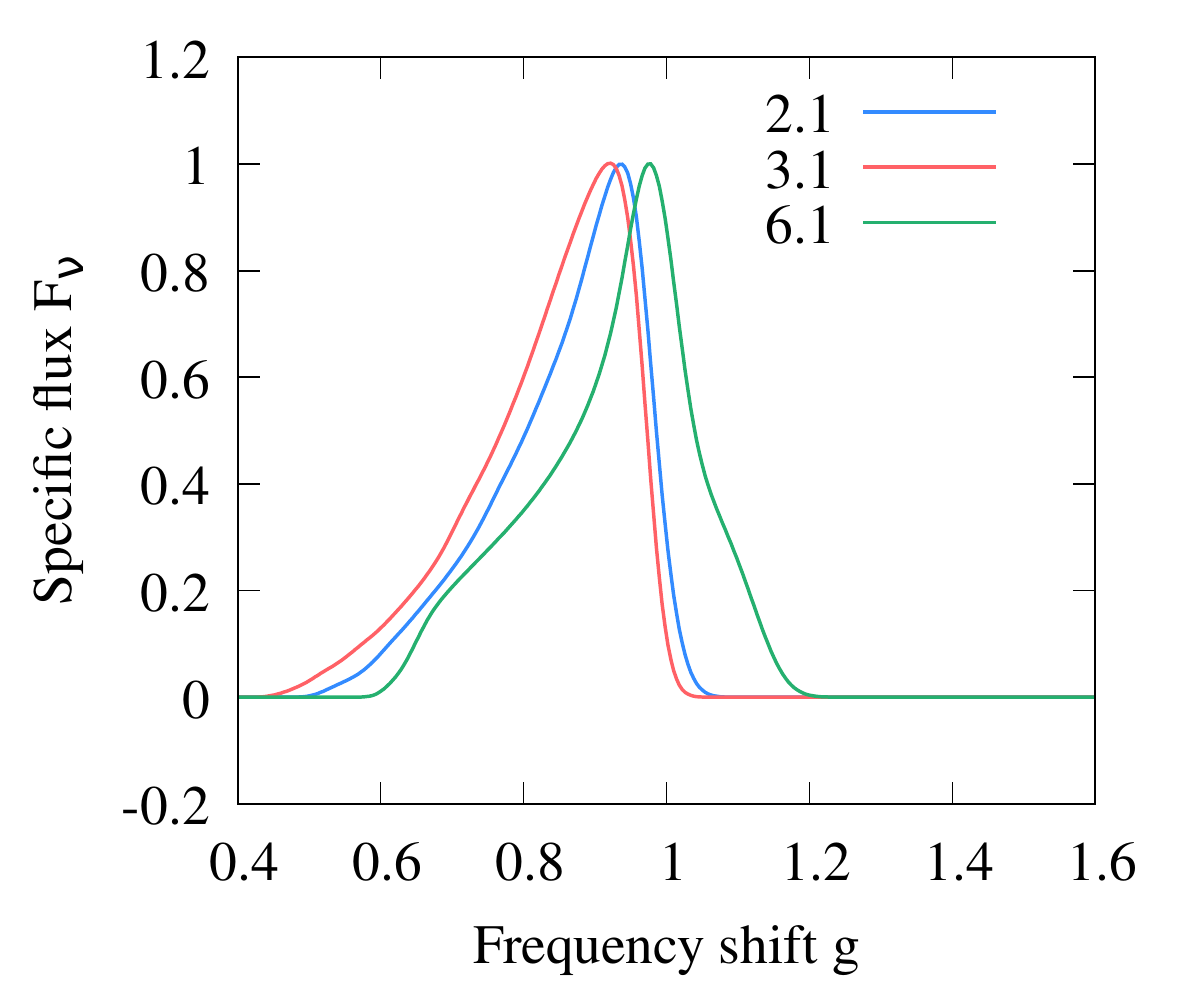}&
			\includegraphics[scale=0.55]{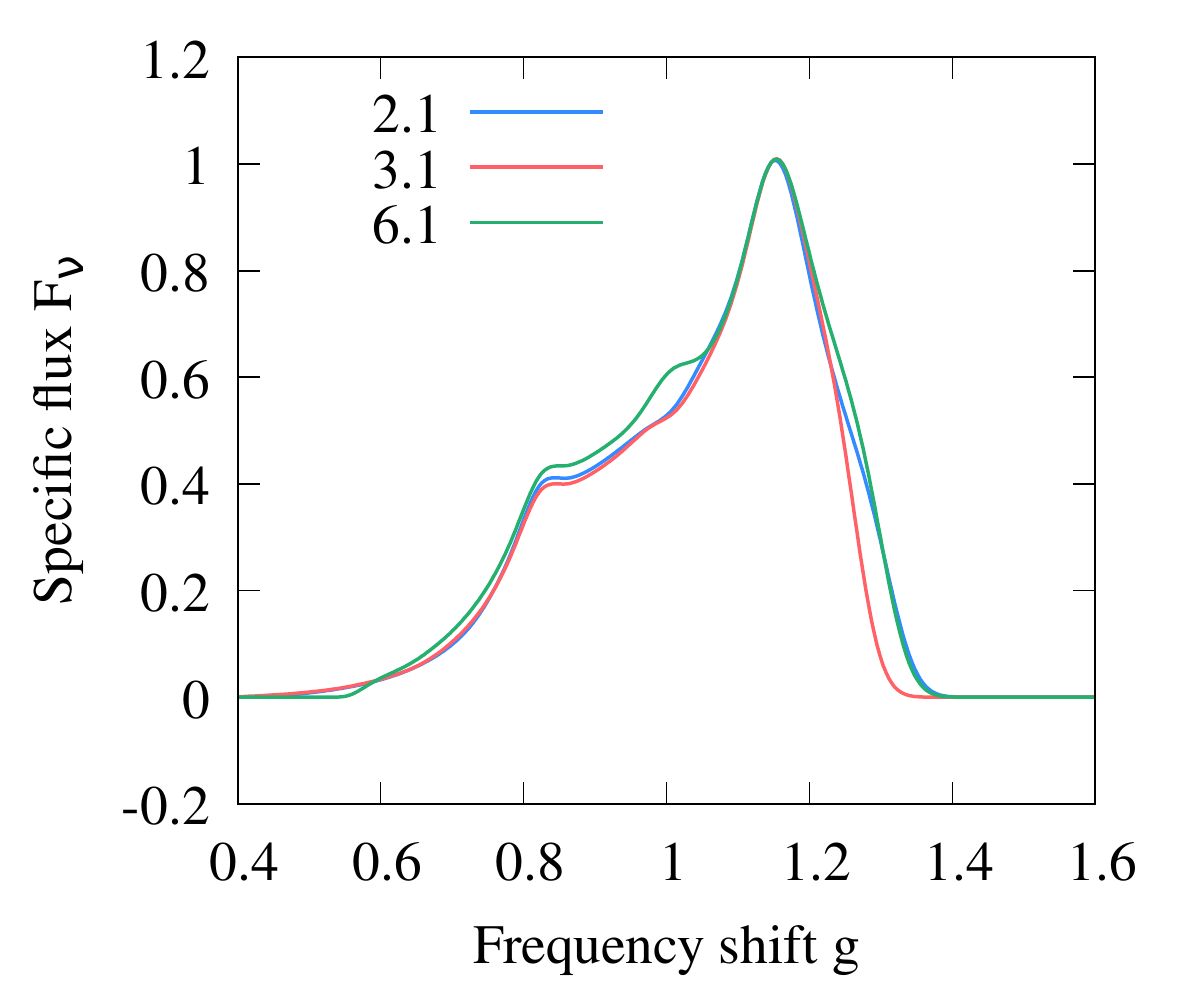}
		\end{tabular}
		\caption{Profiled line of radiation from Keplerian disk in our part of the universe (right) and in the outer part of the universe (left) generated for three representative of observer inclination $\theta_o=30^\circ$ (top), $60^\circ$ (middle), and $80^\circ$ (bottom) and for three representative values of wormhole parameter $a=2.1$ (blue), $3.1$ (red), and $6.1$ (green).\label{pl2}}
	\end{center}
\end{figure}

First, we give the bolometric images of the Keplerian (Novikov-Thorne) discs for three representative values of the wormhole parameter: $a=2.1$ (Fig. \ref{bolo1}), $3.1$ (Fig. \ref{bolo2}), and $6.1$ (Fig. \ref{bolo3}). In the second set of images we present the frequency shift map of the disc surface projected on the image plane, again for the three representative values of wormhole parameter: $a=2.1$ (Fig. \ref{g1img}), $3.1$ (Fig. \ref{g2img}), and $6.1$ (Fig. \ref{g3img}). In the last set of images we give plots of the spectral line profile. We compare the lines for different three representative values of the observer inclination $\theta_o=30^\circ$, $60^\circ$, $80^\circ$, while the wormhole parameter $a$ is fixed, in Fig \ref{pl1}, and the effect of the wormhole parameter $a=2.1$, $3.1$, $6.1$ on the spectral profiled line with the inclination angle $\theta_o$ kept fixed, in Fig. \ref{pl2}. 

\section{Discussion and Conclusions}
In our simulations we constructed image of the Keplerian disc located in the lower (our side, setup A) universe, and in the upper universe (the other side, setup B). In the first set of our results we present the appearance of the Novikov-Thorne disk, assuming its material to be opaque. The distribution of the bolometric flux over the image of the disk is presented in images Figs.\ref{bolo1} - \ref{bolo3}. The left (right) column shows the disc located in the upper (lower) universe. One can clearly distinguish between those two cases. The image of the disc located in the upper universe is located inside the region with boundary corresponding to image of the wormhole throat, while the image of the disk located in the lower universe in outside of this region and the shape correspond to the generic shape of the Keplerian disc in general spherically symmetric spacetimes possessing a photon orbit.

By inspection of the upper universe Keplerian disk image, one finds out that the number, $n$, of the disc images decreases with increasing value of parameter $a$: $(a,n)=(2.1,5),\,(3.1,4),\,(6.1,3)$. This effect is tightly connected with existence of photon circular orbit (photon sphere). In case $a<3M$ there is a circular orbit and we obtain multiple images of the disk while for $a>3M$ there is no photon sphere and therefore there is lack of disk images of higher order.

In the Figs \ref{pl1} - \ref{pl2} we give the plots of profiled lines corresponding to the models of Keplerian discs that are emitting monochromatic radiation. Again, there is clear distinction between the line profiles of radiation emitted by the disc in the upper universe in comparison with the equivalent disc located in the lower universe. In the case A, the lines do not show typical double horn split of the Gaussian bell, and their width is smaller in comparison to what one would observe in case of the profiled line generated by the disc located in the lower universe (which also corresponds to the classical black hole case), making the wormhole easily distinguishable from classical black hole provided there is a disk formed in the outer universe. 

We can thus conclude that even for the reflection-symmetric wormholes, the optical phenomena related to the Keplerian discs located with reflection-symmetry of both sides of the wormhole enable clear distinguishing of the images coming from different sides of the wormhole. The observed significant differences allow for such distinguishing even in the special Simpson-Visser wormhole spacetime where the optical phenomena related to the disc located in our side of the wormhole are very close to those of the standard Schwarzschild black hole. 

\section*{Acknowledgment}
The authors acknowledge support of the Research Centre for Theoretical Physics and Astrophysics, Istitute of Physics, Silesian University in Opava, and of the Czech Science Foundation grant No. 16-03564Y.

\newpage
\appendix

\section{Wormhole Stress-Energy Tensor}
In the static orthonormal frame with vector base $\vec{e}_{\hat{a}}$
\begin{equation}
	\vec{e}_{\hat{0}}=\frac{1}{\sqrt{f(r)}}\vec{e}_t\,,\quad 
	\vec{e}_{\hat{1}}=\sqrt{f(r)}\vec{e}_r\,,\quad
	\vec{e}_{\hat{2}}=\frac{1}{\sqrt{h(r)}}\vec{e}_\theta\,,\quad
	\vec{e}_{\hat{3}}=\frac{1}{\sqrt{h(r)}\sin\theta}
\end{equation}
the Einstein equations read
\begin{equation}
	G_{\hat{a}\hat{b}}=8\pi\,\kappa\,T_{\hat{a}\hat{b}}.\label{EEQ}
\end{equation}
Now, we have a spacetime in our hands and consequently we can determine components of Einstein tensor $G_{\hat{a}\hat{b}}$.
The stress energy tensor, $T_{\hat{a}\hat{b}}$, having the same algebraic structure as Einstein tensor, has four non-zero components $T_{\hat{0}\hat{0}}$, $T_{\hat{1}\hat{1}}$, $T_{\hat{2}\hat{2}}$, $T_{\hat{3}\hat{3}}$ interpreted, in this frame, as $T_{\hat{0}\hat{0}}=\rho(r)$, $T_{\hat{1}\hat{1}}=-\tau(r)$, $T_{\hat{2}\hat{2}}=T_{\hat{3}\hat{3}}=p(r)$ where is $\rho(r)$ the total energy density, $\tau(r)$ is the tension per unit area that is measured in radial direction, and $p(r)$ is the pressure measured in the lateral directions. From Einstein equations (\ref{EEQ}) we obtain the stress-energy tensor components in the form
\begin{equation}
	T_{\hat{a}\hat{b}}=\frac{1}{8\pi\,\kappa}G_{\hat{a}\hat{b}}.
\end{equation}
Functions $\rho(r)$, $\tau(r)$, and $p(r)$ then read
\begin{eqnarray}
	\rho(r)&=&-\frac{a^2}{8\pi \kappa\,h^{5/2}(r)}\left(h^{1/2}(r)-4\right),\label{rho}\\
	\tau(r)&=&\frac{a^2}{8\pi\,\kappa\,h^{2}(r)},\label{tau}\\
	p(r)&=&\frac{a^2}{8\pi\,\kappa\,h^{5/2}}\left(h^{1/2}(r)-1\right).\label{p}
\end{eqnarray}
Recall that metric (\ref{wh}) describes a wormhole provided that spacetime parameter $a>2$, implying $h^{1/2}(r)>2$ for all $r$. The physical quantities $\rho$, $\tau$, $p$ describing material supporting wormhole from its collapse are subjects to  NEC and SEC stating:
\begin{itemize}
	\item \emph{NEC}
		\begin{equation}
			\rho+p\ge 0\quad\mathrm{and}\quad \rho-\tau\ge 0\label{NEC}
		\end{equation}
	\item \emph{SEC}
		\begin{equation}
		\rho-\tau+2p\ge 0.\label{SEC}
		\end{equation}
\end{itemize}
Let's inspect conditions (\ref{NEC}) first. Using (\ref{rho}) - (\ref{p}) in (\ref{NEC}) we obtain inequality
\begin{equation}
	\rho(r)+p(r)=\frac{3a^2}{8\pi\kappa h^{5/2}(r)}>0\quad\forall\, r.
\end{equation}
Now, let's check the second inequality of (\ref{NEC}) and gradually obtain it in the form 
\begin{equation}
	\rho(r)-\tau(r)=-\frac{a^2}{4\pi\kappa h^{5/2}(r)}(h^{1/2}(r)-2)<0\quad \forall\,r.
\end{equation}
On can state that the NEC is not fulfilled by material supporting this wormhole.

Let's turn our attention to validity of SEC (formula (\ref{SEC})). After some algebra one arrives to inequality
\begin{equation}
	\rho(r)+2p(r)-\tau(r)=\frac{a^2}{4\pi\kappa\,h^{5/2}(r)}>0.
\end{equation}
We conclude, that SEC is fulfilled. Of course, one can check for the validity  other energy conditions quantum theory applies to known matter and fields. Here we just demonstrated that in the case of wormhole spacetime (\ref{wh}) there is at least one energy condition fulfilled, what motivated us to study optical phenomena in this spacetime.


\begin{thebibliography}{99}

\bibitem{Abd-Jur-Ahm-Stu:2016:AstroSpaSci:}
{{Abdujabbarov} A., {Juraev} B., {Ahmedov} B., and {Stuchl{\'\i}k} Z.},
\newblock{Shadow of rotating wormhole in plasma environment},
\newblock{\emph{}Astrophys. and Sp. Sci.}, 361, 7, 226 (2016)

\bibitem{Bam-Stoj:UNI:2021:}
Bambi C. and Stojkovic D.,
\newblock{Astrophysical Wormholes},
\newblock{\emph{Universe}}, \textbf{7}, no.5, 136 (2021)


\bibitem{Bla-Stu:2016:PHYSR4:}
  {{Blaschke} M., {Stuchl{\'\i}k} Z.},
	\newblock{Efficiency of the Keplerian accretion in braneworld Kerr-Newman spacetimes and mining instability of some naked singularity spacetimes},
	\newblock{\emph{ Phys. Rev. D}}, 94, 086006 (2016)

\bibitem{Bla-Sal-Kno:2020:EPJC:}
{{Bl{\'a}zquez-Salcedo} J. L. and {Knoll} C.},
\newblock{Constructing spherically symmetric Einstein-Dirac systems with multiple spinors: Ansatz, wormholes and other analytical solutions},
\newblock{\emph{Europ. Phys. Jour. C}}, 80, 2, 174 (2020)

\bibitem{Bla-Sal-Kno-Rad:2021:PhysRevLet:}
{{Bl{\'a}zquez-Salcedo} J. L., {Knoll} Ch., and {Radu} E.},
\newblock{Traversable Wormholes in Einstein-Dirac-Maxwell Theory},
\newblock{\emph{Phys. Rev. Lett.}}, 126, 101102 (2021)

\bibitem{Bro-Mel-Deh:2007:GenRelGrav:}
{{Bronnikov} K.~A., {Dehnen} H., and {Melnikov} V.~N.}, 39,7, p.973-987 (2007)

\bibitem{Bro-Kon:2020:PHYSR4:}
{{Bronnikov}, K. A. and {Konoplya}, R. A.},
\newblock{Echoes in brane worlds: Ringing at a black hole-wormhole transition},
\newblock{\emph{Phys. Rev. D}}, 101, 064004 (2020)
983-1024
\bibitem{Chur-Stu:2020:CLAQG}
{{Churilova} M.~S. and {Stuchl{\'\i}k}, Z.},
\newblock{Ringing of the regular black-hole/wormhole transition}, 37, 075014 (2020)

\bibitem{Dai-Stoj:PhRvD:2019:}
Dai  D.~C. and Stojkovic D.,
\newblock{Observing a Wormhole},
\newblock{\emph{Phys. Rev. D}}, {\bf 100}, no. 8, 083513 (2019)

\bibitem{Doe-etal:Science:2012}
Doeleman S. S., Fish V. L., Schenck D. E.,  Beaudoin C., 
Blundell R.,  Bower G. C., Broderick A. E.,  Chamberlin R., 
Freund R.,  Friberg P.,  Gurwell M. A.,  Ho P. T. P.,  Honma M.,
 Inoue M.,  Krichbaum T. P. et al., 
 \newblock{Jet-launching structure resolved near the supermassive black hole in M87},
\newblock{\emph{Science}} 338, 355 (2012)

\bibitem{EHT-etal:APJ:2019}
Event Horizon Telescope Collaboration et al.,
\newblock{First M87 Event Horizon Telescope Results. I. The shadow of the
supermassive black hole},
\newblock{\emph{Astrophys. J. Lett.}}, 875, L1 (2019)

\bibitem{Elli:1973:JMathPhys:}
Ellis H. G.,
\newblock{Ether flow through a drainhole: A particle model in general relativity},
\newblock{\emph{ Jour. of Math. Phys.}}, 14, 104 (1973)


\bibitem{Gim-Hor:2009:PhysLetB:}
	{{Gimon}  E. G. and {Ho{\v{r}}ava} P.},
	\newblock{Astrophysical violations of the Kerr bound as a possible signature of string theory},
	\newblock{\emph{ Physics Letters B}}, 672, 3, pp.299-302 (2009)

\bibitem{Gra-Wil:2007:PHYSR4:}
Gravanis E. and Willison S.,
\newblock{"Mass without mass” from thin shells in Gauss-Bonnet gravity},
\newblock{\emph{ Phys. Rev. D}}, 75, 084025 (2007)

\bibitem{GRAVITY:2017}
 {{Gravity Collaboration} et al.},
\newblock{First light for GRAVITY: Phase referencing optical interferometry for the Very Large Telescope Interferometer},
\newblock{\emph{Astron. and Astrophys.}}, 602, A94 (2017)

\bibitem{GravityCol:2018:AA:}
 {{Gravity Collaboration}, {Abuter} R., {Amorim} A., etal},
\newblock{Detection of orbital motions near the last stable circular orbit of the massive black hole SgrA*},
\newblock{\emph{Astron. and Astrophys.}}, 618, L10 (2018)

\bibitem{Har-Lob-Mak-Sus:2013:PHYSR4:}
{{Harko} T., {Lobo} F. S.~N., {Mak} M.~K., and {Sushkov} S. V.},
\newblock{Structure of neutron, quark, and exotic stars in Eddington-inspired Born-Infeld gravity},
\newblock{\emph{Phys. Rev. D}}, 88, 044032 (2013)

\bibitem{Mal-Mil-Pop:2018:arXiv:}
{{Maldacena} J., {Milekhin} A., and {Popov} F.},
\newblock{{Traversable wormholes in four dimensions}},
\newblock{\emph{arXiv e-prints}}, arXiv:1807.04726 (2018)


\bibitem{Mor-Tho:1988:AmJPhys:}
{{Morris}, Michael S. and {Thorne}, Kip S.},
\newblock{Wormholes in spacetime and their use for interstellar travel: A tool for teaching general relativity},
\newblock{\emph{Am. Jour. of Phys.}}, 56, 5, p.395-412 (1988)

\bibitem{Mor-Tho-Yur:1988:PhysRevLet:}
{{Morris} M. S., {Thorne} K. S., and {Yurtsever}, U.},
\newblock{Wormholes, time machines, and the weak energy condition},
\newblock{\emph{Phys. Rev. Lett.}}, 61, 13, p.1446-1449 (1988)


\bibitem{Nov-Tho:1973:BlaHol:}
{{Novikov}, I.~D. and {Thorne}, K.~S.},
\newblock{Astrophysics of black holes},
\newblock{ in \emph{Black Holes (Les Astres Occlus)}}, p.343-450 (1973)


\bibitem{Paul-etal:2020:JCAP:}
Paul S., Shaikh R.,  Banerjee P., and  Sarkar T., 
\newblock{Observational signatures of wormholes with thin accretion disks},
\newblock{\emph{Jour. of Cosmo. and Astropart. Phys.}}, 03, 055, 2020

\bibitem{Sim-Vie:JCAP:2019:}
Simpson A. and Visser M., Black-bounce to traversable wormhole, \emph{JCAP} 1902, 042 (2019)


\bibitem{Stu-Bla-Sche:2017:PHYSR4:}
 {{Stuchl{\'\i}k} Z., {Blaschke} M., and {Schee}, J.},
\newblock{Particle collisions and optical effects in the mining Kerr-Newman spacetimes},
\newblock{\emph{Phys. Rev. D}}, 96, 104050 (2017)

\bibitem{Stu-Hle:1999:PHYSR4:}
{{Stuchl{\'\i}k}, Z. and {Hled{\'\i}k}, S.},
\newblock{Some properties of the Schwarzschild-de Sitter and Schwarzschild-anti-de Sitter spacetimes},
\newblock{\emph{Phys. Rev. D}}, 1999, 60, 044006 (1999)

\bibitem{Stu-Hle-Tru:2011:CLAQG:}
 {{Stuchl{\'\i}k} Z.,  {Hled{\'\i}k} S., and {Truparov{\'a}} K.},
\newblock{Evolution of Kerr superspinars due to accretion counterrotating thin discs},
\newblock{\emph{ Class. and Quant. Grav.}}, 28, 155017 (2011)
 
 \bibitem{Stu-Sche:2010:CLAQG:}
   {{Stuchl{\'\i}k} Z. and {Schee} J.},
	\newblock{Appearance of Keplerian discs orbiting Kerr superspinars},
 	\newblock{\emph{ Class. and Quant. Grav.}}, 27, 215017 (2010)
 
\bibitem{Stu-Sche:2012:CLAQG:}
	{{Stuchl{\'\i}k} Z. and {Schee} J.},
\newblock{Observational phenomena related to primordial Kerr superspinars},
\newblock{\emph{ Class. and Quan. Grav.}}, 29, 065002 (2012)

\bibitem{Stu-Sche:2013:CLAQG:}
 {{Stuchl{\'\i}k} Z. and {Schee} J.},
	\newblock{Ultra-high-energy collisions in the superspinning Kerr geometry},
	\newblock{\emph{Classical and Quantum Gravity}}, 30, 075012 (2013)

\bibitem{Sche-Stu:2009:GenRelGrav}
{{Schee} J. and {Stuchl{\'\i}k} Z.},
\newblock{Profiles of emission lines generated by rings orbiting braneworld Kerr black holes},
\newblock{\emph{Gen. Rel. and Grav.}}, 41, 8, p.1795-1818 (2009)


\bibitem{Sche-Stu:2009:IJMPD:}
{{Schee} J. and {Stuchl{\'\i}k} Z.},
\newblock{Optical Phenomena in the Field of Braneworld Kerr Black Holes},
\newblock{\emph{Int. Jour. of Mod. Phys. D}}, 18, 6, p.983-1024  (2009)

\bibitem{Stu:1980:BAC:}
 {{Stuchlik}, Z.},
 \newblock{Equatorial Circular Orbits and the Motion of the Shell of Dust in the Field of a Rotating Naked Singularity},
\newblock{\emph{Bull. of the Astron. Inst. of Czechoslovakia}}, 31, 129 (1980)

\bibitem{stu-etal:2020:UNIV}
Stuchl\'{i}k Z., Kolo\v{s} M., Ková\v{r} J., et al.,
\newblock{Influence of Cosmic Repulsion and Magnetic Fields on Accretion Disks Rotating around Kerr Black Holes},
\newblock{\emph{Universe}}, 6, 2, 26 (2020)

\bibitem{Svi-Tah:2020:EPJC:}
Svítek O. and Tahamtan T.,
\newblock{Nonsymmetric dynamical thinshell wormhole in Robinson-Trautman class},
\newblock{\emph{ Eur. Phys. J. C}}, 78, 167 (2018)

\bibitem{Tho-Page:1974:ApJ:}
{{Page}, Don N. and {Thorne}, Kip S.},
\newblock{Disk-Accretion onto a Black Hole. Time-Averaged Structure of Accretion Disk},
\newblock{\emph{Astrophys J.}}, 191, p.499-506 (1974)

\bibitem{Vis:1989:NuclPhysB:}
 {{Visser} M.},
 \newblock{Traversable wormholes from surgically modified Schwarzschild spacetimes},
 \newblock{\emph{Nuc. Phys. B}}, 328, 1, p.203-212 (1989)

\bibitem{Poi-Vis:1995:PHYSR4:}
 {{Poisson} E. and {Visser} M.},
\newblock{Thin-shell wormholes: Linearization stability},
\newblock{\emph{Phsy. Rev. D}}, 52, 12, p.7318-7321 (1995)

\bibitem{tur-etal:2020:ApJ}
 {{Tursunov} A., {Zaja{\v{c}}ek} M.,  {Eckart} A., {Kolo{\v{s}}} M., {Britzen} S., {Stuchl{\'\i}k} Z., {Czerny} B., and {Karas} V.},
\newblock{Effect of Electromagnetic Interaction on Galactic Center Flare Components},
\newblock{\emph{Astroiphys. J.}}, 897, 1, 99 (2020)

\bibitem{Wie-Hor-Vin-Abr:2020:PHYSR4:}
{{Wielgus} M., {Hor{\'a}k} J., {Vincent} F., and {Abramowicz} M.},
\newblock{Reflection-asymmetric wormholes and their double shadows},
\newblock{\emph{Phys. Rev. D}}, 102, 084044 (2020)

\bibitem{Wiel-etal:ApJ:2020}
Wielgus M., Akiyama K., Blackburn L.,  Chan C.-K., 
Dexter J.,  Doeleman S. S.,  Fish V. L.,  Issaoun S.,
Johnson  M. D.,  Krichbaum T. P.,  Lu R.-S.,  Pesce D. W., 
Wong G. N.,  Bower G. C.,  Broderick A. E. et al., 
\newblock{Monitoring the morphology of M87* in 2009-2017 with the EHT},
\newblock{{Astrophys. J.}}, 901, 67 (2020)


\end{thebibliography}
\end{document}